\documentclass{jfm}
\usepackage{graphicx}
\usepackage{epstopdf, epsfig}
\usepackage{amssymb}
\usepackage{amsmath}

\usepackage{color,soul} 
\graphicspath{{/figures/}} 
\DeclareGraphicsExtensions{.eps,.pdf,.png,.jpg,.jpeg,.xcf}
\usepackage{subcaption}
\usepackage[justification=centering]{caption}
\usepackage{placeins}

\usepackage{adjustbox}
\usepackage{multirow}

\shorttitle{Three-dimensional gap-flow and VIV interference}
\shortauthor{B. Liu and R. K. Jaiman}

\title{Dynamics of gap flow interference in a vibrating side-by-side arrangement of two circular cylinders at moderate Reynolds number}

\author{B. Liu
  \and  R. K. Jaiman
   \corresp{\email{mperkj@nus.edu.sg}}
  }

\affiliation{Department of Mechanical Engineering, National University of Singapore, Singapore.}

\begin{document}

\maketitle

\begin{abstract}
In this work, the coupled dynamics of the gap flow and the vortex-induced vibration (VIV) on a side-by-side (SBS) arrangement of two circular cylinders is numerically investigated at moderate Reynolds number $ 100 \le Re \le 800$. The influence of VIV is incorporated by allowing one of the cylinders to freely vibrate in the transverse direction, which is termed as a vibrating side-by-side (VSBS) arrangement. 
A comparative analysis is performed between the stationary side-by-side (SSBS) and the VSBS arrangements to investigate the characteristics of the complex coupling between the VIV and the gap flow in a three-dimensional flow. The results are also contrasted against the isolated stationary and the vibrating configurations without any proximity and gap flow interference. Of particular interest is to establish a relationship between the VIV, the gap flow and the near-wake instability behind bluff bodies. 
We find that the kinematics of the VIV regulates the streamwise vorticity concentration, which accompanies with a recovery of two-dimensional hydrodynamic responses at the peak lock-in stage. On the other hand, the near-wake instability may develop around an in-determinant two-dimensional streamline saddle point along the interfaces of a pair of imbalanced counter-signed vorticity clusters. The vorticity concentration difference of adjacent vorticity clusters and the fluid momentum are closely interlinked with the prominence of streamwise vortical structures. 
In both SSBS and VSBS arrangements, the flip-flopping frequency is significantly low for the three-dimensional flow, except at the VIV lock-in stage for the VSBS arrangement. A quasi-stable deflected gap flow regime with negligible spanwise hydrodynamic (i.e., two-dimensional) response is found at the peak lock-in stage of VSBS arrangements. Owing to the gap-flow proximity interference, a high streamwise vorticity concentration is observed in its narrow near-wake region. The increase of the gap-flow proximity interference tends to stabilize the VIV lock-in, which eventually amplifies the spanwise correlation length and weakens the streamwise vortical structures. We employ the dynamic mode decomposition procedure to characterize the space-time evolution of the primary vortex wake.   
\end{abstract}


\section{Introduction}\label{sec:intro}
The canonical side-by-side arrangements of circular cylinders are common and 
have a wide range of applications 
in various fields such as the offshore, the wind and the aerospace engineering. 
In addition to their great practical relevance in the engineering applications, 
a side-by-side system has a fundamental value 
due to the richness of nonlinear flow physics associated with the near-wake dynamics and the vortex-to-vortex interactions. There is a considerable difference between the flow dynamics of an isolated cylinder and the multiple-cylinder arrangements. Many comprehensive investigations, e.g., \cite{Zdravkovich1987Jofas,sumner1999,sumner2000,lin2002,Sumner2010JoFaS}, were performed to understand and describe the mutual flow interference in the basic canonical multi-body systems, in which the importance of the wake and proximity interference was discussed. Among them, the flip-flop of gap flow in SBS arrangements has attracted the attention among researchers. Different from the other fundamental flow regimes in a two-dimensional laminar flow, the bi-stable character and chaos-like fluctuation of the flip flop have intrigued the research community over the past few decades.
The flip flop was reported in many experimental works, e.g., 
\cite{ishigai1972,bearman1973interaction,williamson1985evolution,kim1988investigation}. The flip flop was interpreted by \cite{kim1988investigation} as a dynamical system with a bi-stable state, a deflected gap flow regime. \cite{kim1988investigation} reported that the gap flow intermittently switched its direction at a time scale which was few orders of the magnitude greater than the shedding frequency of the primary vortices. Besides the aforementioned experimental works in literature, the flip flop was also observed within a narrow gap ratio range from 0.3 to 1.25 in a two-dimensional laminar flow from various numerical investigations in literature, e.g., \cite{kang2003characteristics,Agrawal2006Cf}. In the deflected gap-flow regime, the narrow near-wake region incorporates an enhanced vortex-wake interaction, which results in a higher vortex shedding frequency and mean drag coefficient value. As a result, the vortex shedding frequency of each cylinder dynamically changes with the gap-flow kinematics in time domain. 

The origin of the flip flop was discussed by many authors. 
\cite{alam2005investigation} reported that a perfect symmetric structure geometry was a critical condition which originated intermittent switching of the gap flow. However, the gap-flow flip flop was also observed in the asymmetric VSBS arrangements from \cite{lbjrk2016pof}. \cite{ishigai1972} also considered the Coanda effect as the origin of the gap-flow flip flop. Nonetheless the flip flop was found in the near-wake region behind a pair of side-by-side flat plates by \cite{bearman1973interaction,williamson1985evolution}. \cite{Peschard1996PRL} modeled the dynamics of the deflected gap-flow regime through a system of two coupled Landau oscillators. The study illustrated that the stable deflected gap-flow regime and the flip flop were formed by different mechanisms. Following the earlier studies, \cite{carini2014origin} reported that the flip flop could be explained as a secondary instability through the coupling between Hopf bifurcation (the in-phase vortex synchronization) and the pitchfork bifurcation (the deflected gap flow regime). This finding was subsequently visualized by \cite{lbjrk2016pof} in which the evolution of the flip flop from the interaction of these two bifurcations were shown in a series of streamline plots as the Reynolds number increased in the laminar flow regime. The exact instants of the flip flop and the instantaneous vortex shedding frequencies were visualized via the Hilbert-Huang Transform (HHT) technique by \cite{huang2014hilbert}. \cite{lbjrk2016pof} also reported that the flip flop was suppressed at the lock-in stage of VIV in VSBS arrangements, in which the time-averaged streamwise velocity profile of the gap flow became asymmetric. On the other hand, the lock-in range with respect to the reduced velocity became narrow, because of the enhanced vortex-to-vortex interaction induced from the gap-flow proximity interference. A topological description based on critical points  has also shown that the in-determinant two-dimensional saddle-point regions intermittently appeared in the middle path of the gap flow in the SSBS arrangements and contributed 
to the near-wake instability. These critical points, where the velocity is zero and the streamline slope is indeterminate, contribute to the shear 
stress and can provide some understanding into the three-dimensionality of 
the flow behind bluff bodies \citep{Zhou1994JoFM}.

A previous study on SBS arrangements by \cite{lbjrk2016pof} was performed for the two-dimensional laminar flow. Herein, the primary focus is to investigate the influence of the three-dimensional (3D) flow characteristics on the submerged SBS system. 
While the complete three-dimensional 
flow field is important for interpreting the topology 
of flow patterns around the critical points, the three-dimensional
information is generally difficult to extract from experiments.
In a 3D flow, the formation of the ribs-like streamwise vortical structures connecting the spanwise K\'arm\'an vortices are one of the characteristic flow features of the organized motion. \cite{Williamson1996ARoFM} reported that there were two types of instabilities in the flow transition, the mode-A and the mode-B. The mode-A instability was believed to be associated with the waviness of the primary K\'arm\'an vortices induced by the elliptic instability. 
The counter-rotating streamwise vortices were formed in the high-strain region between the main spanwise vortices and somewhat irregular with varying size, shape and spatial organizations. The conversion of 
the spanwise vorticity from the K\'arm\'an vortex cores into the streamwise vortices is a result from the elliptic instability and was the central for the mode-A instability of bluff body wakes.
\cite{Williamson1996ARoFM} further mentioned that the onset of mode-A instability exhibited a hysteretic discontinuity of Strouhal number $St$-$Re$ relationship with a spanwise correlation wavelength about $3 \sim 4D$. Mode A naturally triggered a vortex dislocation in the wake of a stationary isolated circular cylinder during the wake transition. Whereas the mode-B with a spanwise wavelength about unity diameter experienced a non-hysteretic transition from the mode A. A relatively high shedding frequency occurred with more organized three-dimensional state of the mode-B and vice versa.

In the context of three-dimensionality associated with the elliptic instability, the hyperbolic critical points were investigated by \cite{Kerswell2002Arofm}, \cite{LeDizes2002JoFM} and \cite{Meunier2005CRP}. From the topological theory of separated flows, the two-dimensional streamline orbitals resemble hyperbolas around a hyperbolic critical point, where its central velocity magnitude is zero and all eigenvalues of velocity gradient have the nonzero real parts. The hyperbolic critical points in the fluid domain had been previously reported as an unstable factor by \cite{lifschitz1991local} and \cite{leblanc1997stability}, where the maximal perturbation growth was found precisely around these hyperbolic points near the vortex wake. 
One of the primary focus in the present study is to interlink the characteristics between the near-wake instability, the vortex wake interaction and the fluid momentum. In spite of the above investigations, many aspects of the proximity interference and the wake interference from the gap flow remain largely unexplored. 

Here, we will present unsteady results from well-resolved numerical simulations of two circular cylinders of SBS arrangements in 3D flow at moderate Reynolds numbers. As another step further, the primary focus is to explore the spanwise characteristics of the gap-flow and VIV kinematics at 3D flow through a systematic numerical analysis using the recently developed variational finite element solver for fluid-structure interaction \citep{jaiman_caf2016,Jaiman2016CMiAMaE}. Of particular interest is to answer the following questions: How do the VIV kinematics, the gap flow instability and the hydrodynamic responses accommodate themselves in a 3D flow? How does the gap-flow kinematics influence the 3D flow features? How does the spanwise correlation response to the cylinder's kinematics and the gap flow instability? In most engineering applications, multiple-body structures subjected to the subtle proximity and flow interference are much more common. The in-depth analysis of such complex nonlinear coupling is essential in various engineering design and operations in which the SBS submerged arrangements are prevailing and subjected to the influence from a 3D flow. In particular, the incorporation of VIV in the investigation is crucial to reflect the practical aspects of structural motion and the interference on the hydrodynamic forces.  

The manuscript is organized as follows.
The numerical formulation, the problem setup and the verification are briefly presented in Section~\ref{sec:Form_Val}. Following that, the regulation effect of VIV kinematics on the three-dimensional feature is discussed in Section~\ref{sec:3D_viv}. We next investigate the mutual interference between the 3D flow and the gap-flow kinematics in SSBS arrangements in Section~\ref{sec:3D_gap_flow}. The triple-coupling flow regime characteristics of the VSBS arrangements are presented in Section~\ref{sec:3D_VIV_gap}. The primary focus is on the VIV lock-in phenomenon in VSBS arrangements for a range of reduced velocity at representative gap ratios in the flip-flopping regime. We investigate the flow physics of the flip-flop phenomenon and the VIV kinematics in terms of the wake topology, the 
response characteristics, the force components, the phase and frequency characteristics.  The spanwise correlation and the three-dimensional modal analysis are discussed in Section~\ref{lamda} and Section~\ref{3D_DMD_VD},
respectively. 
Finally, we present the concluding remarks in Section~\ref{sec:conclusion}.

\section{Numerical methodology}\label{sec:Form_Val}
\begin{table*}
\setlength{\tabcolsep}{12pt}
\renewcommand{\arraystretch}{2}
\begin{center}
\begin{tabular}{l l l}
\hline
Parameter & Value & Description\\
\hline
$l^* = L/D$ & $5$-$10$ & Dimensionless spanwise length\\
$g^*=g/D$ & $0.3$-$3$ & Gap ratio\\ 
$m^*=\frac{4 m}{ \rho \pi D^2 l }$ & $10$ & Mass ratio\\ 
$\zeta = \frac{C}{4 \pi m f_{n}} $ & $0.01$ & Damping ratio\\
$U_{r} = \frac{U}{f_{n} D}$ & $0$-$10$ & Reduced velocity\\
$Re = \frac{\rho U D}{\mu}$ & $100$-$800$ & Reynolds number\\
\hline
\end{tabular}
\caption{Dimensionless parameters. Here $l$, $g$, $D$, $m$, $\rho$, $f_{n}$, $U$ and $\mu$ are respectively the spanwise distance, the gap distance, the cylinder diameter, the cylinder mass, the structural frequency, the free-stream velocity and the dynamic viscosity. }
\label{tab:para}

\end{center}
\end{table*}

\begin{table*}
\setlength{\tabcolsep}{12pt}
\renewcommand{\arraystretch}{2}
\begin{center}
\begin{tabular}{l l l}
\hline
Parameter & Description\\
\hline
$St=\frac{f_{vs}D}{U}$ & Strouhal number\\
$A_{y}^{max}=\sqrt{2}A^{rms}_y$ & Dimensionless transverse displacement\\
$\Delta \phi = \phi_{A_y}-\phi_{C_l}$ & Phase angle difference\\
$\lambda^*=\lambda/D$ & Dimensionless spanwise wavelength\\
$\lambda = \frac{1}{n} \sum \limits^{n}_{t_i = 1} \lambda^m (t_i)$ & Averaged spanwise wavelength\\ 
$C_{l} = \frac{F_x}{\frac{1}{2} \rho U^2 D l}$ & Lift coefficient\\
$C_{d} = \frac{F_y}{\frac{1}{2} \rho U^2 D l}$ & Drag coefficient\\
$C_{z} = \frac{F_z}{\frac{1}{2} \rho U^2 D l}$ & Spanwise hydrodynamic coefficient\\
$C_{e} = \int_{T} C_l v^* d\tau$ & Energy transfer coefficient\\
\hline
\end{tabular}
\caption{Derived dimensionless quantities for post-processing. Here $f_{vs}$, $A^{rms}_y$, $\phi_{A_y}$, $\phi_{C_l}$, $F_x$, $F_y$, $F_z$ and $v^* = v/U$ are respectively the vortex shedding frequency, the root-mean-square transverse vibration amplitude, the instantaneous phase angle of $A_y$, the instantaneous phase angle of $C_l$, the streamwise hydrodynamic force, the transverse hydrodynamic force, the spanwise hydrodynamic force and dimensionless transverse velocity of a vibrating cylinder. $\lambda^m (t_i)$ is the mean value of instantaneous spanwise wavelength along cylinder span at time $t_i$ and $n$ is the number of sample locations in a shedding cycle.}
\label{tab:quan}
\end{center}
\end{table*}
\subsection{Formulation of coupled fluid-structure system}\label{sec:Form}
A Petrov-Galerkin finite element formulation is employed to investigate the fluid-structure interaction problem where the body interface is tracked accurately by the arbitrary Lagrangian-Eulerian technique. The traction and the velocity continuity conditions are imposed on the body-conforming fluid-solid interface via the non-linear iterative force correction procedure \citep{jaiman_caf2016,Jaiman2016CMiAMaE}. 
The coupling scheme relies on a dynamic interface force sequence parameter to stabilize the coupled fluid-structure dynamics with strong inertial effects of incompressible flow on immersed solid bodies.
The temporal discretizations of both the fluid and structural equations are formulated in the variational generalized-$\alpha$ framework and the systems of linear equations are solved via the Generalized Minimal Residual (GMRES) solver \citep{jaiman_ficf2015,Jaiman2016CMiAMaE}. The validation results of the two- and three-dimensional simulations are reported in \cite{lbjrk2016pof}, \cite{ravi2016} and \cite{li2016vortex}. These validation results and convergence investigations provide the validity and reliability of the fluid-structure solver to simulate the gap flow and VIV interaction. The corresponding dimensionless simulation parameters and the post-processing quantities are listed in table~\ref{tab:para} and table~\ref{tab:quan}, respectively.

\begin{figure} \centering
	\begin{subfigure}[b]{0.6\textwidth}	
	\centering
	\hspace{-25pt}\includegraphics[trim=0.2cm 5cm 1.2cm 5cm,scale=0.36,clip]{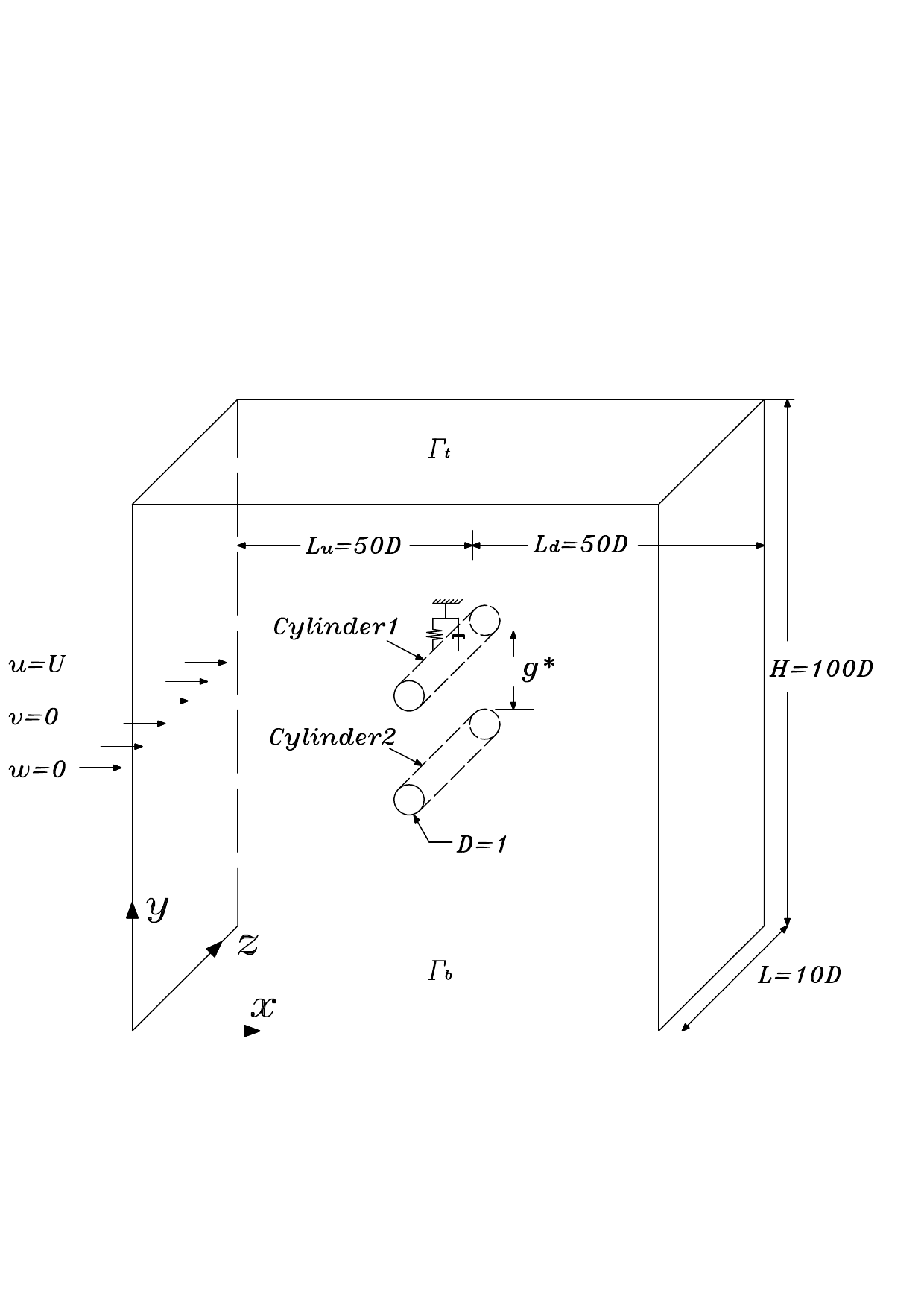}
    \caption{$\qquad$ $\qquad$}
	\label{fig:setup_3D}
	\end{subfigure}%
	\begin{subfigure}[b]{0.5\textwidth}
	\centering
	\hspace{-25pt}\includegraphics[trim=1cm 0.1cm 0.1cm 0.1cm,scale=0.42,clip]{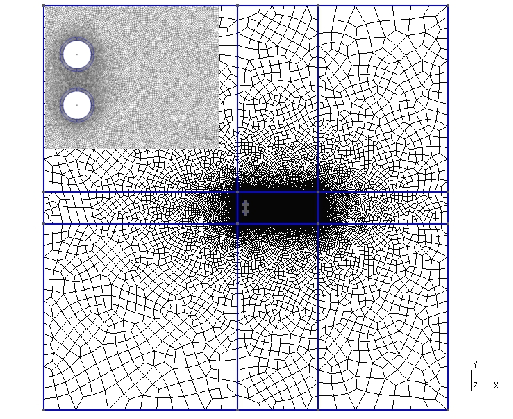}
    \caption{$\qquad$ $\qquad$}
	\label{fig:mesh}
    \end{subfigure}
\caption{Three-dimensional computational setup of SBS arrangements: (a) schematic diagram of the fluid domain and the boundary conditions; 
 (b) representative unstructured mesh distribution in $(x,y)$-plane at $g^*=0.8$. Here \emph{Cylinder1} is a freely transverse vibrating cylinder 
 and \emph{Cylinder2} is stationary in the VSBS arrangements. }
\label{fig:scheme}
\end{figure}

\begin{table*}
	\setlength{\tabcolsep}{12pt}
	\renewcommand{\arraystretch}{2}
	\begin{center}
		\def\arraystretch{1.0}
		\begin{tabular}{c p{1.5cm} p{1.5cm} p{1.5cm} p{1.5cm}}
			\hline
			Spanwise resolution & $C^{mean}_{d}$ &  $C^{rms}_{l}$ & $St$ \\
			\hline
			$\Delta z=0.4$ & 1.341 (12.2$\%$) & 0.763 (118.6$\%$) & 0.2197  (7.1$\%$)\\
			$\Delta z=0.15$ & 1.196 (0.8$\%$) & 0.358 (2.5$\%$) & 0.2051  (0.0$\%$)\\
			$\Delta z=0.075$ & 1.195 & 0.349 & 0.2051 \\
			\hline
		\end{tabular}
	\end{center}
	\caption{Convergence of the global flow quantities at different spanwise mesh resolutions for a stationary isolated circular cylinder at $Re=500$ and $l^*=10$}
	\label{tab:mesh_con}
\end{table*}
\begin{table*}
	\setlength{\tabcolsep}{12pt}
	\renewcommand{\arraystretch}{2}
	\begin{center}
		\def\arraystretch{1.0}
		\begin{tabular}{c p{1.5cm} p{1.5cm} p{1.5cm} p{1.5cm}}
			\hline
			$x$-$y$ plane element Number & $C^{mean}_{d}$ &  $C^{rms}_{l}$ & $St$ \\
			\hline
			$50$ $\times$ $10^3$ & 1.29 (7.9$\%$) & 0.637 (80$\%$) & 0.2051  (0.0$\%$)\\
			$81$ $\times$ $10^3$ & 1.196 (0.5$\%$) & 0.358 (1.1$\%$) & 0.2051  (0.0$\%$)\\
			$110$ $\times$ $10^3$ & 1.1954 & 0.354 & 0.2051 \\
			\hline
		\end{tabular}
	\end{center}
	\caption{Convergence of the global flow quantities at different $x$-$y$ plane mesh for a stationary isolated circular cylinder at $Re=500$, $\Delta z = 0.15$ and $l^*=10$}
	\label{tab:2Dmesh_con}
\end{table*}
\subsection{Dynamic mode decomposition}\label{sec:dmd}
While the proper orthogonal decomposition (POD) modes may not necessarily provide 
a description of a dynamically evolving flow driven by a momentum input, 
the dynamic modal decomposition allows to extract the dominant spatial and temporal 
information about the flow \citep{Schmid2010JoFM}.
Therefore, we employ the dynamic modal decomposition to fit a discrete-time linear 
system to a set of snapshots from three-dimensional wake data. 
The goal of the DMD technique is to approximate the system on a low dimensional subspace and 
to construct a set of approximated eigenvectors 
and eigenvalues to identify the spatial and temporal modes. 
For more robust and effective implementation of the DMD technique, 
the modal amplitude ($\alpha$) is computed through the best-fit 
between the linearized modes and original snapshot data in a 
least-squares sense \citep{Jovanovic2014PoF1}. 
The amplitude can be physically interpreted as the strength of a particular mode to 
the dynamical response of a flow system. This analysis 
is useful to examine the nonlinear dynamical behaviour with evolving frequencies. 
Hence one is able to analyze a particular characteristics or mechanism 
through an appropriate selection of modes based on their spatial distribution, 
frequency and growth or decay rate. Among various DMD algorithms, the singular value 
decomposition (SVD)-based DMD is the most popular and robust algorithm 
against round-off errors.  In literature \citep{Schmid2010JoFM,Schmid2011TaCFD,lbjrk2016pof}, 
DMD technique was well applied in fluid dynamics problems, from which the decomposed DMD modes 
represent the fluid field characteristics in the space, the time and the frequency domains.   
Herein, the focus is on the application of one of the variants of DMD technique, sparsity-promoting dynamic mode decomposition (SP-DMD) by \cite{Jovanovic2014PoF1}, to analyze 
the near-wake stability and to decompose the complex flow dynamics in 
the near-wake region behind the SBS system. Details of the SP-DMD formulation
can be found in \cite{Jovanovic2014PoF1}.
\subsection{Problem setup and verification}\label{sec:Val}
The problem setup for the 3D flow analysis is a spanwise extension of the two-dimensional setup implemented in \cite{lbjrk2016pof}, where the upstream distance, the downstream distance and the overall height of the fluid domain are respectively $50D$, $50D$ and $100D$. A schematic diagram of the three-dimensional SBS arrangement is shown in figure \ref{fig:setup_3D}. The traction-free boundary conditions are respectively implemented along the domain boundaries $\Gamma_t$, $\Gamma_b$ and $\Gamma_o$.  The top cylinder, \emph{Cylinder1}, is elastically-mounted in the transverse direction for the VSBS arrangements. The blockage ratio is taken as $2 ~\%$.
A pair of equal diameter $D$ cylinders
is placed in a three-dimensional hexahedron domain, where a uniform free-stream flow with 
velocity $U$ is along the streamwise $x$-axis, 
while the axis of the cylinder is along the spanwise $z$-axis.
A representative $(x,y)$-plane sectional mesh configuration is exhibited in figure~\ref{fig:mesh}. 
Based on the mesh convergence analysis in \cite{lbjrk2016pof}, the spatial discretization error is less than $1~\%$ in the $(x,y)$-plane mesh. For the 3D flow at $Re=500$, the $x$-$y$ sectional mesh is further refined, particularly the mesh within the boundary layer and the near-wake regions. The dimensionless wall distance $y^{+}$ is kept less than one (within the viscous sublayer) for the first layer of the structural mesh around bluff bodies. The incremental ratio of element size from the boundary layer to the near-wake region and far field is less than $1.1$ to reduce the effect of element skewness. Overall, there are approximately $80 \times 10^3$ elements and $120 \times 10^3$ elements on each $x$-$y$ section for the isolated cylinder cases and SBS arrangement cases, respectively.

\begin{table*}
\setlength{\tabcolsep}{12pt}
\renewcommand{\arraystretch}{2}
\begin{center}
\def\arraystretch{1.5}
\begin{tabular}{l p{1cm} p{1cm} p{1cm} p{1cm}}
\hline
&  & $C^{mean}_d$ &  $C^{rms}_l$ & $S_t$\\
\hline
\multirow{4}{*}{Simulation}  & \multicolumn{1}{l}{\cite{zhang1995transition}} & \multicolumn{1}{l}{1.44} & \multicolumn{1}{l}{0.68} & \multicolumn{1}{l}{0.216}\\
& \multicolumn{1}{l}{\cite{persillon1998physical}} & \multicolumn{1}{l}{1.366} & \multicolumn{1}{l}{0.477} & \multicolumn{1}{l}{0.206} \\
 & \multicolumn{1}{l}{\cite{behara2010wake}} &  \multicolumn{1}{l}{1.390} & \multicolumn{1}{l}{0.594} & \multicolumn{1}{l}{0.210}  \\
 & \parbox[t]{5cm}{Present} & \multicolumn{1}{l}{1.26} &\multicolumn{1}{l}{0.5} & \multicolumn{1}{l}{0.205}  \\
\cline{1-5}
\multirow{3}{*}{Experiment} & \multicolumn{1}{l}{Wieselsberger (1921)} & \multicolumn{1}{l}{1.208} & \multicolumn{1}{l}{-} & \multicolumn{1}{l}{-} \\
 & \multicolumn{1}{l}{\cite{williamson1996three}} & \multicolumn{1}{l}{-} & \multicolumn{1}{l}{-} & \multicolumn{1}{l}{0.203} \\
\hline
\end{tabular}
\end{center}
\caption{Comparison of numerical and experimental results for a stationary isolated circular cylinder at $Re = 300$, where $C^{mean}_d$ is the mean drag coefficient, $C^{rms}_l$ is the root-mean-square of lift coefficient fluctuation and $St$ is the Strouhal number.}
\label{tab:3D_val}
\end{table*}
\begin{table*}
\setlength{\tabcolsep}{12pt}
\renewcommand{\arraystretch}{2}
\begin{center}
\def\arraystretch{1.0}
\begin{tabular}{l p{1.5cm} p{2.5cm} p{2.5cm} p{2.5cm}}
\hline
$Re$ & $U_r$ & Simulation \citep{Blackburn2001JoFaS} & Experiment \citep{Blackburn2001JoFaS} &  Present\\
\hline
606.1 & 5.51 & 0.460 & 0.550 & 0.525\\
713.9 & 6.49 & 0.433 & 0.485 & 0.462\\
848.1 & 7.71 & 0.420 & 0.430 & 0.424\\
\hline
\end{tabular}
\end{center}
\caption{Validation of transverse amplitude $A^{max}_{y}$ for a freely transverse vibrating cylinder in three-dimensional flow at $m^*=5.08$ and $\zeta=0.024$}
\label{tab:3D_viv}
\end{table*}

The spanwise length is taken as $l^*=10D$, based on the aspect ratio analysis in the numerical simulations from \cite{lei2001spanwise} and the experiments from \cite{Szepessy1992JoFM}. A periodic boundary condition is employed at the ends of the cylinder span to eliminate the end-plate effect. 
The mesh convergence study along the $z$-axis is shown in table~\ref{tab:mesh_con}. The spanwise resolution $\Delta z =0.15$ is chosen such that the spanwise spatial discretization error is controlled within $2.5\%$ while maintaining the computational efficiency for our parametric study. Furthermore, the $x$-$y$ plane mesh convergence analysis in table~\ref{tab:2Dmesh_con} shows that the spatial discretization error is within $1 \%$ at chosen $x$-$y$ plane mesh resolution $81$ $\times$ $10^3$. 
Since the gap flow instability is the key concern of the present investigation, the majority of the investigations are performed at two representative gap ratios $g^*=0.8$ and $g^*=1.0$. However, the investigations on the boundary circumstances, e.g., around $g^* \approx 0.3$ and $g^* \approx 1.5$ where the gap flow is significantly suppressed and weakened, are still incorporated to facilitate the generality of our analysis. The temporal convergence study of the current numerical solver has been performed in \cite{Jaiman2016CMiAMaE}. The $L_{2}$ norm error was reported at about 1\% at a constant time step $\Delta t=0.05$.
Except stated otherwise, all positions and length scales are normalized by the cylinder diameter $D$, velocities with the free stream velocity $U$, and frequencies with $U/D$.
To validate the numerical formulation in a three-dimensional flow, a comparison of a stationary isolated circular cylinder at $Re=300$ are presented in table~\ref{tab:3D_val}. The comparison of the overall VIV response with previous results from \cite{Blackburn2001JoFaS} are shown in table~\ref{tab:3D_viv}.  
The results are in close agreement with the previous studies, and thus the computational setup is adequate for our investigation.
\begin{figure}
\begin{center}
\centering
\includegraphics[trim=0.1cm 0.1cm 0.1cm 0.1cm,scale=0.32,clip]{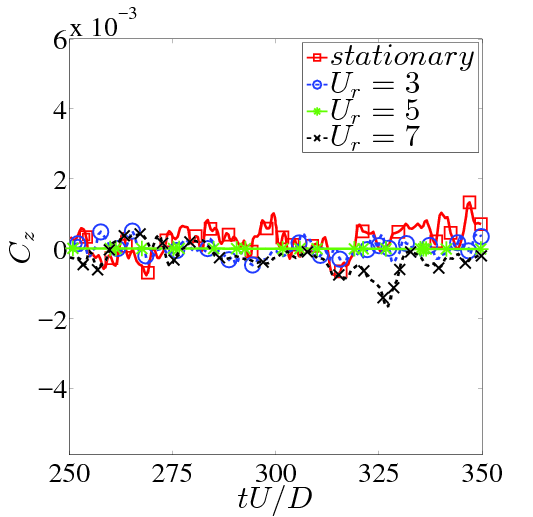}
\end{center}
\caption{Time series of the spanwise force $C_z$ for an isolated cylinder at $Re=500$, $m^*=10$, $\zeta=0.01$ and $U_r \in[0$, $7]$. The VIV kinematics possesses a regulation effect to the spanwise force fluctuation, whereby it is suppressed at the peak lock-in stage $U_r=5$}
 \label{fig:czs_re500}
\end{figure}
A total of eighty-five simulations is performed in the present investigation, 
comprising seven simulations for the validation of the three-dimensional FSI solvers, 
fifty-two cases for the principal investigation of the isolated, SSBS and VSBS arrangements;
and twenty-six two-dimensional cases  to investigate the relationship 
between the near-wake instability, the fluid shearing ratio and the fluid momentum. 
By taking into the consideration of large number of three-dimensional simulations 
and the involved computational resources, the selected time window is constrained at
$tU/D \in[250$, $350]$, in which the fluid flow is already fully-developed for 
the extraction of flow statistics. 
In the selected time window, all fluid features such as the hydrodynamic responses 
and the vibration amplitude, undergo at least twenty cycles. In particular, we are interested 
in the behaviour of the flip flop subjected to the influence of VIV and three-dimensionality 
within a short time window. 

\section{Vortex-induced vibration in three-dimensional flow}\label{sec:3D_viv}
\begin{figure} \centering
	\begin{subfigure}[b]{0.5\textwidth}	
	\centering
	\hspace{-25pt}\includegraphics[trim=0.2cm 0.2cm 0.2cm 0.3cm,scale=0.26,clip]{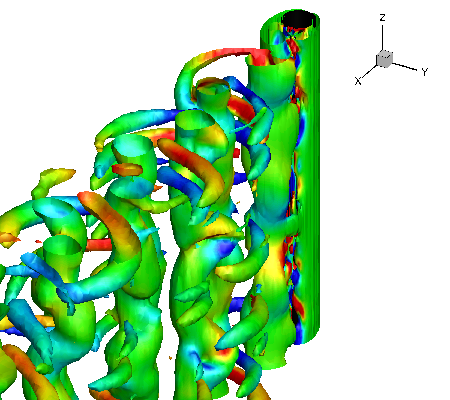}
    \caption{$\qquad$}
	\label{fig:con_re500}
	\end{subfigure}%
	\begin{subfigure}[b]{0.5\textwidth}
	\centering
	\hspace{-25pt}\includegraphics[trim=0.2cm 0.2cm 0.2cm 0.3cm,scale=0.26,clip]{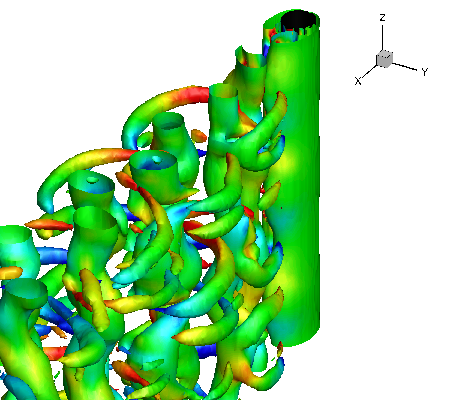}
    \caption{$\qquad$}
    \label{fig:con_re500r3}
    \end{subfigure}
    \begin{subfigure}[b]{0.5\textwidth}
	\centering
	\hspace{-25pt}\includegraphics[trim=0.2cm 0.2cm 0.2cm 0.25cm,scale=0.26,clip]{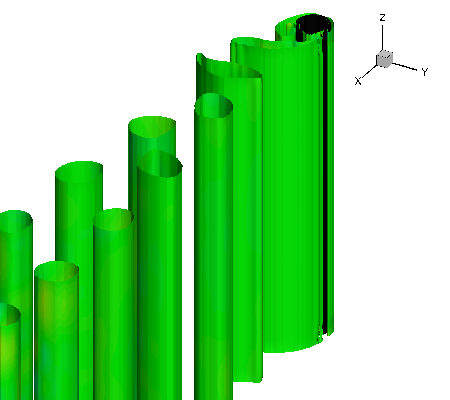}
    \caption{$\qquad$}
    \label{fig:con_re500r5}
    \end{subfigure}%
    \begin{subfigure}[b]{0.5\textwidth}
	\centering
	\hspace{-25pt}\includegraphics[trim=0.2cm 0.2cm 0.2cm 0.25cm,scale=0.26,clip]{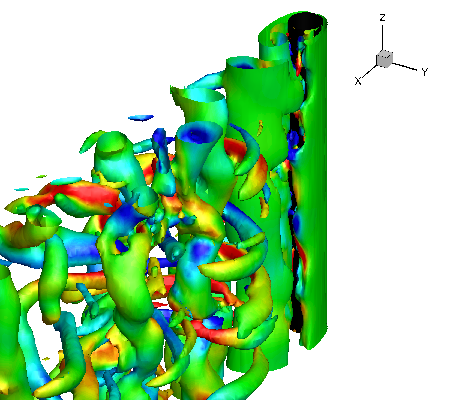}
    \caption{$\qquad$}
    \label{fig:con_re500r7}
    \end{subfigure}
\caption{Instantaneous vortical structures using the Q-criterion for an isolated cylinder at $Re=500$, $Q=0.2$, $\omega_y = \pm 1$ (\emph{contours}) and $tU/D=300$: (a) stationary; $U_r=$ (b) 3; (c) 5; and (d) 7 at $m^*=10$ and $\zeta=0.01$ for a freely transverse vibrating cylinder. The streamwise vorticity clusters vanish at the peak lock-in stage $U_r=5$.}
\label{fig:cons_re500}
\end{figure}
Before we proceed to further investigation on the complex coupling between the 3D flow, the VIV and the gap-flow kinematics in the SBS arrangements,  we systematically examine the interference of the VIV on the 3D flow dynamics.
Similar to the work of \cite{Borazjani2009Jofm}, the spanwise hydrodynamic coefficient $C_z$ is considered to quantify the overall spanwise fluctuation of the hydrodynamic forces induced from the 3D flow. Owing to  the symmetrically-imposed periodic end-wall boundary conditions along the cylinder span, the magnitude of $C_z$ is nearly two orders of magnitudes smaller than $C_d$ and $C_l$. This small value of the spanwise force arises from the intrinsic 3D flow dynamics along the cylinder, and is not related to any numerical errors.
In figure~\ref{fig:czs_re500}, the fluctuation of $C_z$ is found negligible at the lock-in stage $U_r = 5$, in contrast to its counterpart at the off-peak stage.  Such weakening effect of the spanwise force suggests that there exists a particular regulation mechanism which causes a recovery of two-dimensional  (2D) hydrodynamic responses along the cylinder. 
This regulation or stabilization effect is further visualized by the iso-surfaces of the vortical structures using 
a vortex-identification based on $Q$-criterion \citep{hunt}  in figure~\ref{fig:cons_re500}. In agreement with the aforementioned observations, the streamwise vorticity clusters at the lock-in stage are completely invisible in figure~\ref{fig:con_re500r5}. On the other hand, an additional 2D simulation at the identical problem setup is performed. The resultant hydrodynamic responses from the 2D configuration are identical with its 3D counterpart in the present investigation. Because there is no external energy source to perturb the flow field, we can deduce that this recovery of two-dimensionality at the peak lock-in stage is an intrinsic characteristic of the fluid response, instead of an artificially restricted fluid behaviour.
 
 \begin{figure} \centering
 	\begin{subfigure}[b]{0.5\textwidth}	
 	\centering
 	\hspace{-25pt}\includegraphics[trim=0.1cm 0.2cm 0.15cm 0.2cm,scale=0.125,clip]{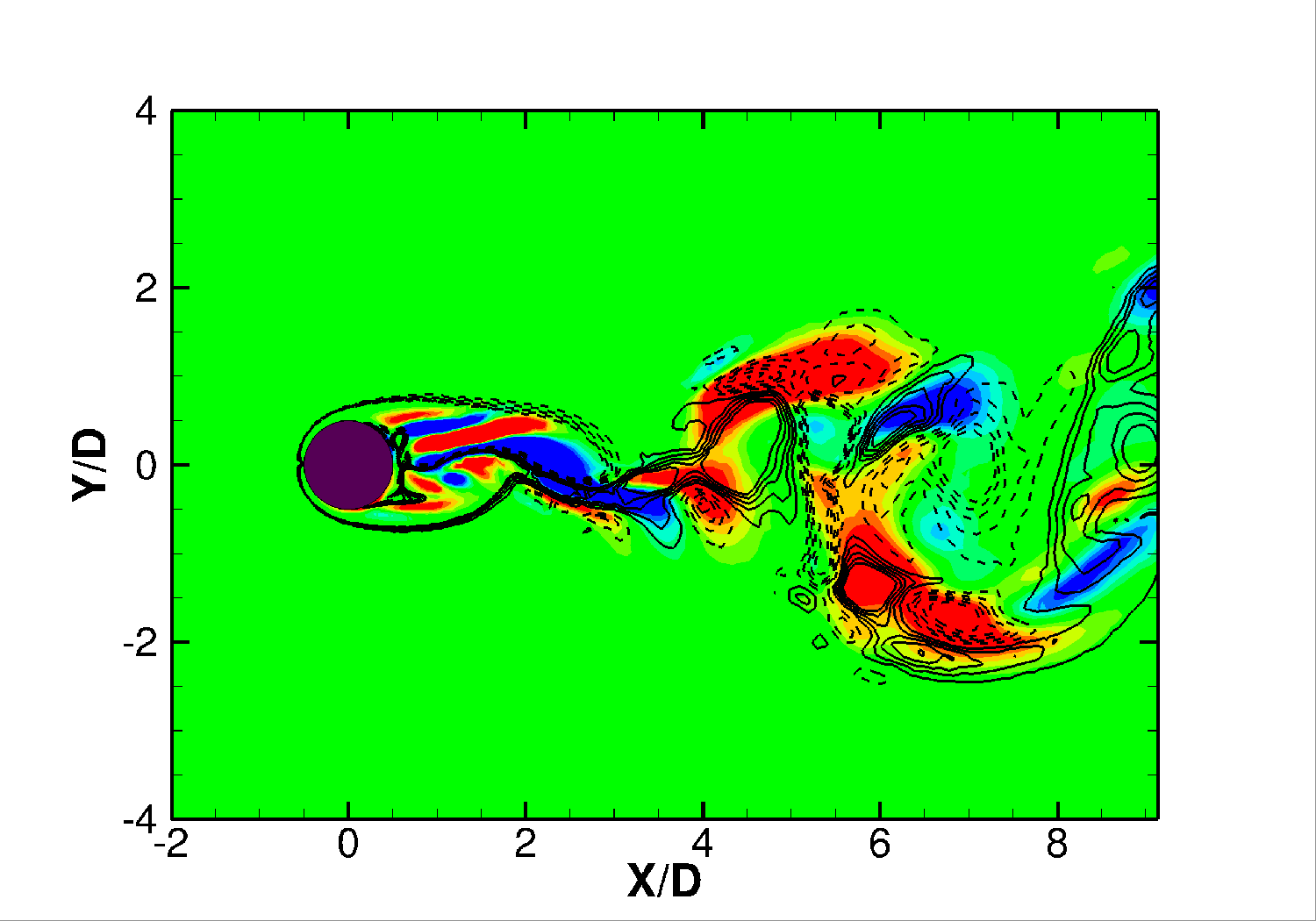}
     \caption{• $\qquad$}
 	\label{fig:iso_z4_wzy}
 	\end{subfigure}%
 	\begin{subfigure}[b]{0.5\textwidth}
 	\centering
 	\hspace{-25pt}\includegraphics[trim=0.1cm 0.2cm 0.15cm 0.2cm,scale=0.125,clip]{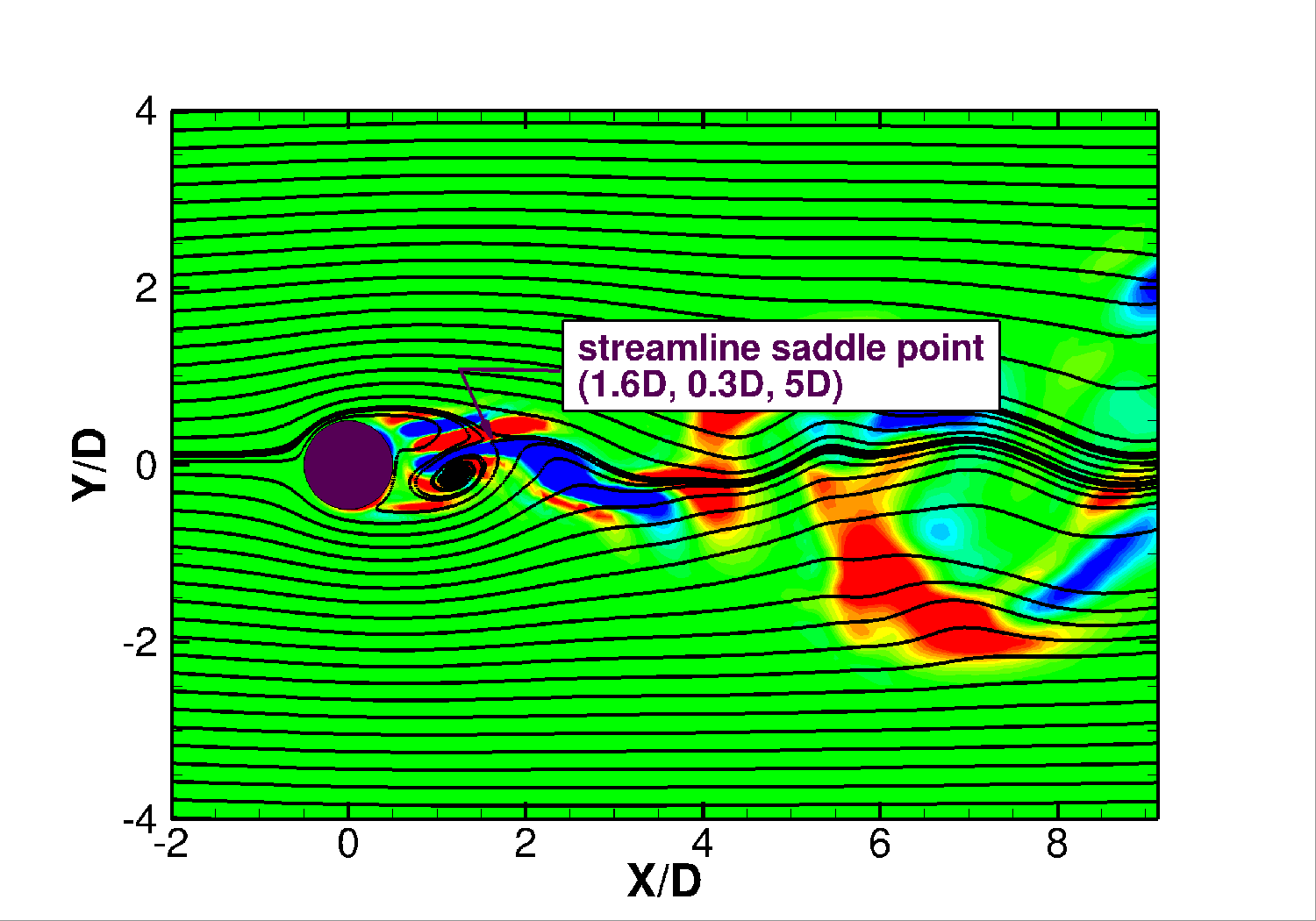}
     \caption{• $\qquad$}
     \label{fig:iso_z4_swy}
     \end{subfigure}
     \begin{subfigure}[b]{0.5\textwidth}	
 	\centering
 	\hspace{-25pt}\includegraphics[trim=0.1cm 0.2cm 0.15cm 0.2cm,scale=0.125,clip]{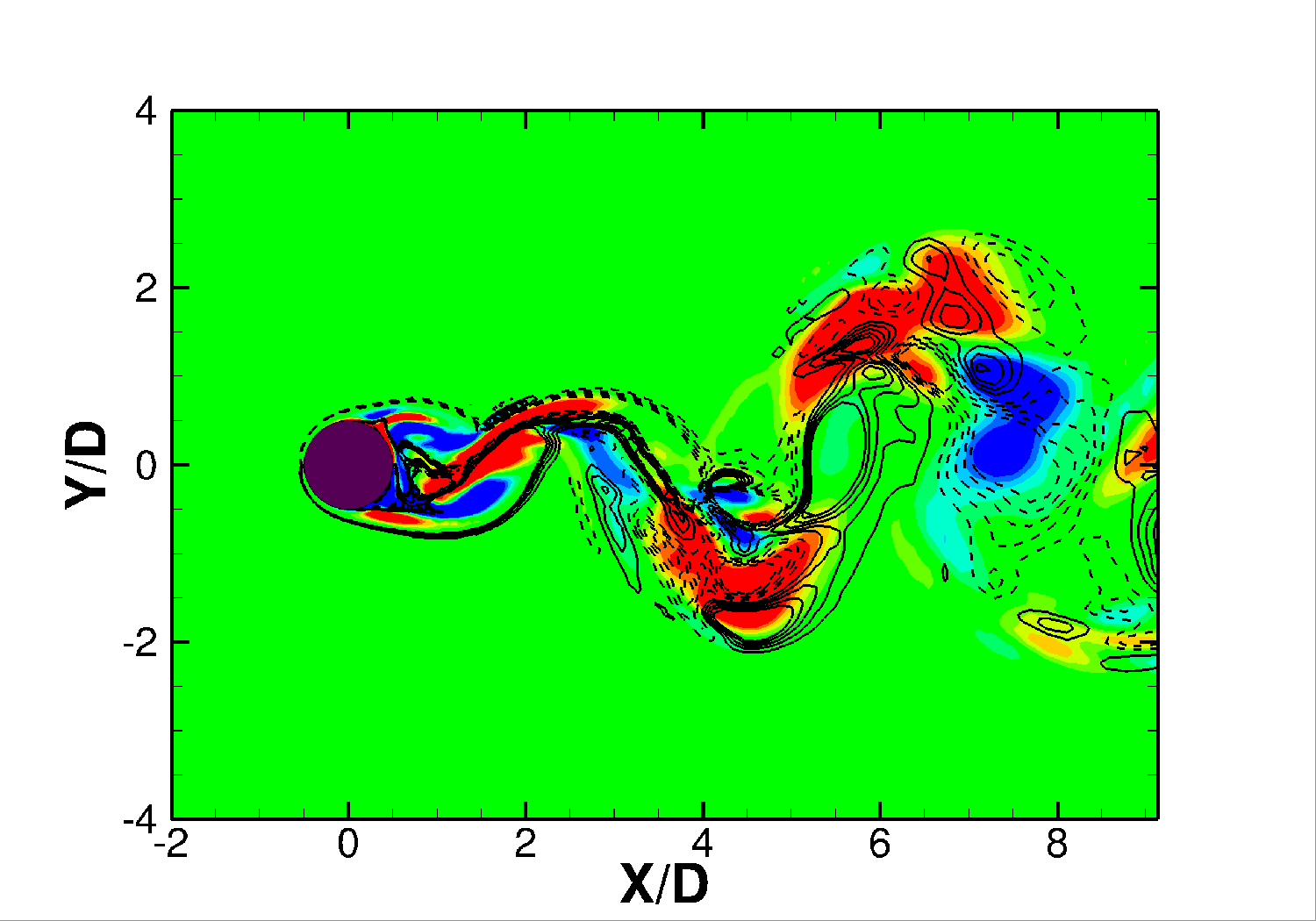}
     \caption{• $\qquad$}
 	\label{fig:iso_z8_wzy}
 	\end{subfigure}%
 	\begin{subfigure}[b]{0.5\textwidth}
 	\centering
 	\hspace{-25pt}\includegraphics[trim=0.1cm 0.2cm 0.15cm 0.2cm,scale=0.125,clip]{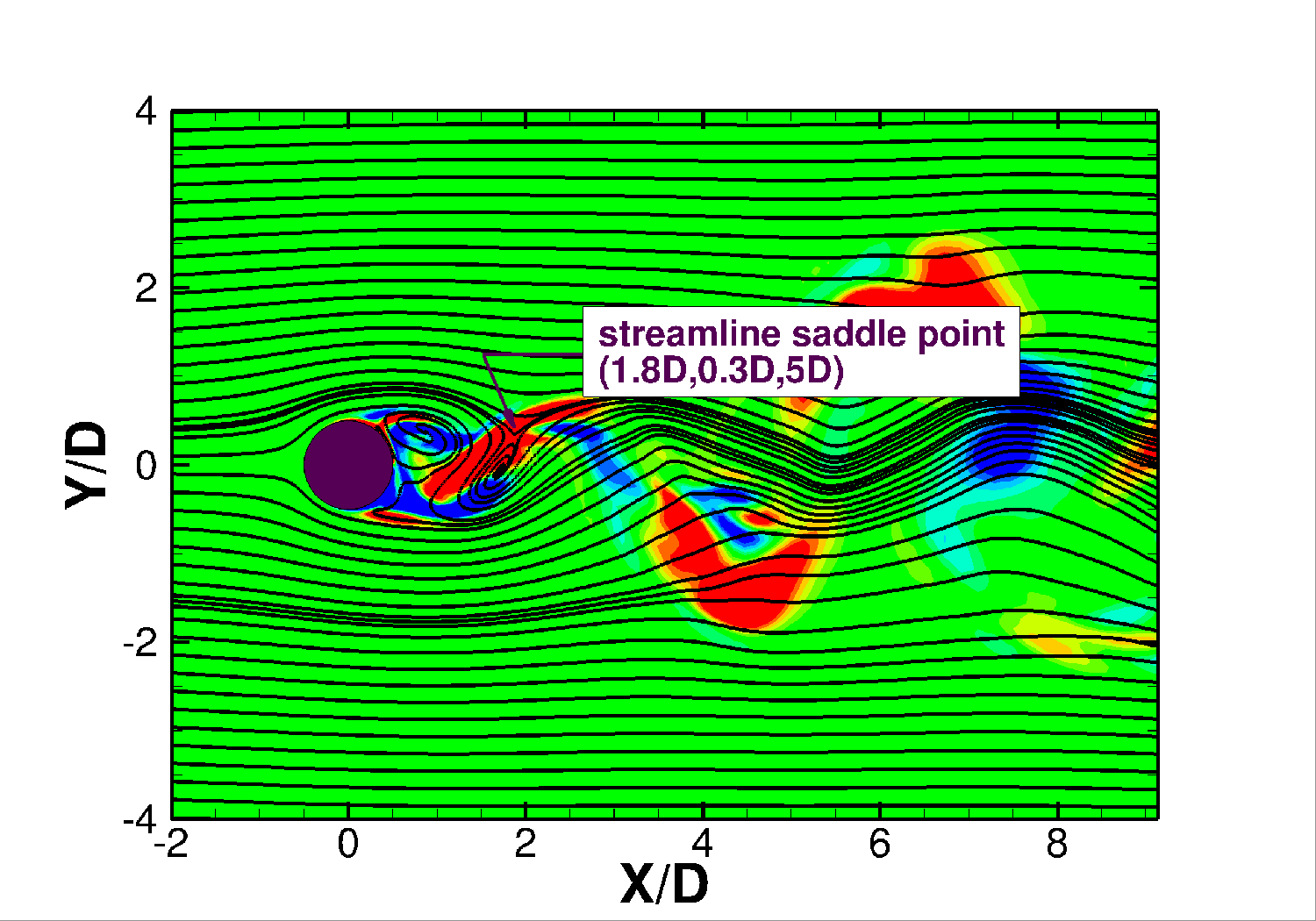}
     \caption{• $\qquad$}
     \label{fig:iso_z8_swy}
     \end{subfigure}
 \caption{Illustration of  saddle-point regions along the interface of imbalanced counter-signed vorticity clusters for a stationary cylinder at $tU/D=300$ and $l^*=5$ for $Re=$ (a,b) $500$; (c,d) $800$ whereby $\omega_x=\pm 1.0$  \emph{contours}, $\omega_z=\pm 1.0$ \emph{the solid and dash lines in (a, c)} and sectional streamlines in (b, d)}
 \label{fig:iso_snapshots}
 \end{figure}
\begin{figure} \centering
	\begin{subfigure}[b]{0.5\textwidth}	
	\centering
	\hspace{-25pt}\includegraphics[trim=0.2cm 0.2cm 0.2cm 0.6cm,scale=0.13,clip]{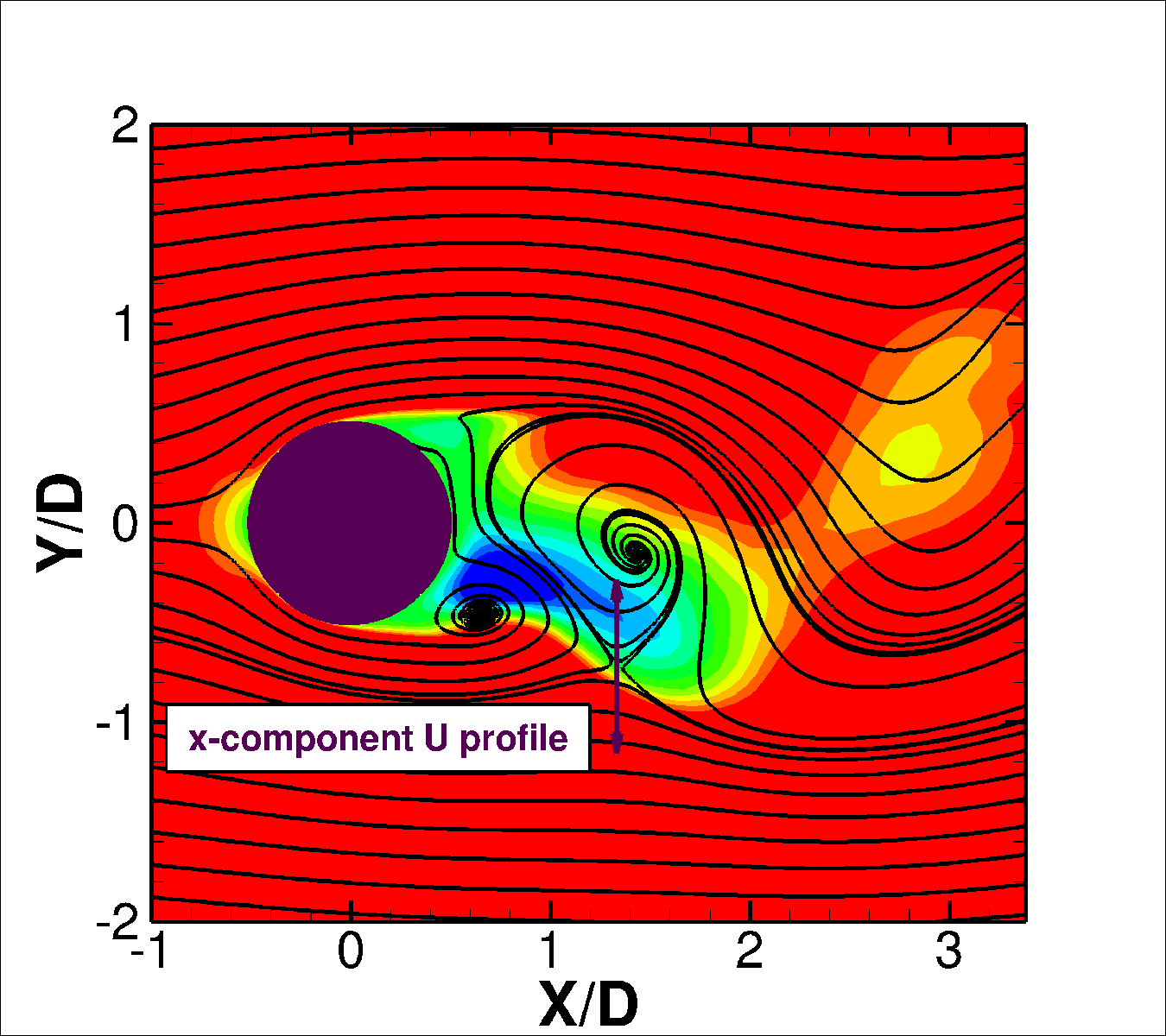}
    \caption{$\qquad$}
	\label{fig:con_283_us}
	\end{subfigure}%
	\begin{subfigure}[b]{0.5\textwidth}
	\centering
	\hspace{-25pt}\includegraphics[trim=0.1cm 0.4cm 0.1cm 0.6cm,scale=0.3,clip]{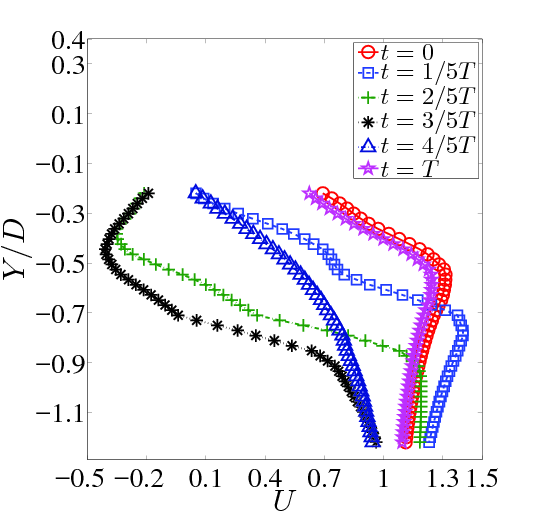}
    \caption{$\qquad$}
    \label{fig:u_profile}
    \end{subfigure}
    \begin{subfigure}[b]{0.5\textwidth}
	\centering
	\hspace{-25pt}\includegraphics[trim=0.2cm 0.2cm 0.2cm 0.6cm,scale=0.13,clip]{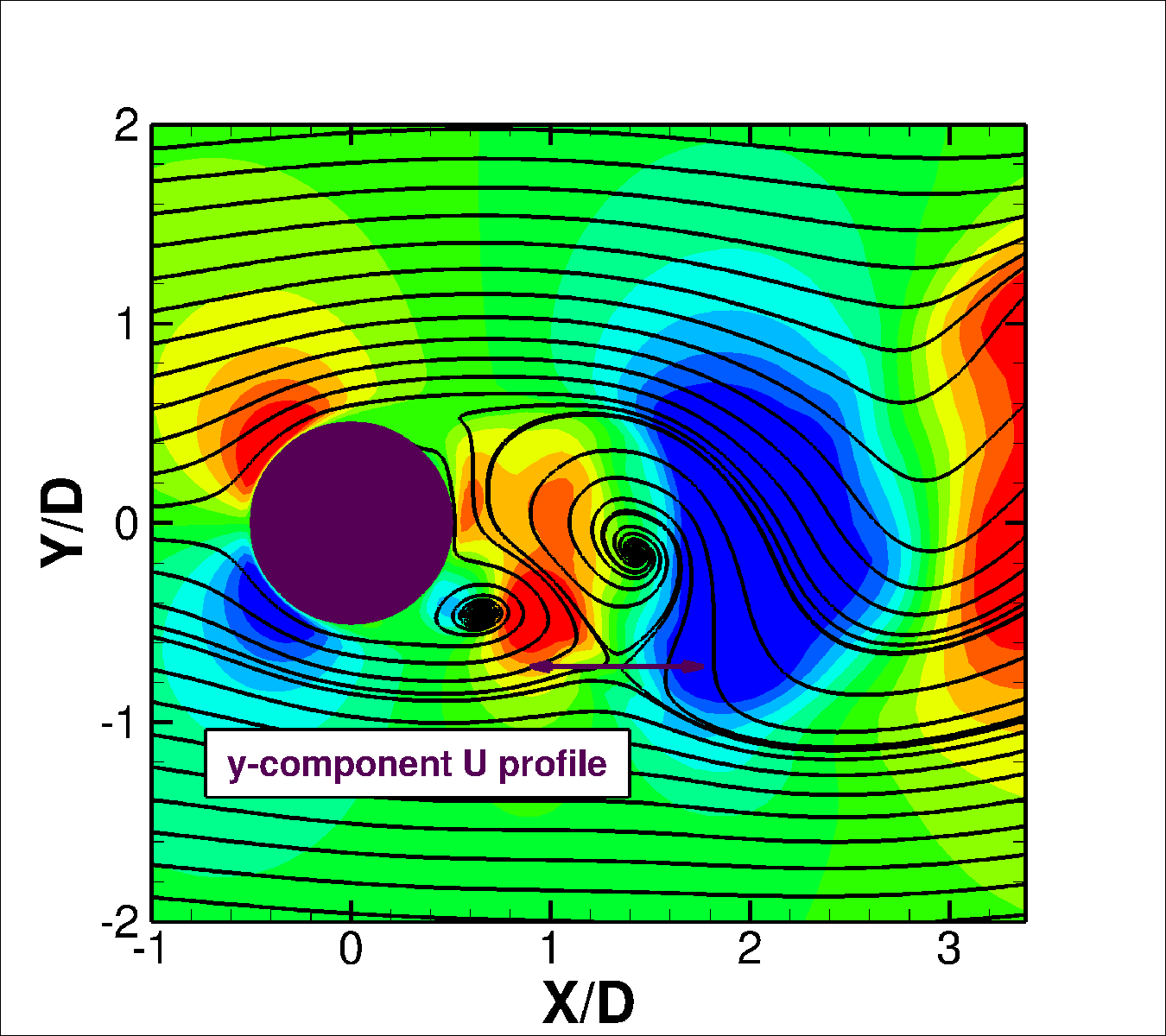}
    \caption{$\qquad$}
    \label{fig:con_283_vs}
    \end{subfigure}%
    \begin{subfigure}[b]{0.5\textwidth}
	\centering
	\hspace{-25pt}\includegraphics[trim=0.1cm 0.4cm 0.1cm 0.6cm,scale=0.3,clip]{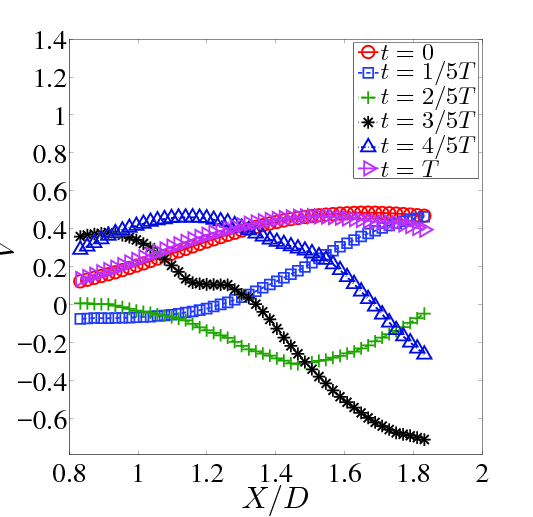}
    \caption{$\qquad$}
    \label{fig:v_profile}
    \end{subfigure}
    \begin{subfigure}[b]{0.5\textwidth}	
    \centering
    \hspace{-25pt}\includegraphics[trim=0.2cm 0.2cm 0.2cm 0.6cm,scale=0.13,clip]{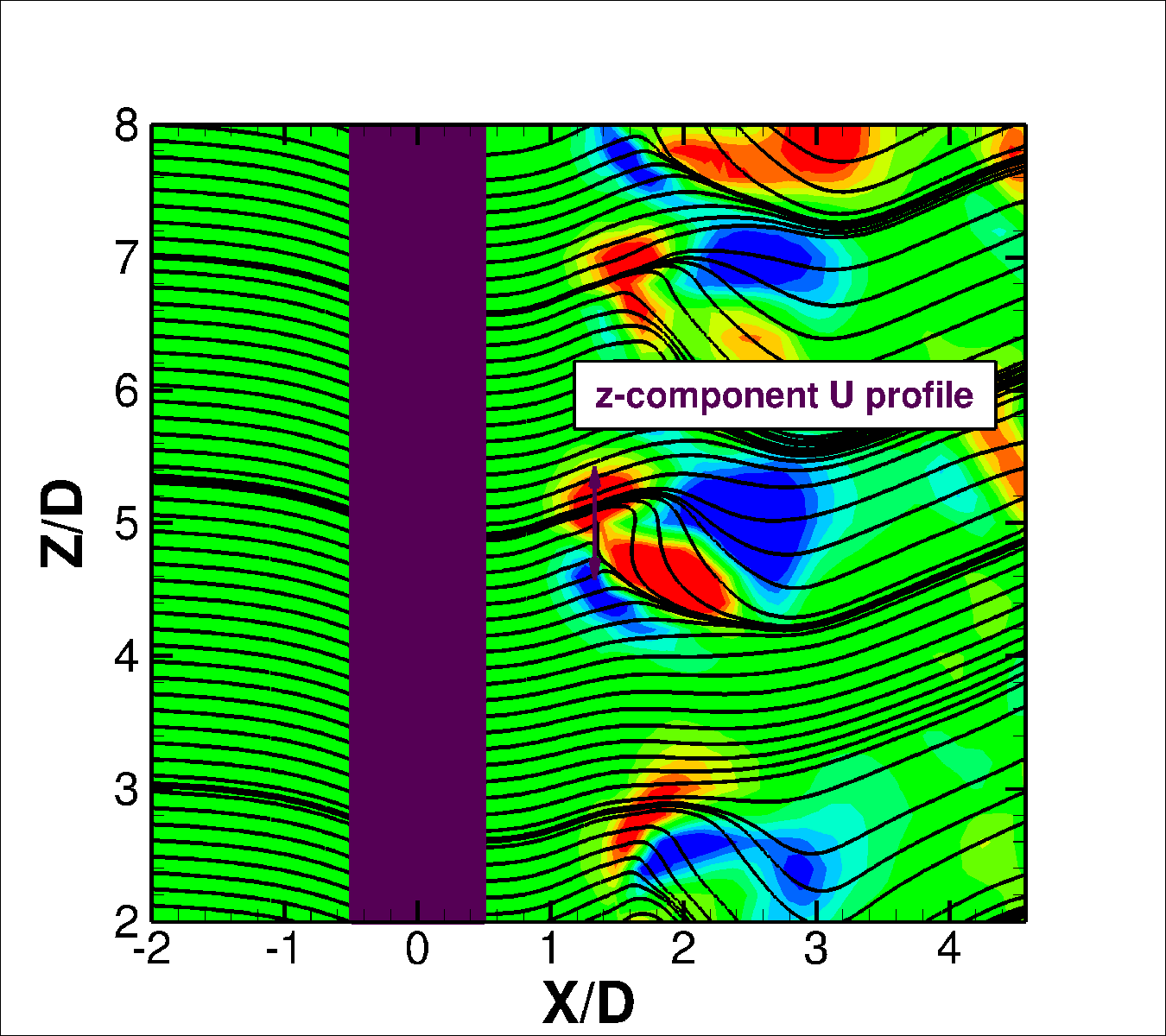}
    \caption{$\qquad$}
    \label{fig:con_283_ws}
    \end{subfigure}%
    \begin{subfigure}[b]{0.5\textwidth}
    \centering
    \hspace{-25pt}\includegraphics[trim=0.1cm 0.4cm 0.1cm 0.6cm,scale=0.3,clip]{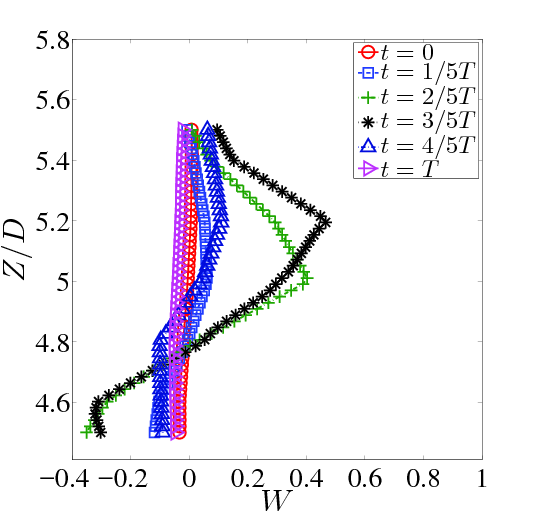}
    \caption{$\qquad$}
   \label{fig:w_profile}
   \end{subfigure}
\caption{Instantaneous velocity profiles across a streamline saddle-point at (1.33D,-0.72D, 5D) for a stationary isolated circular cylinder at $Re=500$: (a,b) $u$ vs. $y$ from (1.33D, -0.22D to -1.22D, 5D), $u \in [-0.5$, $0.5]$ contour; (c,d) $v$ vs. $x$ from (0.83D to 1.83D, -0.72D, 5D), $v \in [-0.5$, $0.5]$ contour; (e,f) $w$ vs $z$ from (1.33D, -0.72D, 4.5D to 5.5D), $w \in [-0.25$, $0.25]$ contour. $T$ is one period of primary vortex shedding cycle. The inflection points are in the saddle-point region.}
\label{fig:saddle_profile}
\end{figure}
\begin{figure} \centering
	\begin{subfigure}[b]{0.5\textwidth}	
	\centering
	\hspace{-25pt}\includegraphics[trim=0.2cm 0.2cm 0.2cm 0.2cm,scale=0.135,clip]{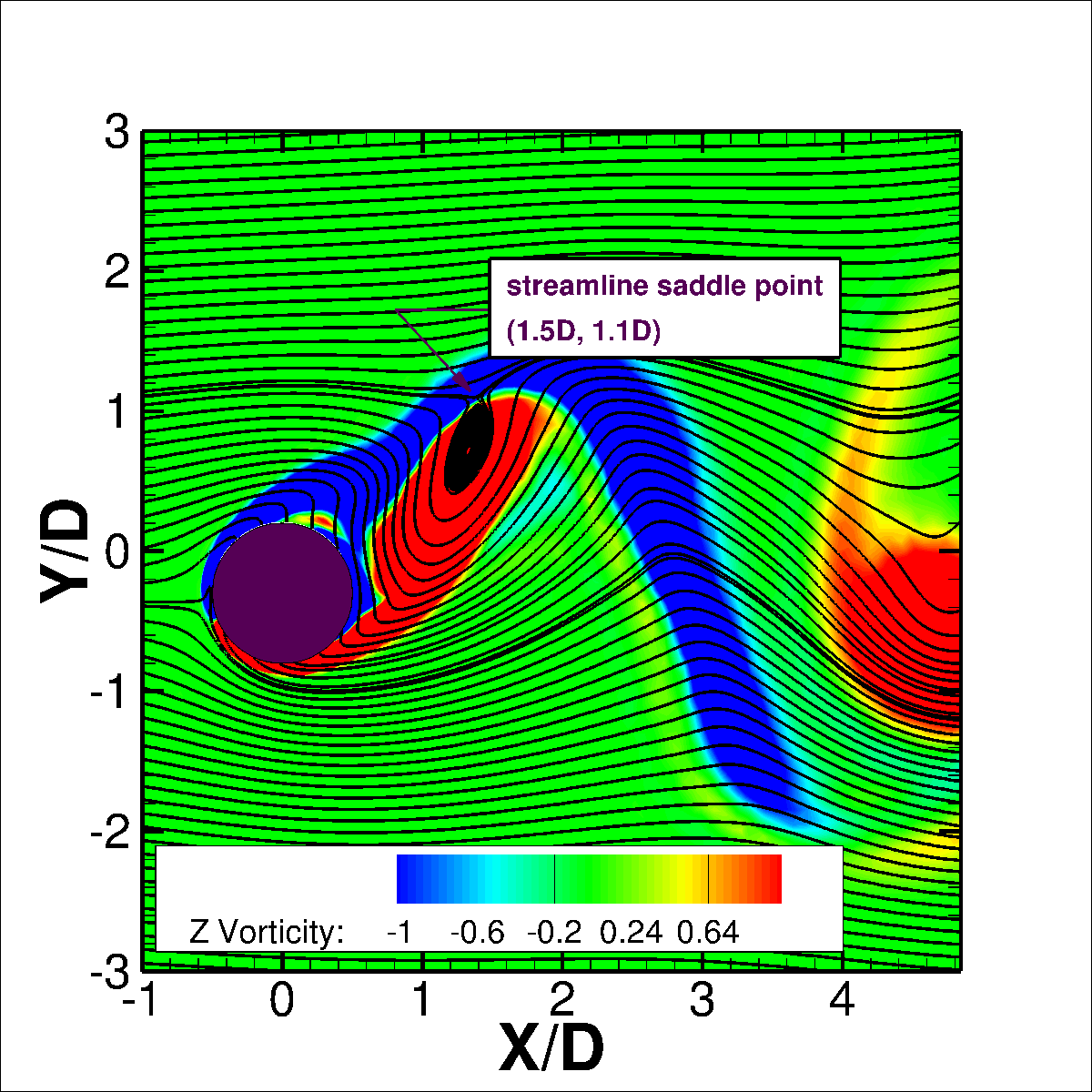}
    \caption{• $\qquad$}
	\label{fig:500r5_1}
	\end{subfigure}%
	\begin{subfigure}[b]{0.5\textwidth}
	\centering
	\hspace{-25pt}\includegraphics[trim=0.2cm 0.2cm 0.2cm 0.2cm,scale=0.135,clip]{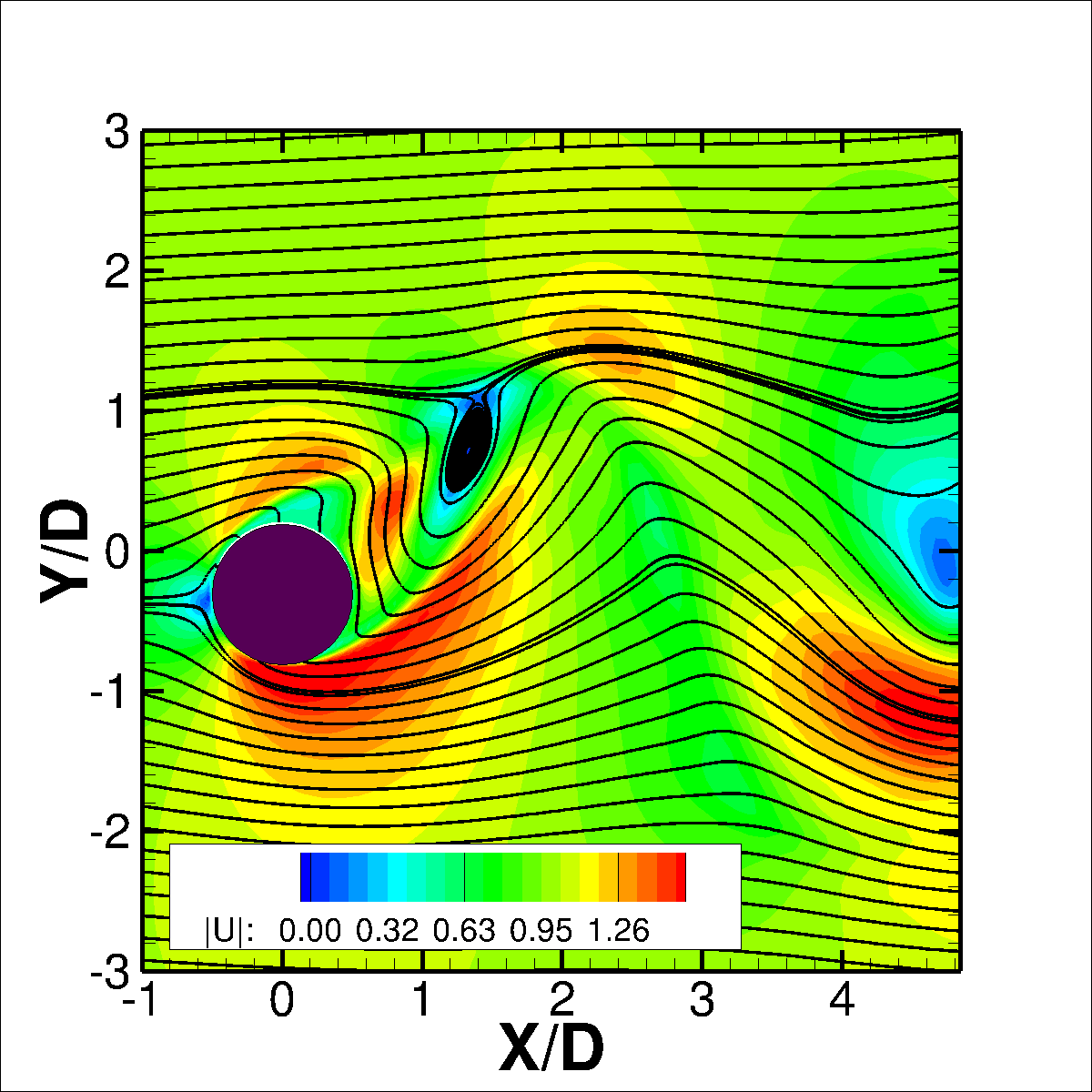}
    \caption{•$\qquad$}
    \label{fig:500r5_1T}
    \end{subfigure}
    \begin{subfigure}[b]{0.5\textwidth}	
	\centering
	\hspace{-25pt}\includegraphics[trim=0.2cm 0.2cm 0.2cm 0.2cm,scale=0.135,clip]{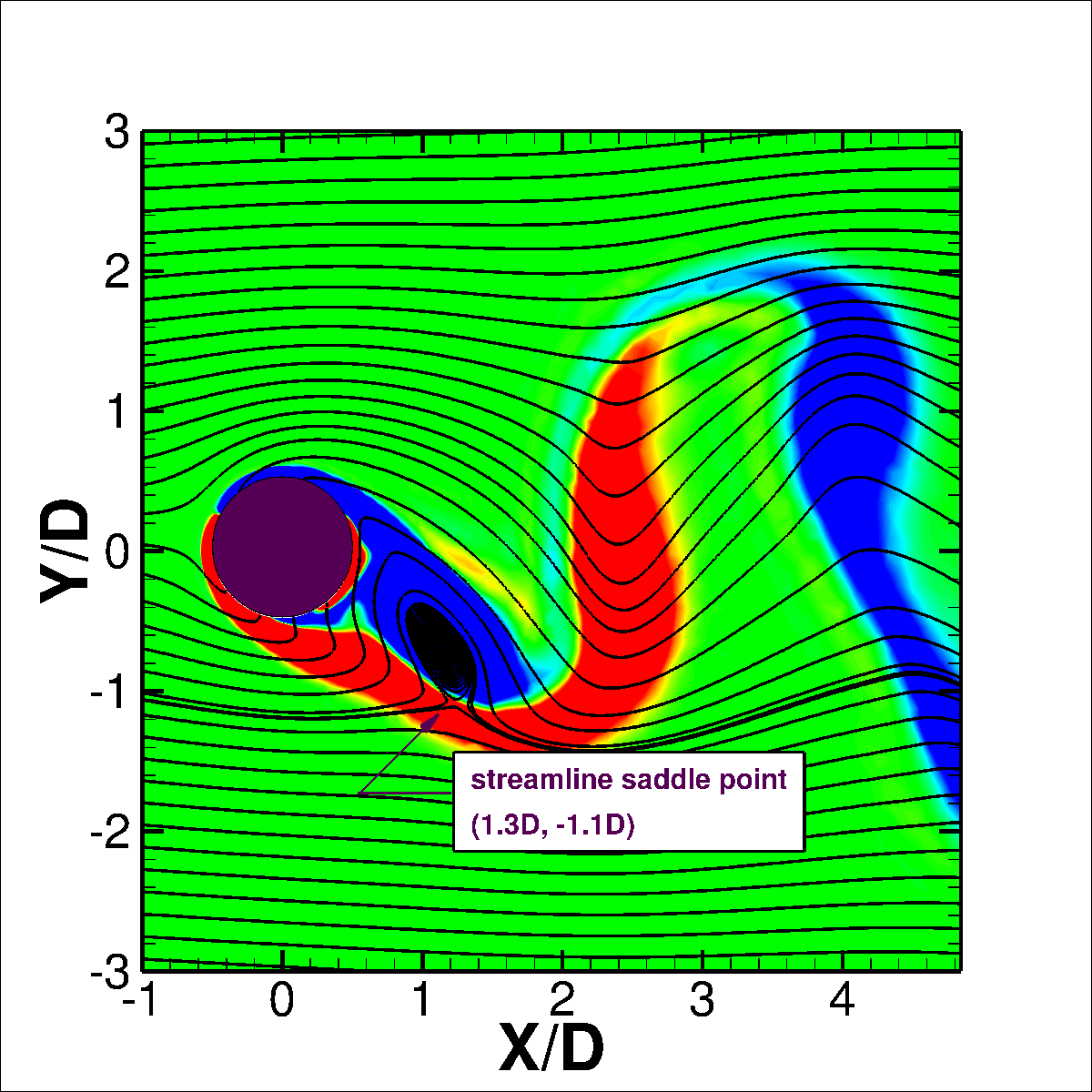}
    \caption{•$\qquad$}
	\label{fig:500r5_3}
	\end{subfigure}%
	\begin{subfigure}[b]{0.5\textwidth}
	\centering
	\hspace{-25pt}\includegraphics[trim=0.2cm 0.2cm 0.2cm 0.2cm,scale=0.135,clip]{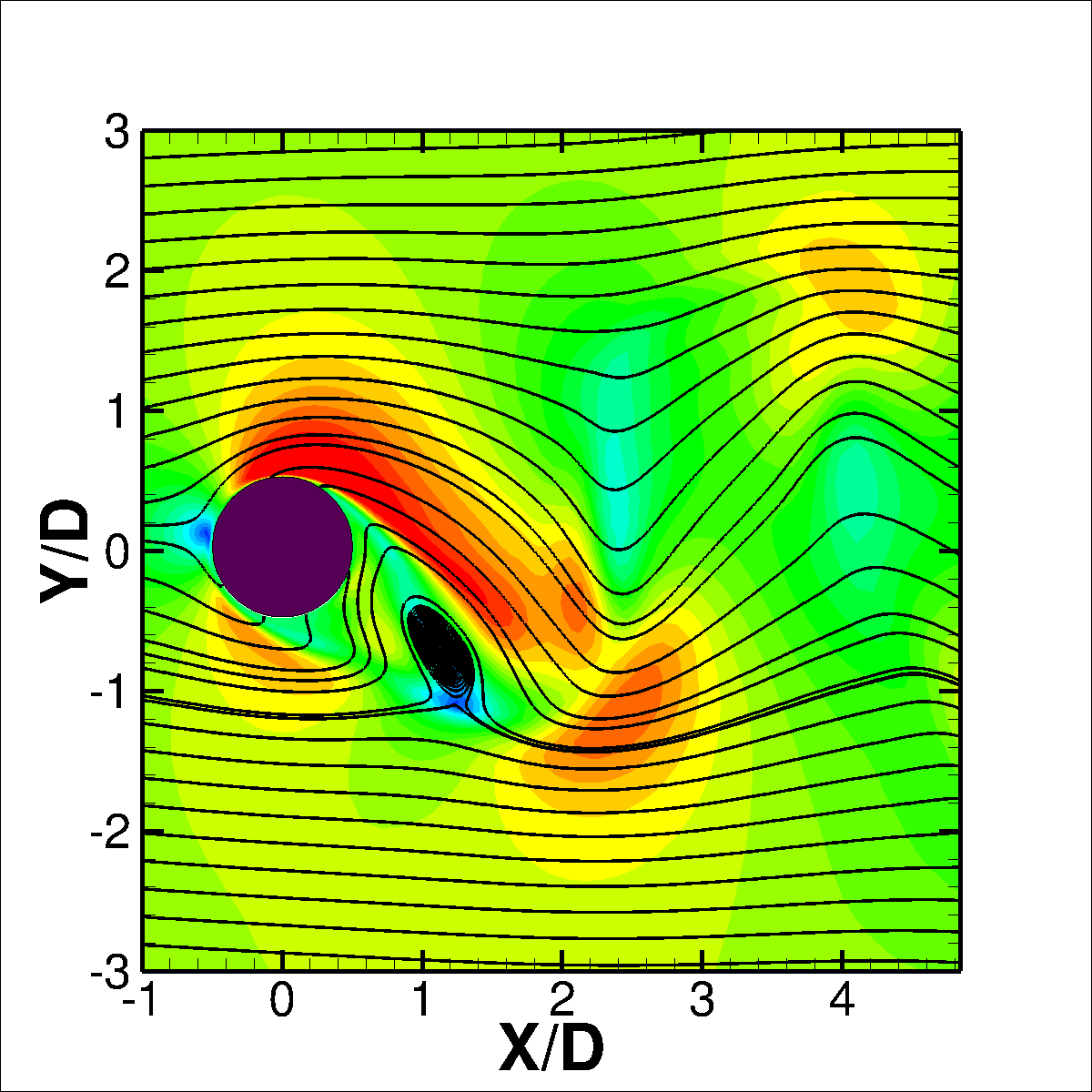}
    \caption{•$\qquad$}
    \label{fig:500r5_3T}
    \end{subfigure}
    \begin{subfigure}[b]{0.5\textwidth}	
	\centering
	\hspace{-25pt}\includegraphics[trim=0.2cm 0.2cm 0.2cm 0.2cm,scale=0.135,clip]{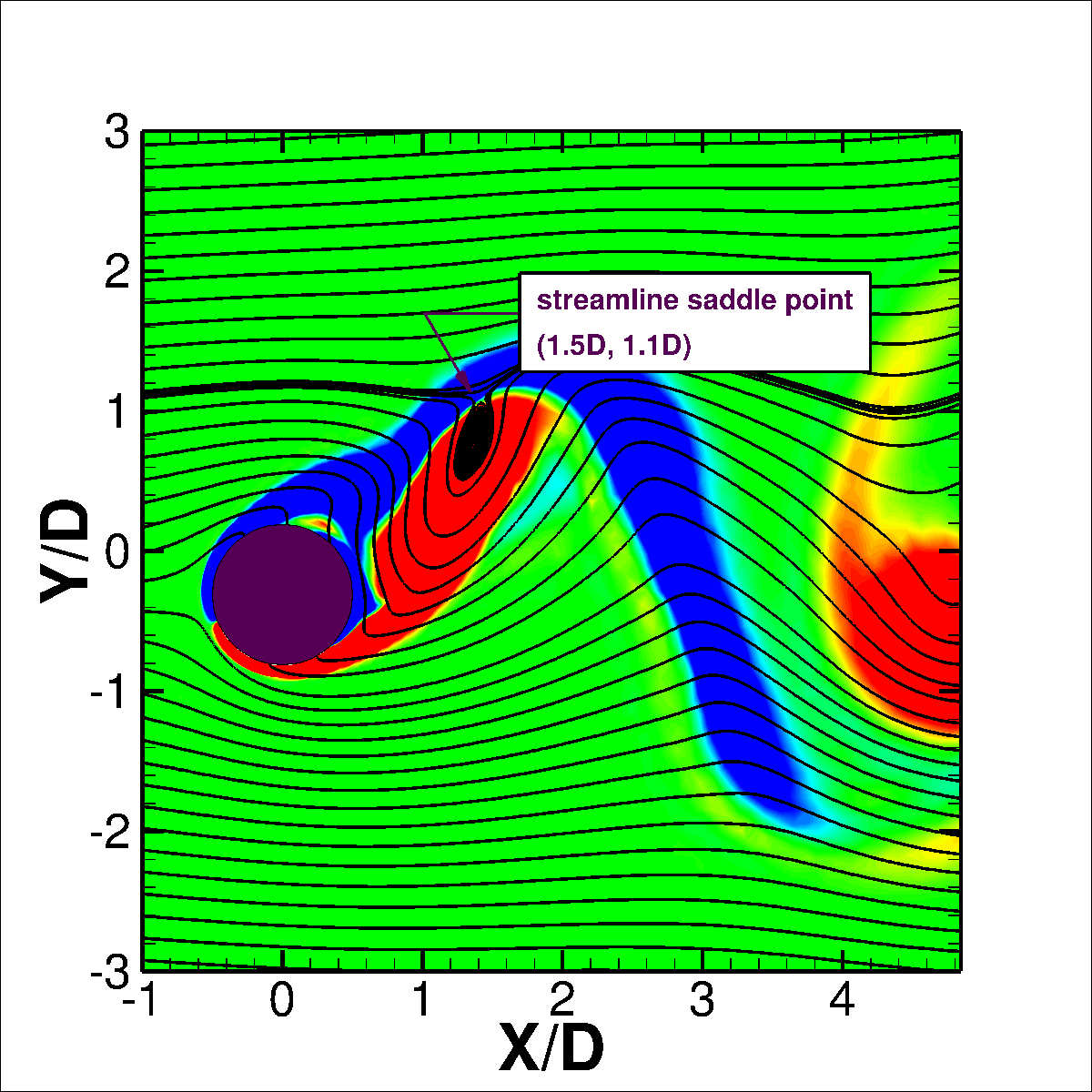}
    \caption{•$\qquad$}
	\label{fig:500r5_6}
	\end{subfigure}%
	\begin{subfigure}[b]{0.5\textwidth}
	\centering
	\hspace{-25pt}\includegraphics[trim=0.2cm 0.2cm 0.2cm 0.2cm,scale=0.135,clip]{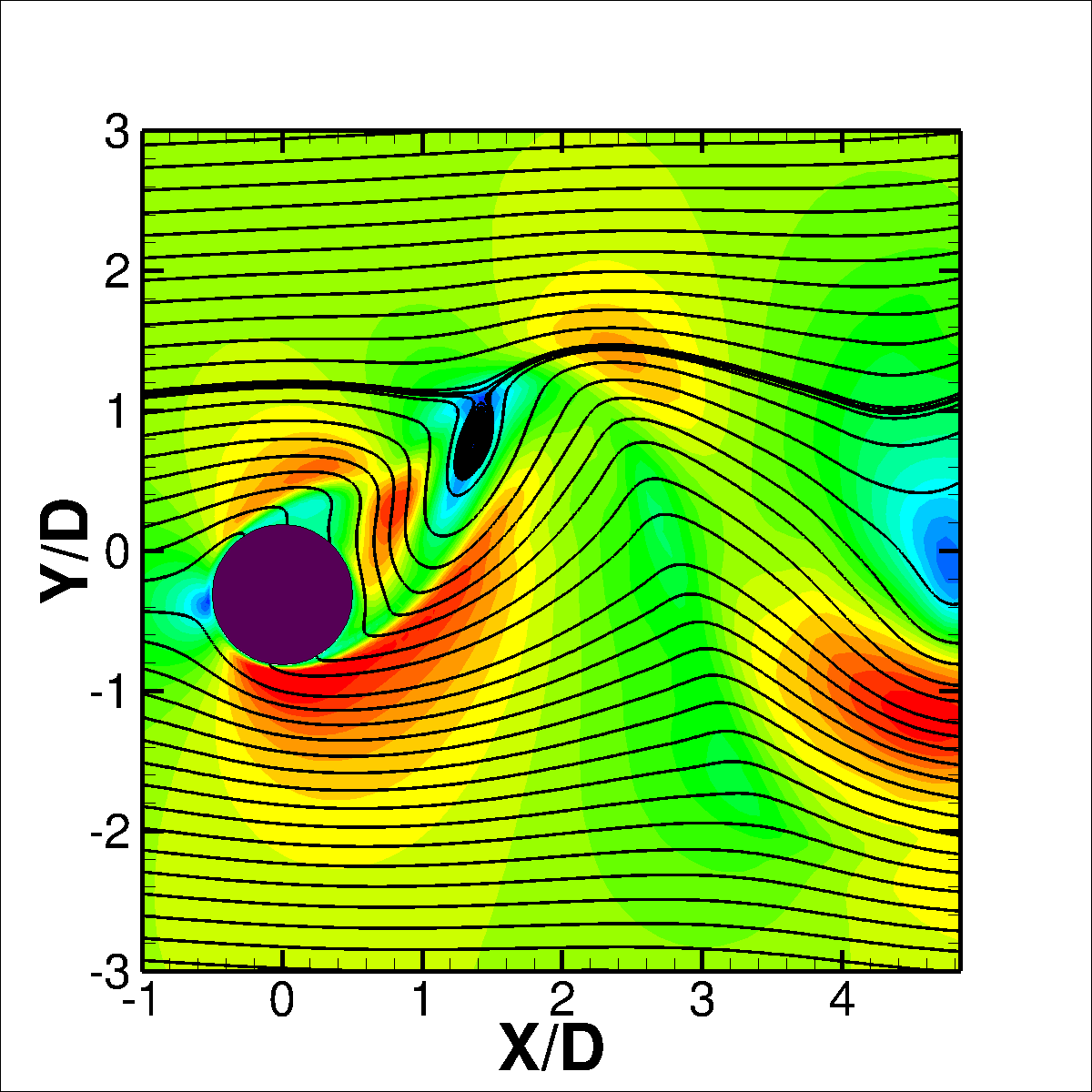}
    \caption{•$\qquad$}
    \label{fig:500r5_6T}
    \end{subfigure}
\caption{Flow contours and sectional streamline topology at $(x,y)$-plane for a transversely vibrating isolated cylinder at the peak lock-in stage in one shedding cycle of the primary vortex (anti-phase): $Re=500$, $m^*=10$, $\zeta=0.01$, $U_r=5$, $l^*=5$, $\omega_z = \pm 1.0$, \emph{contours in (a,c,e)}, and velocity amplitude $|U| = \sqrt{U^2 + V^2} \in [0,$ $1.5]$, \emph{contours in (b,d,f)}}  
\label{fig:scon_500r5}
\end{figure}
\begin{figure} \centering
	\begin{subfigure}[b]{0.5\textwidth}	
	\centering
	\hspace{-25pt}\includegraphics[trim=0.1cm 0.1cm 0.1cm 0.5cm,scale=0.3,clip]{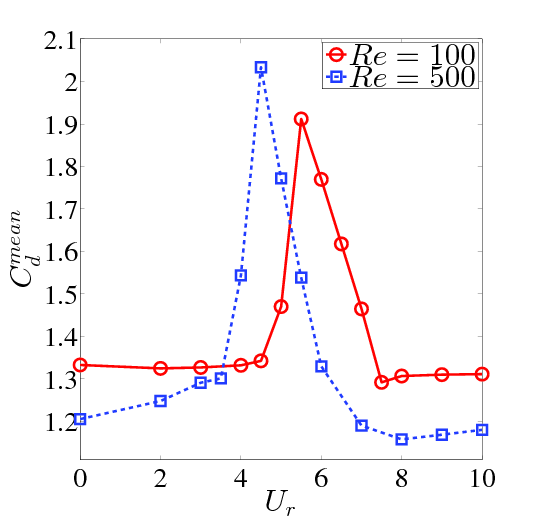}
    \caption{$\qquad$}
	\label{fig:100500_cd}
	\end{subfigure}%
	\begin{subfigure}[b]{0.5\textwidth}
	\centering
	\hspace{-25pt}\includegraphics[trim=0.1cm 0.1cm 0.1cm 0.5cm,scale=0.3,clip]{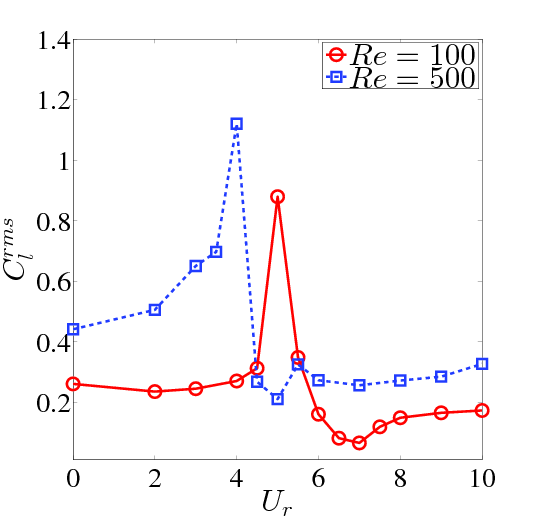}
    \caption{$\qquad$}
    \label{fig:100500_cl}
    \end{subfigure}
\caption{Hydrodynamic forces as a function of the reduced velocity for a transversely-vibrating isolated cylinder at $m^*=10$ and $\zeta=0.01$: (a) the mean drag coefficient $C_d^{mean}$ with respect to the reduced velocity $U_r$; (b) the r.m.s. lift coefficient $C_l^{rms}$ with respect to the reduced velocity $U_r$.}
\label{fig:100500_cdcl}
\end{figure}
\begin{figure} \centering
	\begin{subfigure}[b]{0.5\textwidth}	
	\centering
	\hspace{-25pt}\includegraphics[trim=0.1cm 0.1cm 0.1cm 0.5cm,scale=0.3,clip]{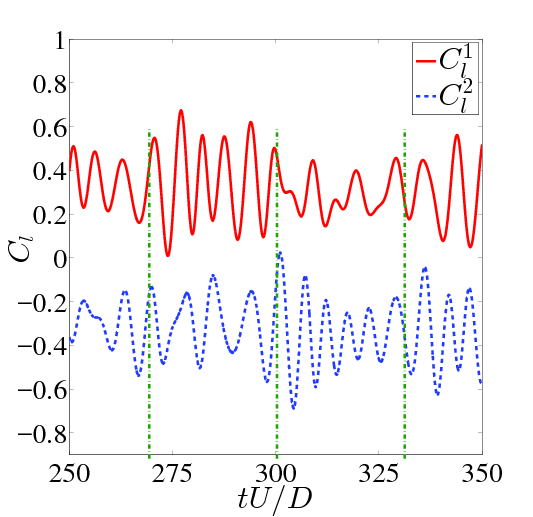}
    \caption{$\qquad$}
	\label{fig:cl_re100td18}
	\end{subfigure}%
	\begin{subfigure}[b]{0.5\textwidth}
	\centering
	\hspace{-25pt}\includegraphics[trim=0.1cm 0.1cm 0.1cm 0.5cm,scale=0.097,clip]{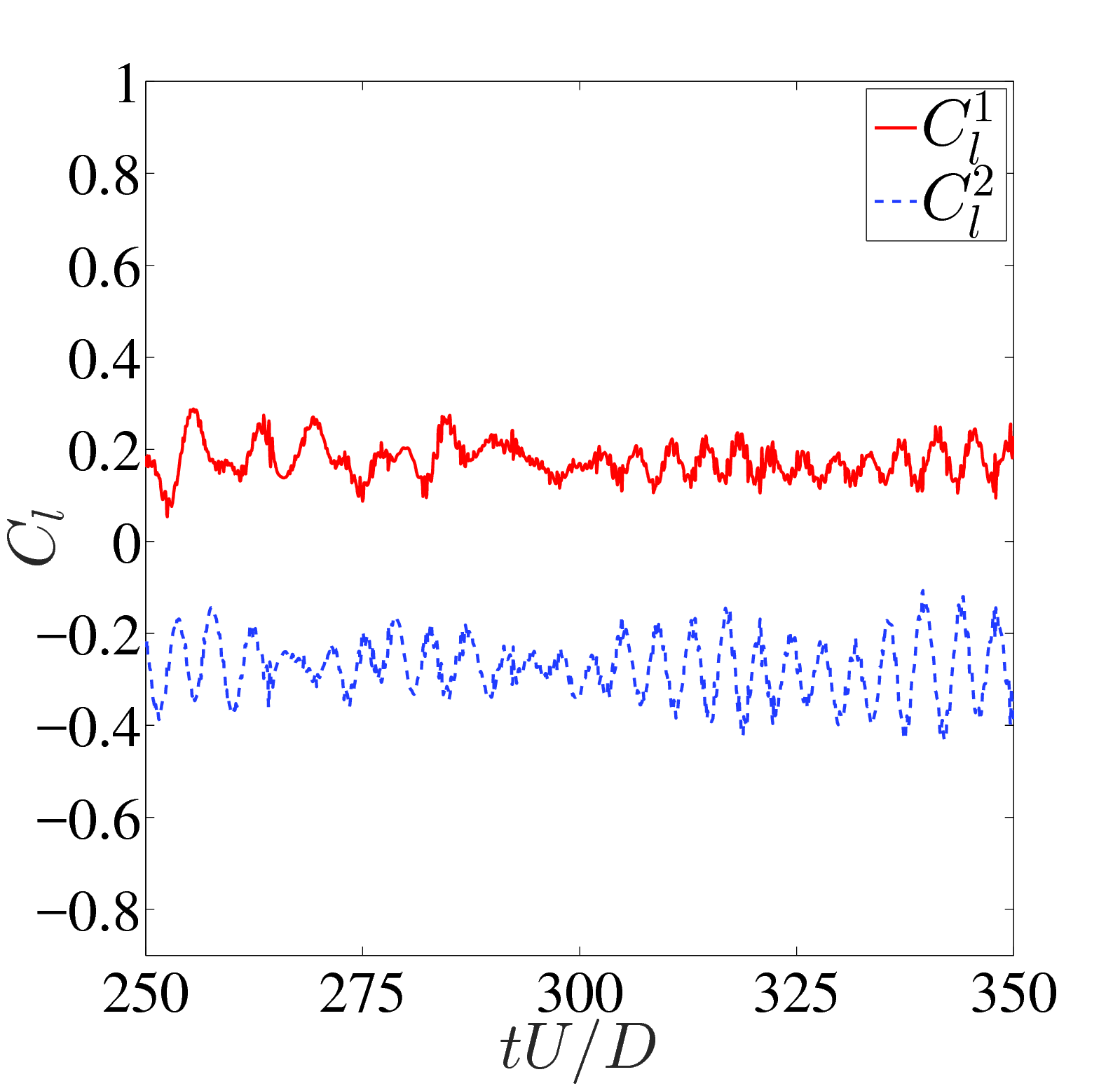}
    \caption{$\qquad$}
    \label{fig:cl_re500td18}
    \end{subfigure}
	\begin{subfigure}[b]{0.5\textwidth}	
	\centering
	\hspace{-25pt}\includegraphics[trim=0.1cm 0.1cm 0.1cm 0.5cm,scale=0.3,clip]{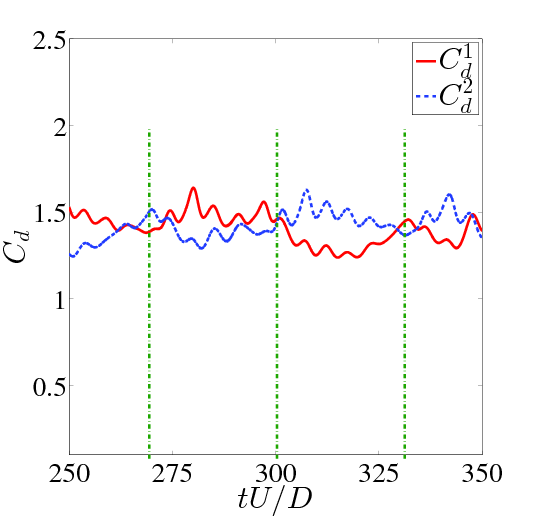}
    \caption{$\qquad$}
	\label{fig:cd_re100td18}
	\end{subfigure}%
	\begin{subfigure}[b]{0.5\textwidth}
	\centering
	\hspace{-25pt}\includegraphics[trim=0.1cm 0.5cm 0.1cm 0.5cm,scale=0.097,clip]{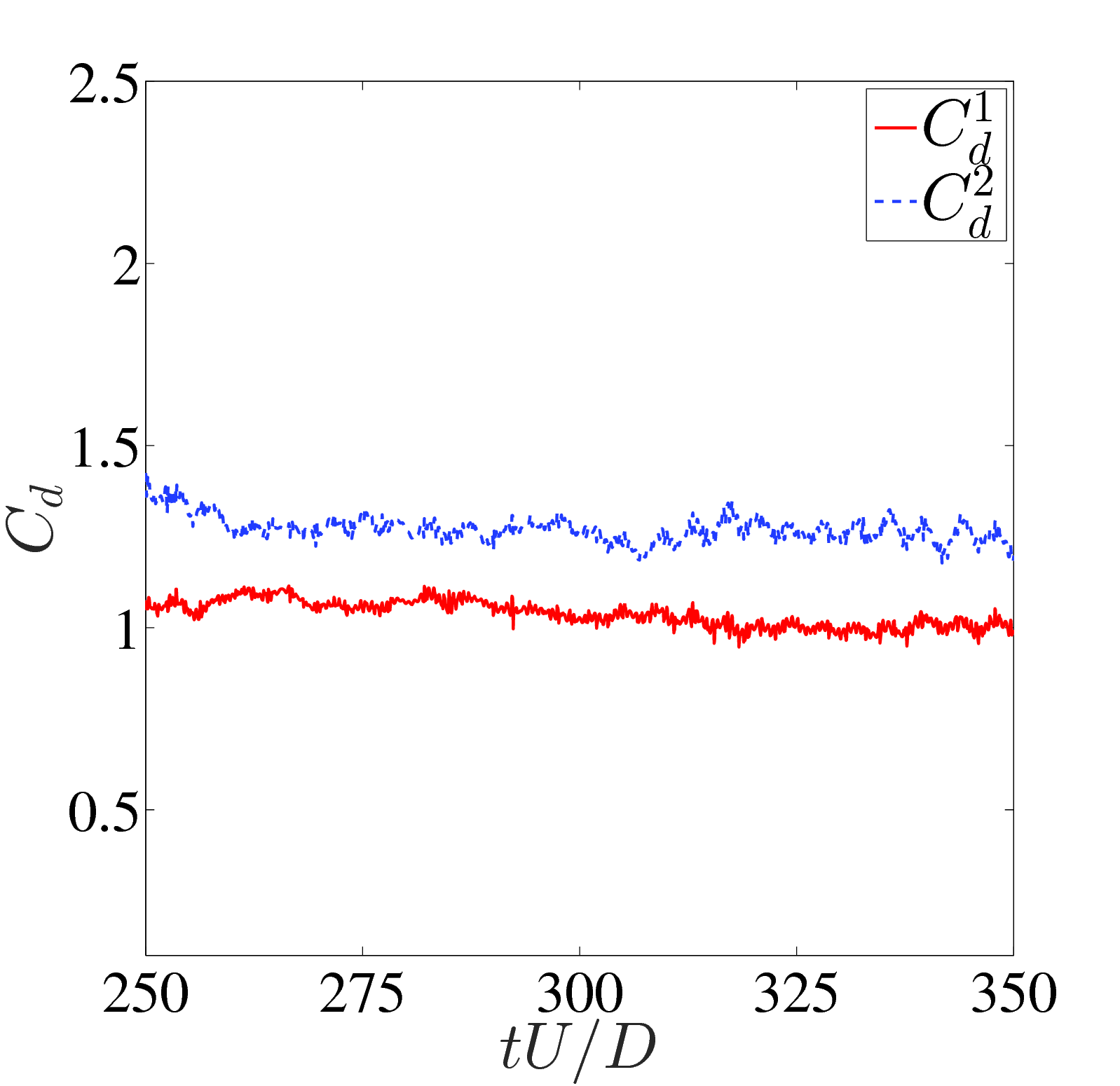}
    \caption{$\qquad$}
    \label{fig:cd_re500td18}
    \end{subfigure}
\caption{Time series of the hydrodynamic forces for SSBS arrangement at $g^*=0.8$: (a,c) $Re=100$, the flip-flop are marked at $tU/D = $ 270, 300 and 330; (b,d) $Re=500$}
\label{fig:clcd_100500td18}
\end{figure}
\begin{figure}
\begin{center}
\centering
\includegraphics[trim=0.1cm 0.1cm 0.1cm 0.1cm,scale=0.32,clip]{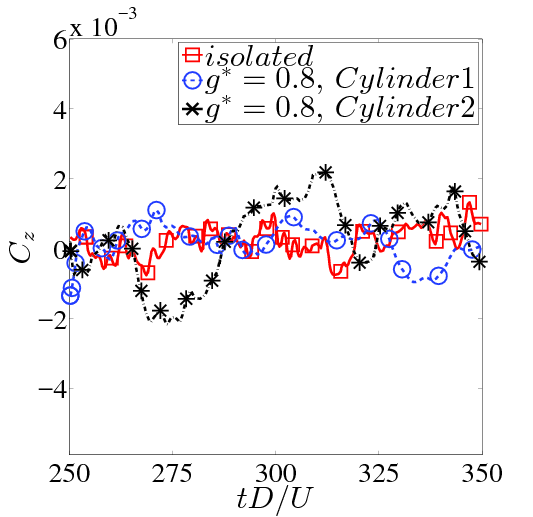}
\end{center}
\caption{Comparison of the spanwise hydrodynamic force $C_z$ between a stationary cylinder and SSBS arrangement at $Re=500$ and $g^*=0.8$. When the gap flow deflects to \emph{Cylinder2} for $tU/D \in [250$, $350]$, $C_z$ amplitude (\emph{the solid-dot star}) is amplified.}
 \label{fig:cz_re500_isosbs}
\end{figure}
\begin{figure} \centering
	\begin{subfigure}[b]{0.5\textwidth}	
	\centering
	\hspace{-25pt}\includegraphics[trim=0.2cm 0.2cm 0.2cm 0.3cm,scale=0.26,clip]{figures/re500/contour300/re500_6000_iso.png}
    \caption{•}
	\label{fig:con_re500a}
	\end{subfigure}%
	\begin{subfigure}[b]{0.5\textwidth}
	\centering
	\hspace{-25pt}\includegraphics[trim=0.15cm 0.2cm 0.2cm 0.1cm,scale=0.26,clip]{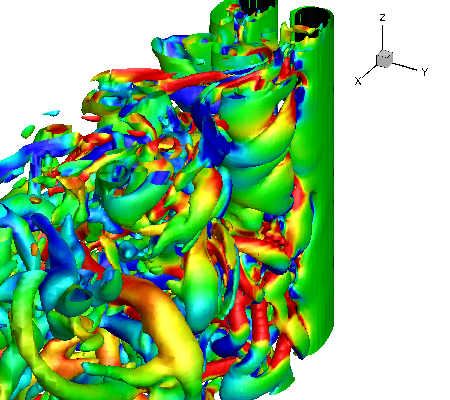}
    \caption{•}
    \label{fig:con_re500td18}
    \end{subfigure}
\caption{Instantaneous vortical structures using the Q-criterion at $Re = 500$, $tU/D=300$, $Q = 0.2$ and $\omega_y = \pm 1$ (\emph{contours}): (a) stationary cylinder; (b) SSBS arrangement at $g^*=0.8$. The streamwise vorticity concentration is higher in a narrow near wake region}
\label{fig:con_iso&td18}
\end{figure}
\begin{figure}
\begin{center}
\centering
\includegraphics[trim=0.1cm 0.1cm 0.1cm 0.1cm,scale=0.32,clip]{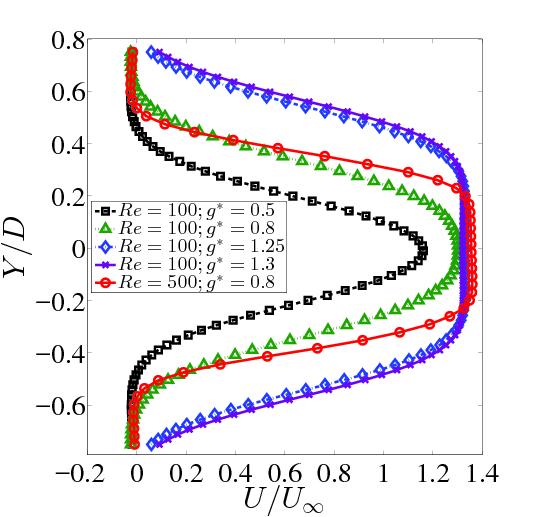}
\end{center}
\caption{Horizontal velocity profile of the gap flow for the SSBS arrangements at $Re=100$ and $Re=500$. The velocity profiles are extracted at the fluid jet location ($0.6D$, $0.7D$ to $-0.7D$) where ($0D$, $0D$) is the center between the cylinders. The time-averaging is performed from $tU/D \in [250$, $350]$. }
\label{fig:profile_1}
\end{figure}
\begin{figure} \centering
	\begin{subfigure}[b]{0.5\textwidth}	
	\centering
	\hspace{-25pt}\includegraphics[trim=0.1cm 0.1cm 0.15cm 0.1cm,scale=0.125,clip]{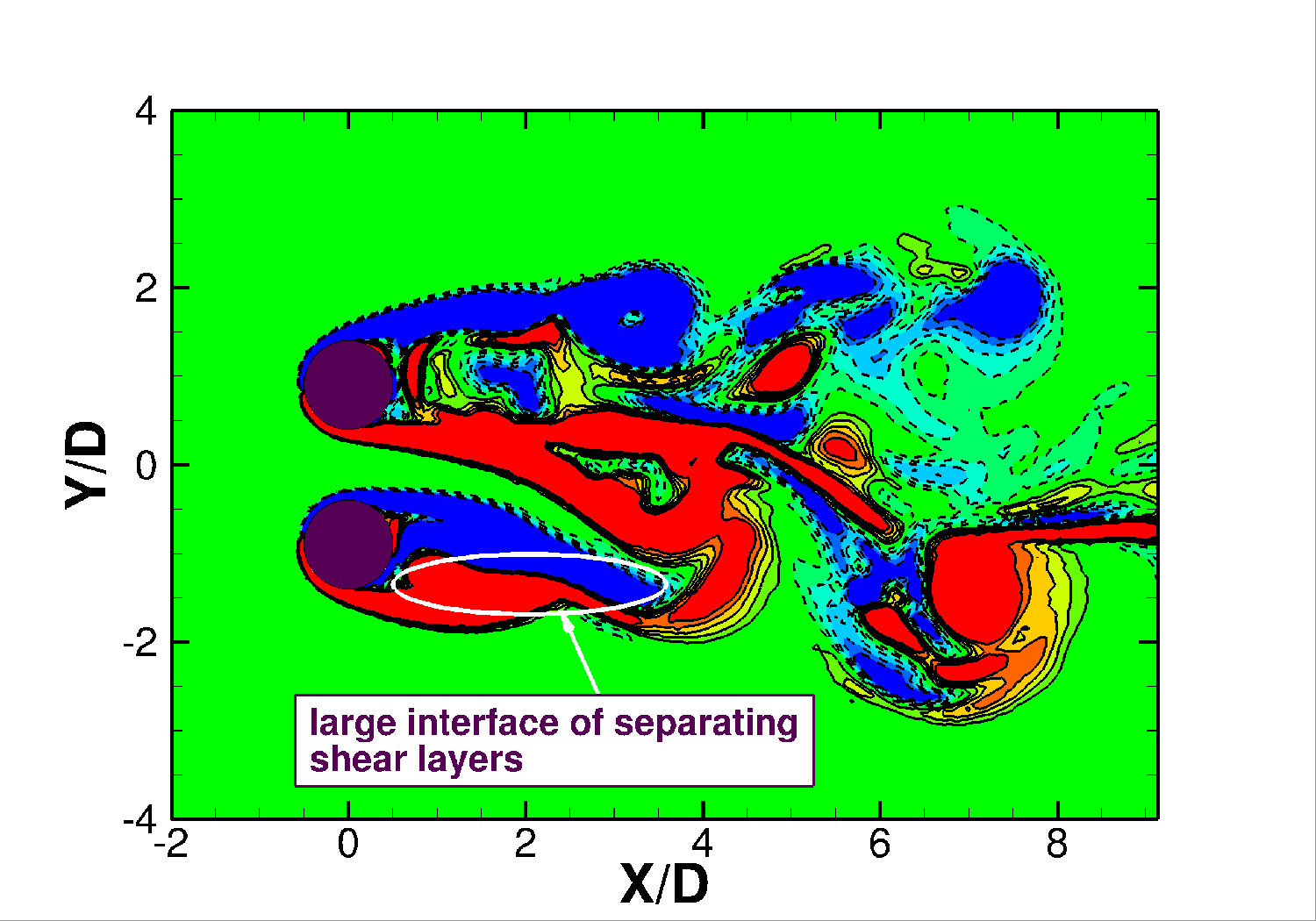}
    \caption{$\qquad$}
	\label{fig:con_td18_xy_305}
	\end{subfigure}%
	\begin{subfigure}[b]{0.5\textwidth}
	\centering
	\hspace{-25pt}\includegraphics[trim=0.1cm 0.1cm 0.15cm 0.1cm,scale=0.125,clip]{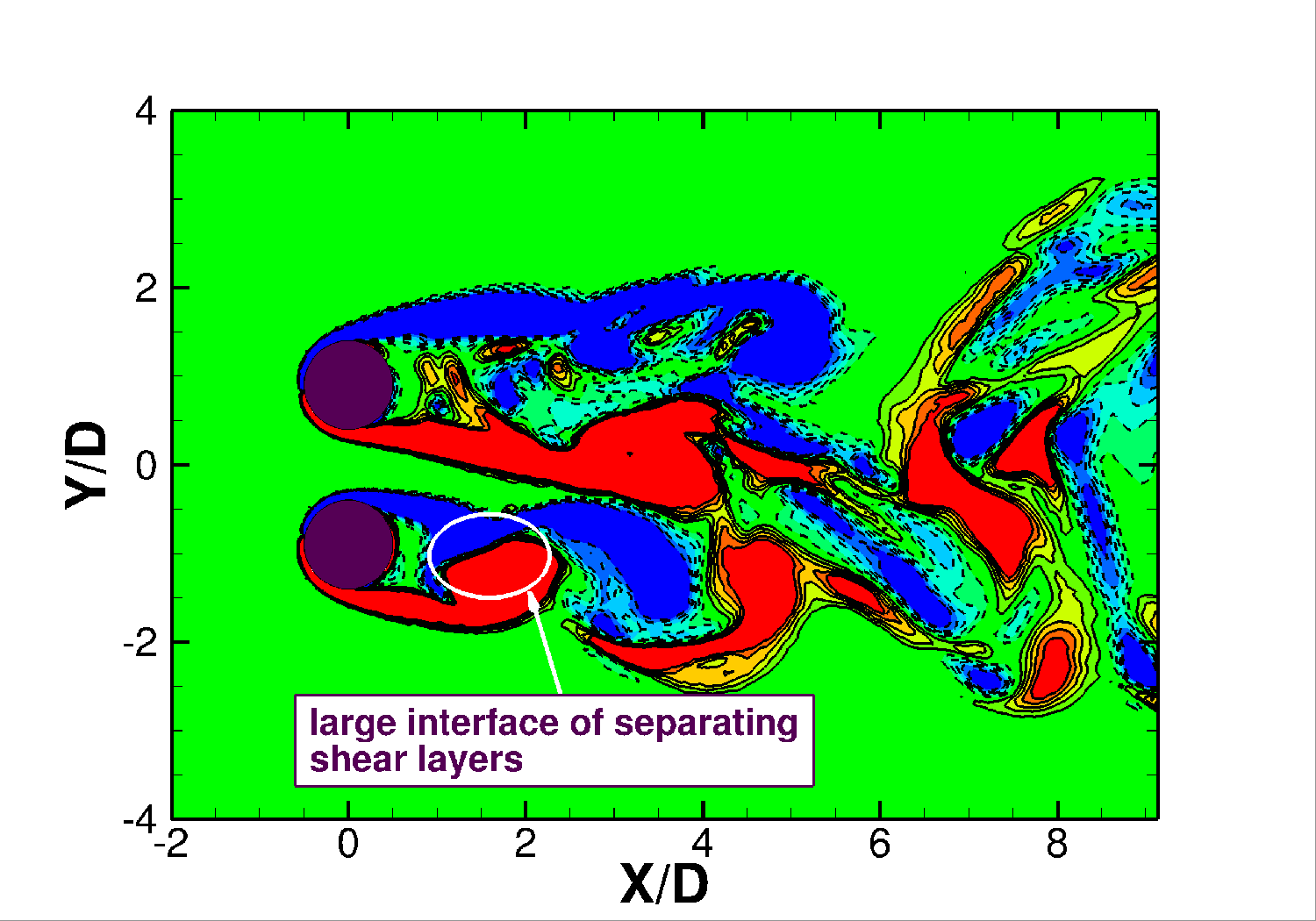}
    \caption{$\qquad$}
    \label{fig:cz_td18_xy_317}
    \end{subfigure}
\caption{Spanwise vorticity $\omega_z$ contours of cylinders in SSBS arrangement at $Re=500$, $g^*=0.8$, $l^*=5$ and $\omega_{z}=\pm 1.0$ (\emph{contours}): (a) $tU/D= 303$; (b) $tU/D=317$. Large interfaces of different vorticity concentrations are observed in the narrow near-wake regions}
\label{fig:con_td18}
\end{figure}
\begin{figure} \centering
    \begin{subfigure}[b]{0.5\textwidth}	
	\centering
	\hspace{-25pt}\includegraphics[trim=0.1cm 0.1cm 0.15cm 0.1cm,scale=0.125,clip]{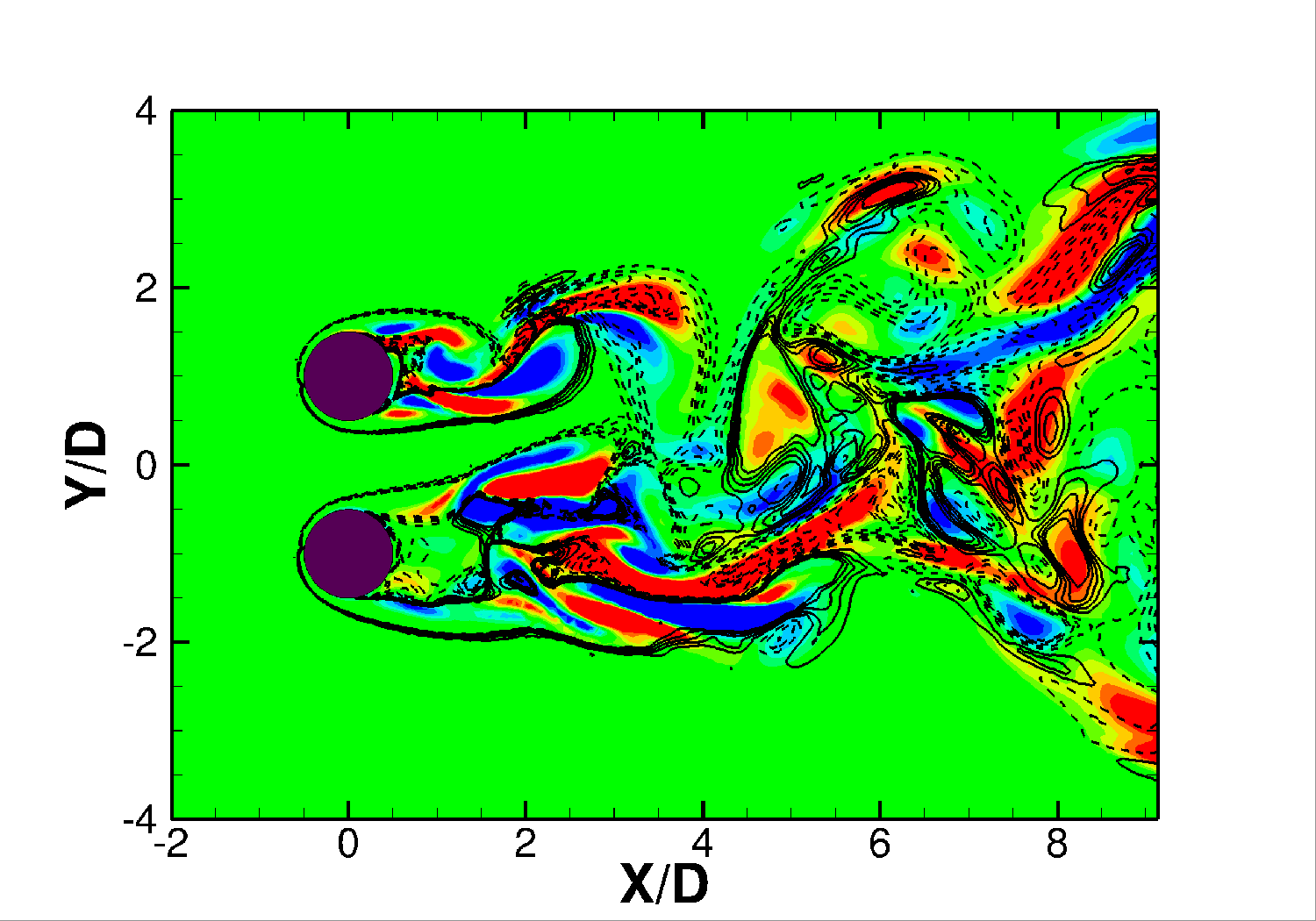}
    \caption{$\qquad$}
	\label{fig:td20_z4_wxz}
	\end{subfigure}%
	\begin{subfigure}[b]{0.5\textwidth}
	\centering
	\hspace{-25pt}\includegraphics[trim=0.1cm 0.1cm 0.15cm 0.1cm,scale=0.125,clip]{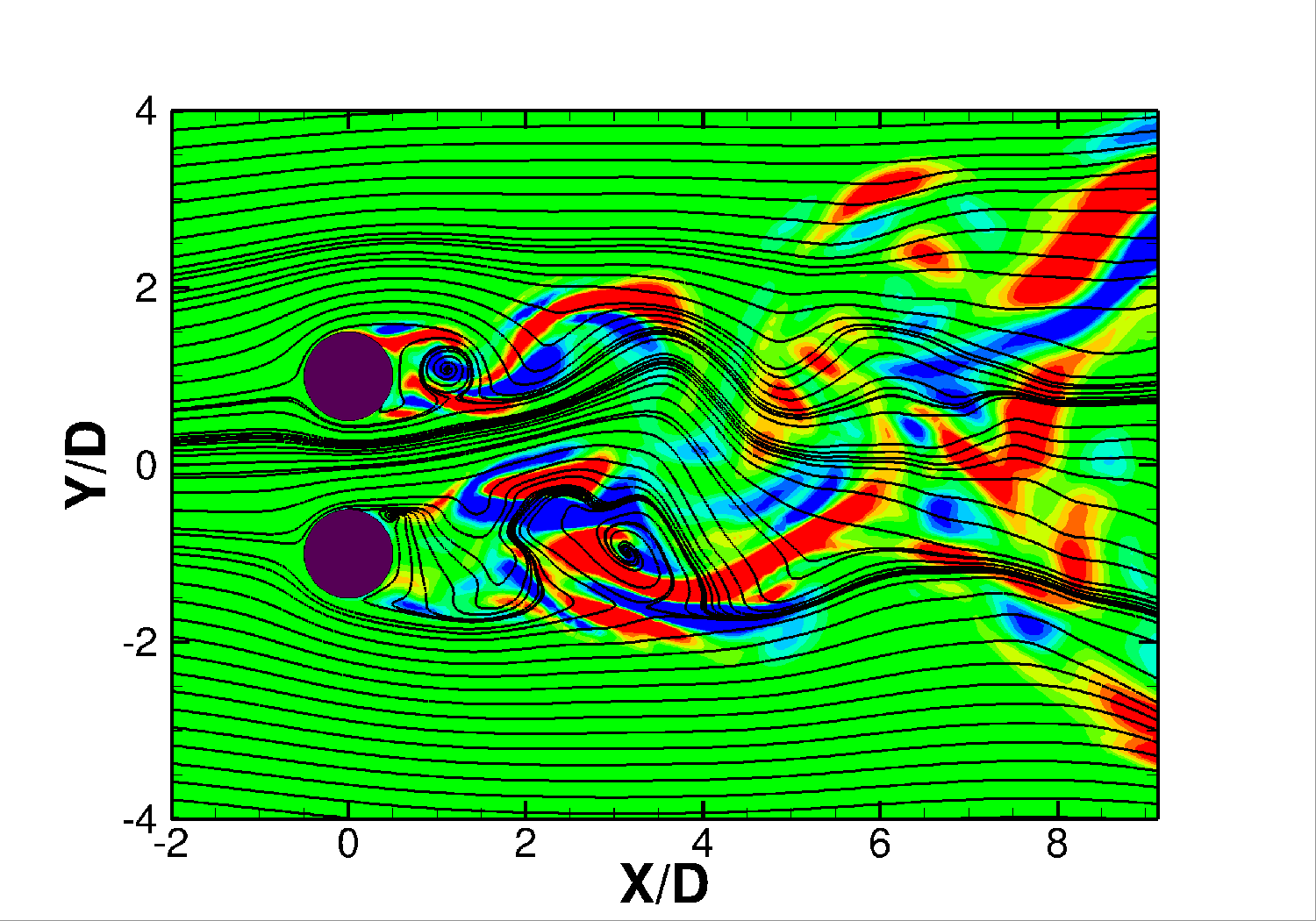}
    \caption{$\qquad$}
    \label{fig:td20_z4_swx}
    \end{subfigure}
    \begin{subfigure}[b]{0.5\textwidth}	
	\centering
	\hspace{-25pt}\includegraphics[trim=0.1cm 0.1cm 0.15cm 0.1cm,scale=0.125,clip]{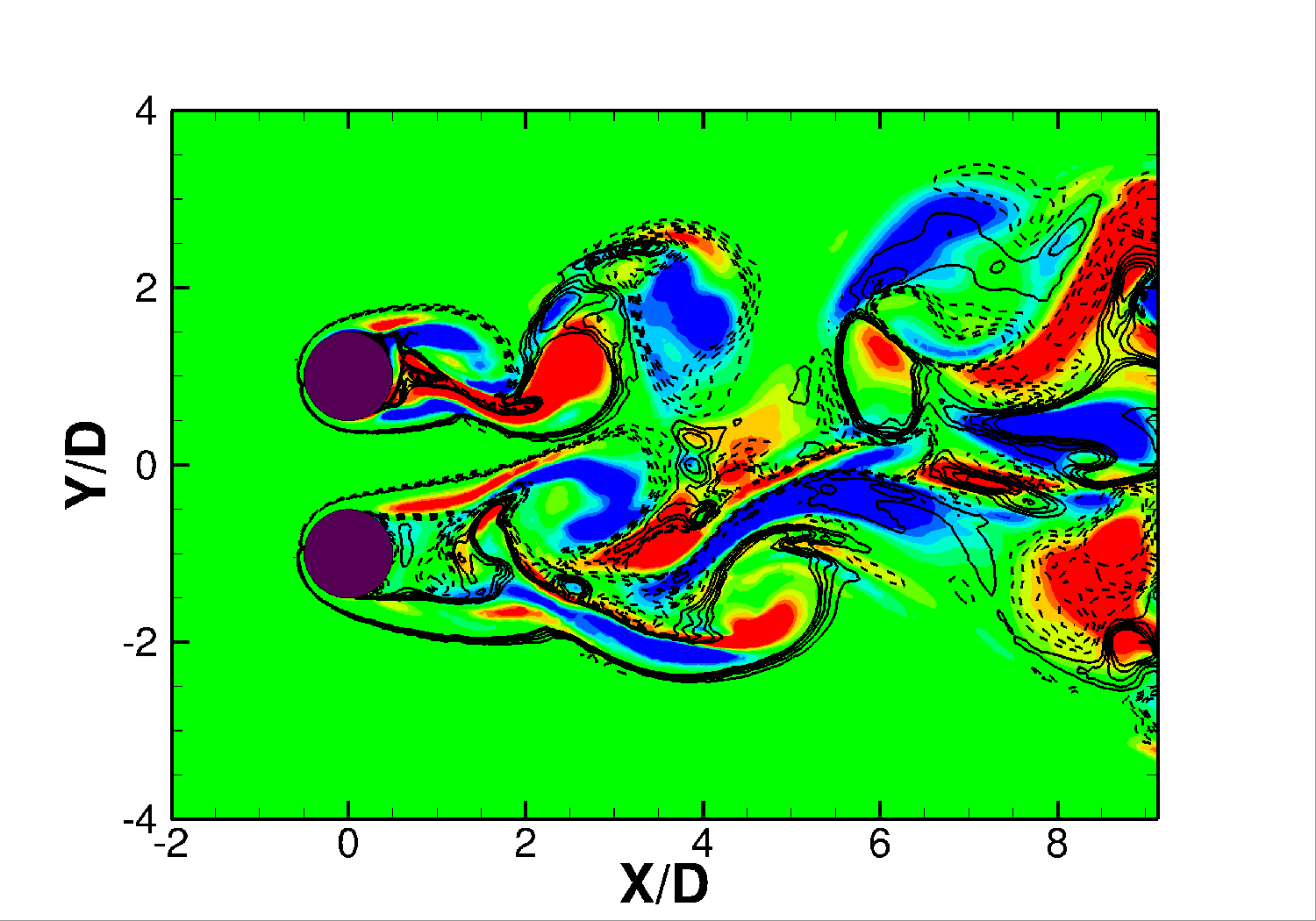}
    \caption{$\qquad$}
	\label{fig:td20_z8_wxz}
	\end{subfigure}%
	\begin{subfigure}[b]{0.5\textwidth}
	\centering
	\hspace{-25pt}\includegraphics[trim=0.1cm 0.1cm 0.15cm 0.1cm,scale=0.125,clip]{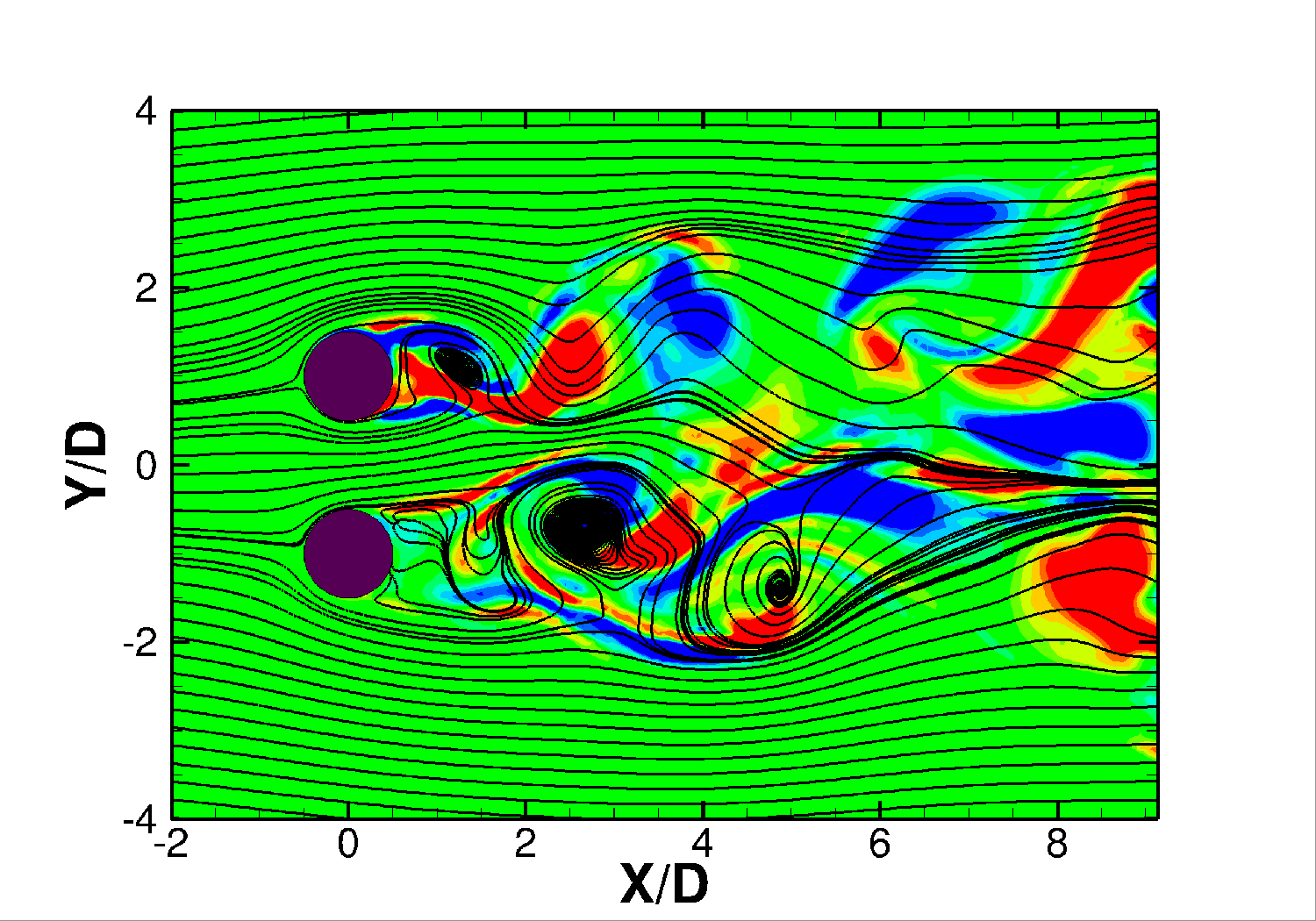}
    \caption{$\qquad$}
    \label{fig:td20_z8_swx}
    \end{subfigure}
\caption{Streamline and contours of the streamwise vorticity $\omega_x$ and the spanwise vorticity $\omega_z$ of cylinders in SSBS arrangement at $(x,y)$-plane at $Re=500$, $g^*=1.0$, $tU/D=300$, $\omega_x=\pm 1.0$ \emph{(contours)}, $\omega_z=\pm 1.0$ \emph{(solid-dash lines)} in (a,c) and sectional streamlines in (b,d): (a,b) $l^*=4$; (c,d) $l^*=8$}
\label{fig:td20_snapshots}
\end{figure}
\begin{figure} \centering
	\begin{subfigure}[b]{1.0\textwidth}	
	\centering
	\hspace{-25pt}\includegraphics[trim=0.1cm 0.1cm 0.1cm 0.1cm,scale=0.215,clip]{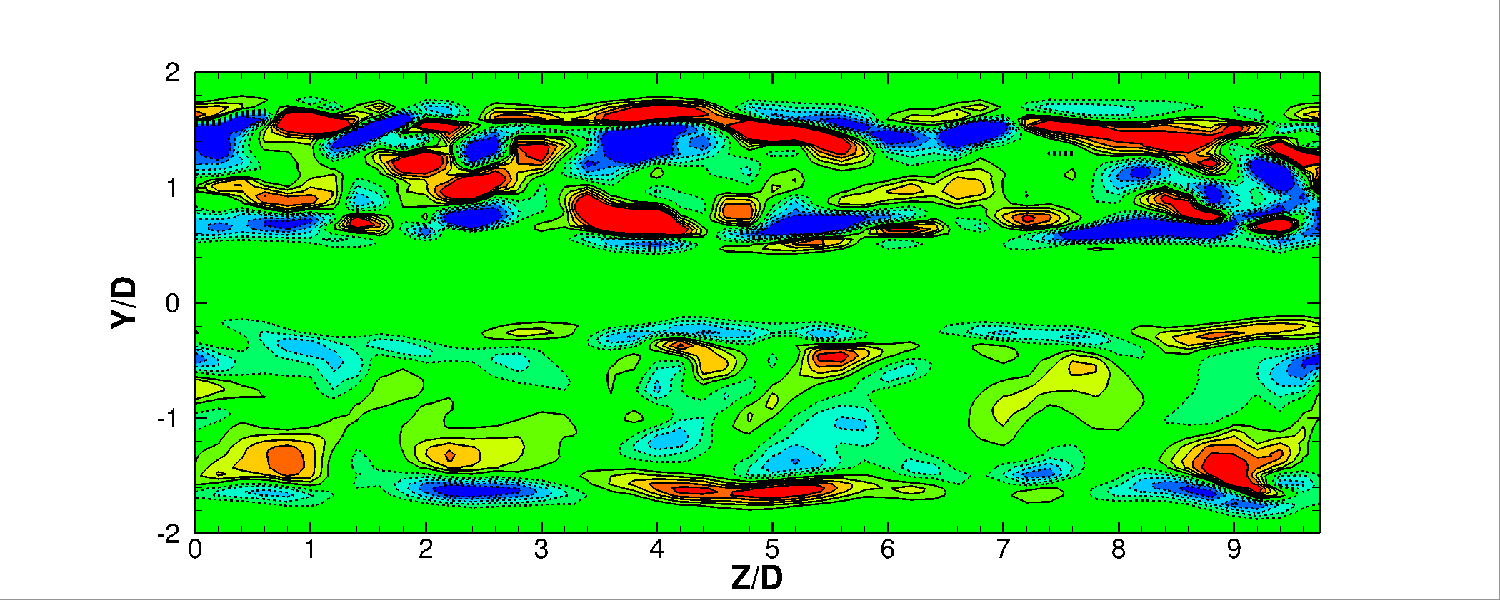}
    \caption{$\qquad$}
	\label{fig:con_td18_175a}
	\end{subfigure}
	\begin{subfigure}[b]{1.0\textwidth}
	\centering
	\hspace{-25pt}\includegraphics[trim=0.1cm 0.1cm 0.1cm 0.1cm,scale=0.215,clip]{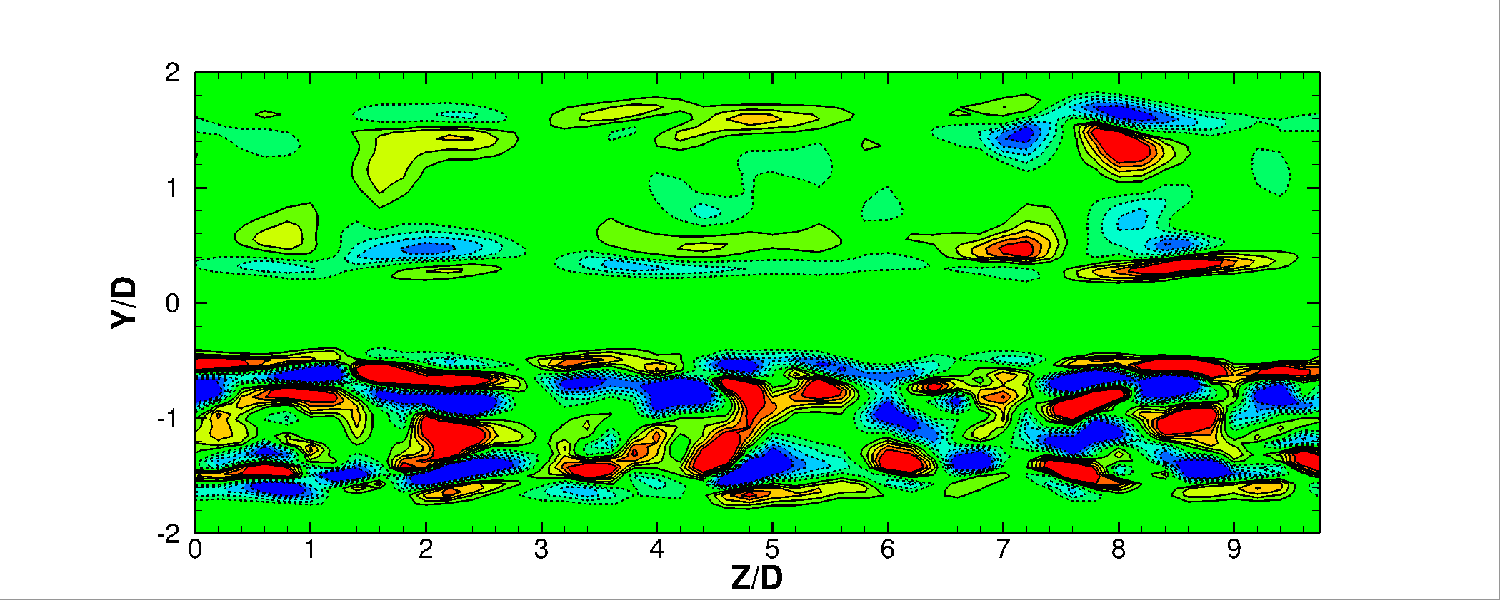}
    \caption{$\qquad$}
    \label{fig:con_td18_350a}
    \end{subfigure}
\caption{$\omega_x$ contours in $(y,z)$-plane for the cylinders in SSBS arrangement at $Re=500$, $g^*=0.8$, $\omega_x=\pm 1.0$ and $x/D=1$: (a) $tU/D=175$, the gap flow momentarily deflects to \emph{Cylinder1} (top section); (b) $tU/D=350$ the gap flow momentarily deflects to \emph{Cylinder2} (bottom section).}
\label{fig:con_td18_175_350}
\end{figure}

To understand the aforementioned behaviour, we consider a stationary isolated cylinder at representative moderate Reynolds numbers. Figure~\ref{fig:iso_snapshots} shows the $(x,y)$-sectional snapshots at $l^*=5D$ in the near-wake regions at $Re=500$ and $Re=800$. The intense streamwise vortex rollers are formed along the interface of the counter-signed vortex separating layers and across 
the saddle-point regions. This observation conforms with the topological description of 
the turbulent flow pattern from \cite{Perry1987ARoFM}. 
Based on the definition of streamline saddle point (SSP), Appendix~\ref{appA} shows that the inflection velocity profiles are indeed in the saddle-point region. These inflection velocity profiles are subsequently visualized in figures~\ref{fig:saddle_profile} (a,b) and (c,d) at $t=3/5T$ when the SSP appears at the investigated location, ($1.33D$, $-0.72D$, $5D$). In particular, a high streamwise vorticity concentration of the same sign appears 
on both sides of the saddle point along the $z$-axis in figure~\ref{fig:con_283_ws}, which conforms with the observation 
from \cite{Zhou1994JoFM}, whereby the streamwise vortical structures  
are inclined and crossed approximately at the saddle-point region on the $(x,y)$-plane. 
Hence the presence of the high-strain rates reflects the significance 
of the two-dimensional SSP to the three-dimensionality of this flow.

Following that, a stability analysis using the DMD technique is performed in a saddle-point region at low Reynolds number in Appendix~\ref{appB}. 
The near-wake instability around the saddle-point is found to be dependent upon the intensity 
of the fluid momentum and the vorticity concentration difference of adjacent voritcity clusters. Recently, \cite{Huang2010PoF} also reported that a planar shear flow could enhance the three-dimensionality in the wake behind a circular cylinder. \footnote{Here a planar shear flow refers to an inflow with a constant velocity gradient along $y$-axis in the present problem configuration.} In the present investigation, we further analyze the relationship between the near-wake instability and the resultant imbalanced vortex-to-vortex interaction from the planar shear flow. 
The fluid shearing is known to be critical to the Kelvin-Helmholtz instability. 
As the Reynolds number increases, the fluid instability can develop around the 
inflection points along a velocity profile which have a direct connection with the fluid shear stresses. 
Since significant shear stresses are observed on the interface of the imbalanced counter-signed vorticity clusters, the interaction between different vorticity clusters is believed to be crucial to the near-wake instability. A supportive example could be the saddle-point at the tip of the formation region behind a stationary isolated circular cylinder at $Re \lesssim 48$, where the symmetric counter-signed circulations are interacting and no instability is observed. It is when the perturbation approaches to the brink of a critical value, the perturbation becomes non-negligible and induces the need for the extra dimension to quantify itself e.g., the introduction of new dimension in the Hopf bifurcation and the flow transition from the laminar flow to the chaotic turbulent flow. These two factors associated with the near-wake instability facilitate the understanding of the proximity interference induced from the gap flow behaviour in a three-dimensional flow in Section~\ref{sec:3D_gap_flow}. 

To link the observations on the saddle-point regions to the recovery of two-dimensional hydrodynamic responses, a series of streamline-contour plots of a locked-in cylinder are investigated for one primary vortex shedding cycle in figure~\ref{fig:scon_500r5}. Based on the discussion about a two-dimensional streamline saddle point in Appendix~\ref{appA}, the saddle-point generates a local stagnant region which inhibits the transfer of kinetic energy from the mean flow. The saddle-point moves with the kinematics of the separating shear layer along the interface and represents a communication barrier. As the vortex wake reaches its maximum growth, it breaks up and sheds downstream. 
The strict periodic motion of the cylinder at the peak lock-in stage also generated a well segregated vortex wakes downstream with relatively benign interactions from the vortex wakes\footnote{The focus is to investigate the influence of a saddle-point region to the vortex shedding process. Hence the vortex shedding modes are not discussed here.}. 
This vortex shedding mode inhibits the formation of the fluid shearings along the interface of vortex wakes. 
Consequently,  the SSP vanishes and the corresponding streamwise vorticity concentration 
weakens downstream. A direct consequence is a recovery of the two-dimensional hydrodynamic response
along the cylinder. The intense turbulent flow at the off lock-in stage results 
into a smaller $C^{mean}_d$ value than its laminar flow counterpart 
in figure~\ref{fig:100500_cdcl}, where $C^{mean}_d$ is over-predicted by the relatively 
large two-dimensional vortex wakes at the lock-in. Overall, the response of $C_l$ 
shows an increment in the transverse fluctuating lift force and an earlier onset of the VIV lock-in. As reported by~\cite{lbjrk2016pof} for the similar problem setups with two-dimensional laminar flow, the earlier onset of VIV is attributed to the enhanced vortex interaction which leads to higher vortex shedding frequency.  
Both $C_d$ and $C_l$ at $Re=500$ show respectively about 6\% and 22\% 
amplification compared to their laminar counterparts from~\cite{lbjrk2016pof} at the peak lock-in. 
These results indicate that the VIV regulation effect has a profound influence to 
the transverse hydrodynamic response, compared with its streamwise one.
A similar phenomenon was observed by~\cite{Zhao2014JoFaS}, in which the spanwise correlations were discussed at the VIV lock-in stage and a uniformity of $C_l$ was observed along the span of a rigid circular cylinder. 
Here, the primary focus is to understand the complex near-wake flow physics in the SBS arrangements. 

\section{Three-dimensional flow interference to the gap flow}\label{sec:3D_gap_flow}
In this section, the relationship between the gap-flow proximity and 
the near-wake instability is discussed for the cylinders with SBS arrangements.
An incorporation of the interference from the VIV kinematics is important to analyze the practical applications and operations in side-by-side systems. So far the three-dimensional numerical investigation of the gap flow instability in the SBS arrangements is rarely documented. Hence, the present investigation on the VSBS arrangements in a 3D flow is deemed as another step further to understand the gap flow and the VIV kinematics. From a systematic analysis viewpoint, it is desirable to first focus merely on the interaction between the gap-flow kinematics and the 3D flow by eliminating the motion of the structure.
\begin{figure} \centering
	\begin{subfigure}[b]{0.5\textwidth}	
	\centering
	\hspace{-25pt}\includegraphics[trim=0.1cm 0.1cm 0.1cm 0.1cm,scale=0.3,clip]{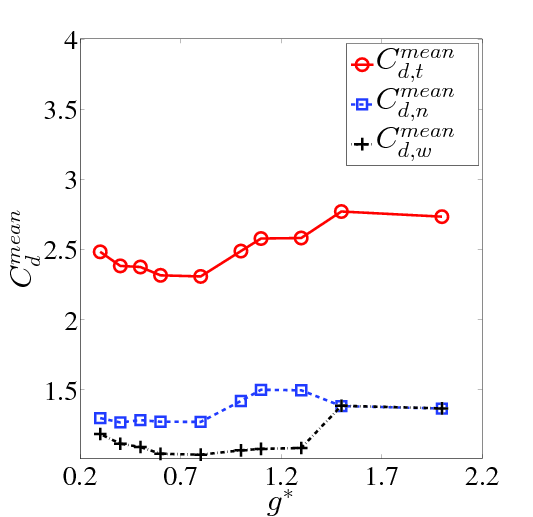}
    \caption{$\qquad$}
	\label{fig:cd_vs_g}
	\end{subfigure}%
	\begin{subfigure}[b]{0.5\textwidth}
	\centering
	\hspace{-25pt}\includegraphics[trim=0.1cm 0.1cm 0.1cm 0.1cm,scale=0.3,clip]{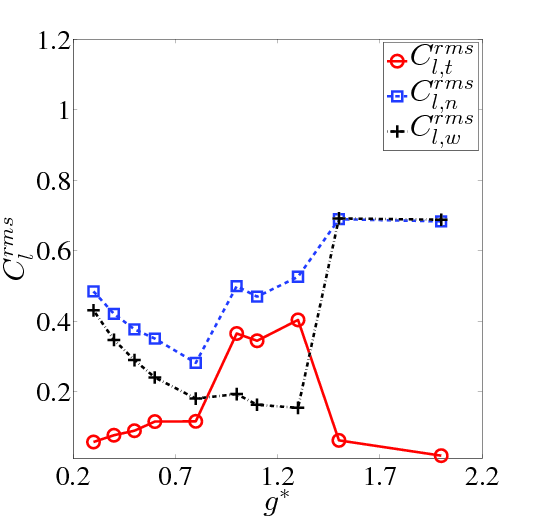}
    \caption{$\qquad$}
    \label{fig:cl_vs_g}
    \end{subfigure}
\caption{Hydrodynamic forces as a function of the gap ratio: (a) mean drag coefficient with respect to the gap ratio, (b) r.m.s. lift coefficient with respect to the gap ratio. The subscripts $t$, $n$ and $w$ denote respectively the total, the narrow and the wide near-wake regions. $C^{mean}_{d}$ and $C^{rms}_{l}$ are higher along the cylinder with a narrow near-wake region, where a force modulation is observed.}
\label{fig:cdcl_vs_g}
\end{figure}
\begin{figure} \centering
	\begin{subfigure}[b]{0.5\textwidth}
	\centering
	\hspace{-25pt}\includegraphics[trim=0.1cm 0.1cm 0.1cm 0.1cm,scale=0.097,clip]{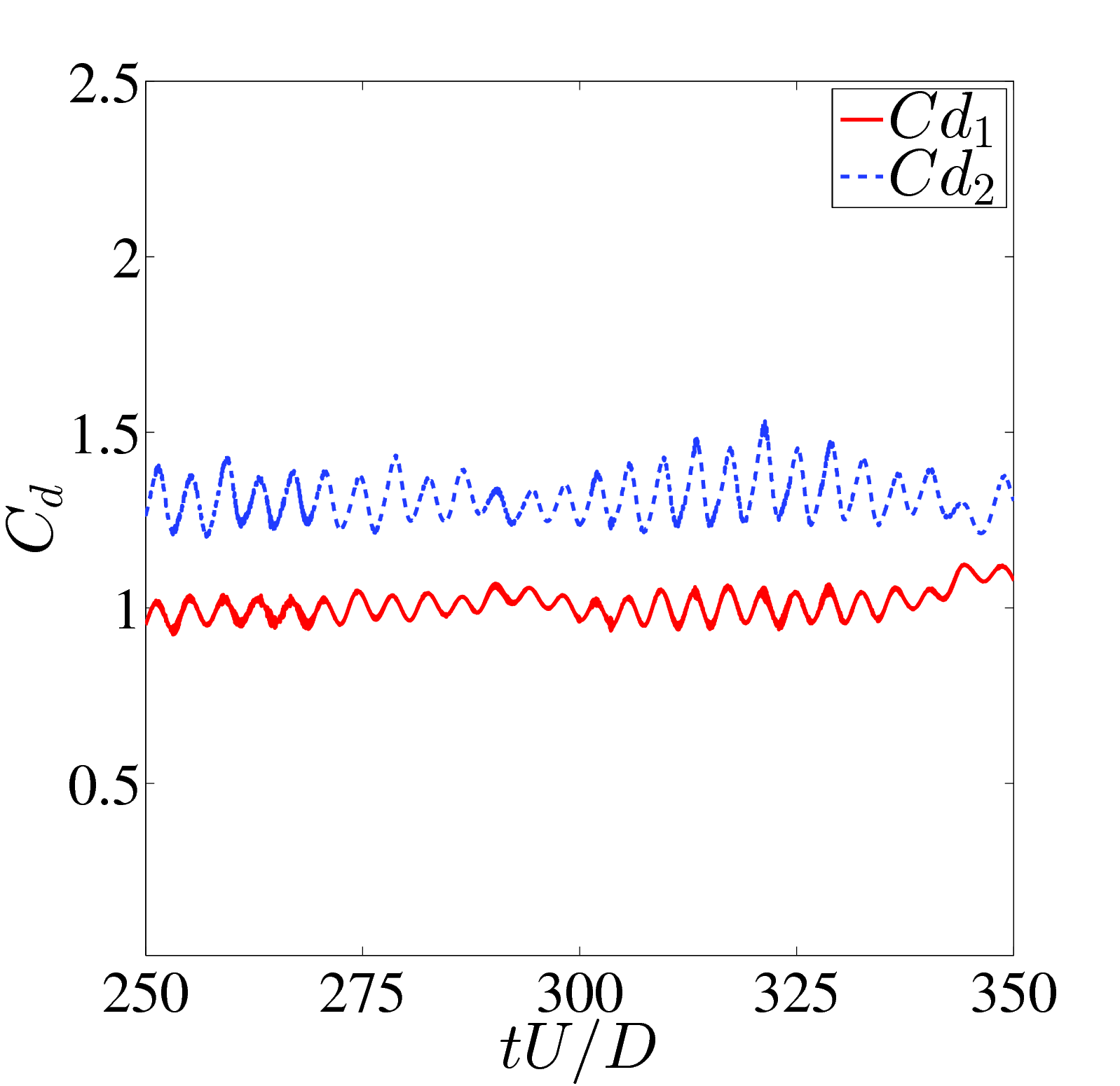}
    \caption{$\qquad$}
    \label{fig:td18r3_cd}
    \end{subfigure}%
	\begin{subfigure}[b]{0.5\textwidth}
	\centering
	\hspace{-25pt}\includegraphics[trim=0.1cm 0.1cm 0.1cm 0.1cm,scale=0.097,clip]{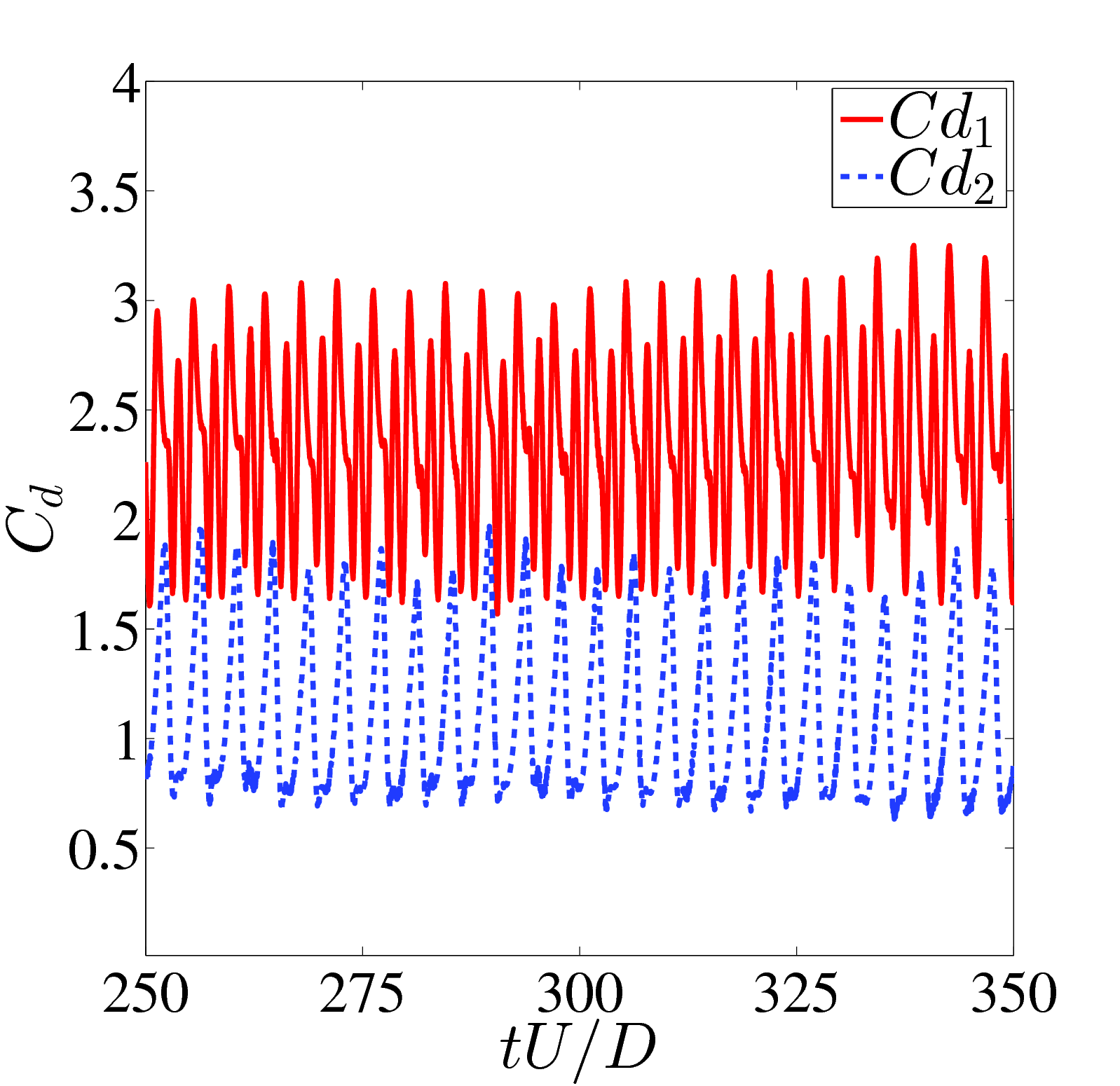}
    \caption{$\qquad$}
    \label{fig:500td18r4_cd}
    \end{subfigure}
    \begin{subfigure}[b]{0.5\textwidth}	
	\centering
	\hspace{-25pt}\includegraphics[trim=0.1cm 0.1cm 0.1cm 0.1cm,scale=0.097,clip]{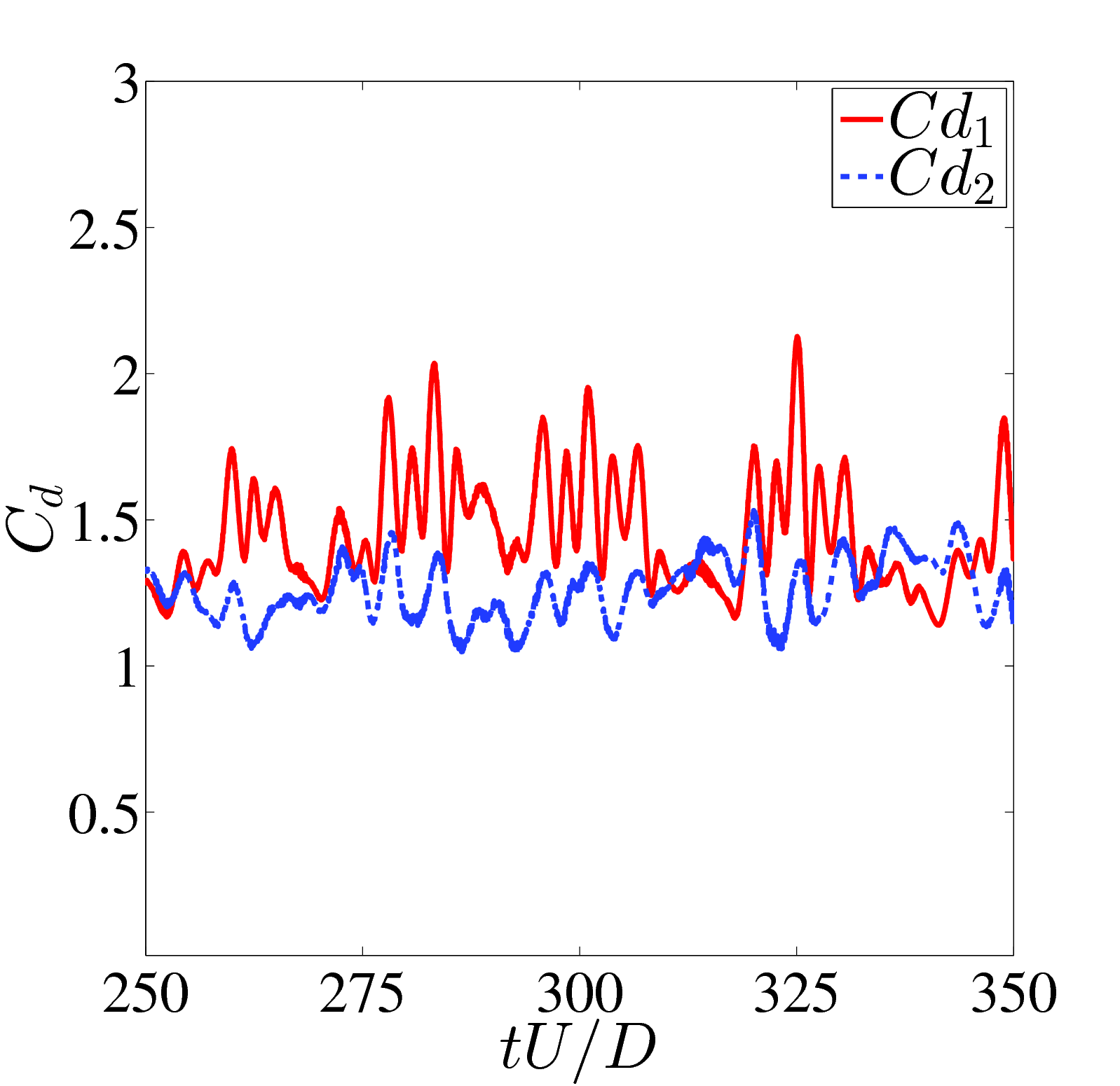}
    \caption{$\qquad$}
	\label{fig:td18r6_cd}
	\end{subfigure}%
	\begin{subfigure}[b]{0.5\textwidth}
	\centering
	\hspace{-25pt}\includegraphics[trim=0.1cm 0.1cm 0.1cm 0.1cm,scale=0.097,clip]{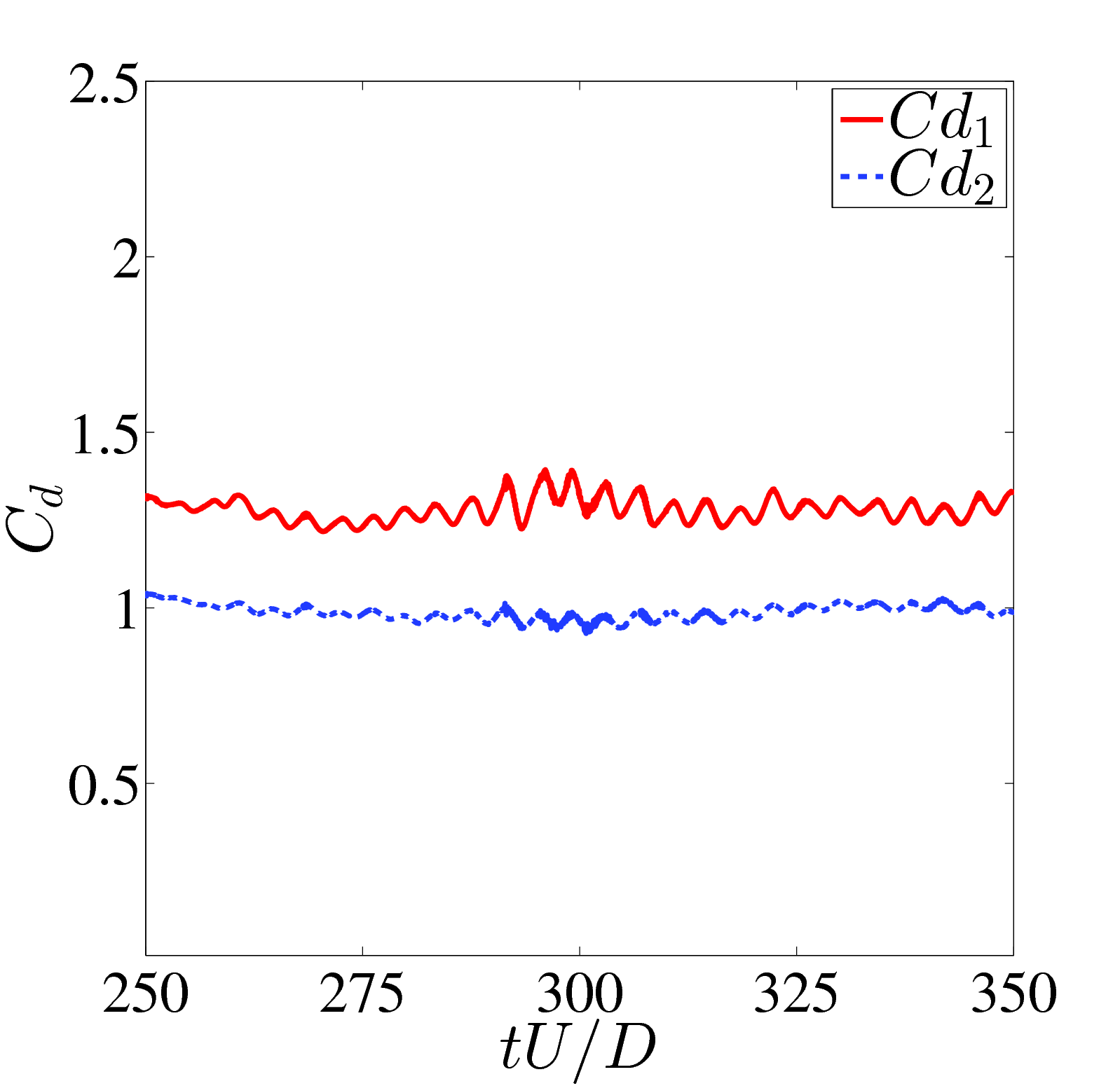}
    \caption{$\qquad$}
    \label{fig:500td18r8_cd}
    \end{subfigure}
\caption{Time series of the drag coefficient of cylinders in the VSBS arrangements where \emph{Cylinder1} vibrates in the cross-flow direction at $Re=500$, $g^*=0.8$, $m^*=10$ and $\zeta=0.01$: $U_r = $ (a) $3$; (b) $4$; (c) $6$ and (d) $8$}
\label{fig:td18_cds}
\end{figure}
\begin{figure}
\begin{center}
\centering
\includegraphics[trim=0.1cm 0.1cm 0.1cm 0.1cm,scale=0.32,clip]{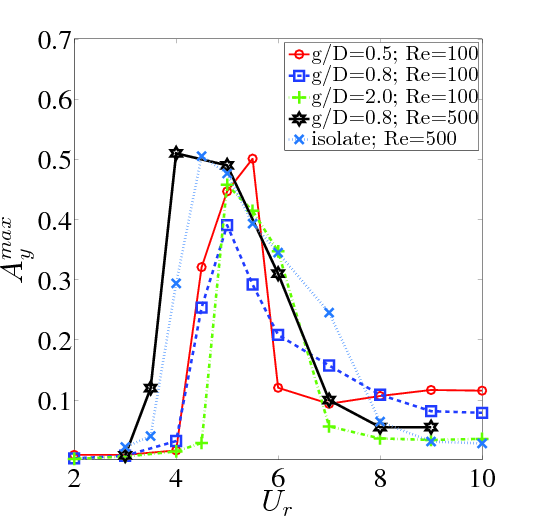}
\end{center}
\caption{Time-averaged maximum transverse vibration amplitude as a function of the reduced velocity in the VSBS arrangements at $Re=100$ and $500$, $m^*=10$ and $\zeta=0.01$. \emph{Cylinder1} vibrates in the transverse direction. The onset of the VIV lock-in occurs earlier at a smaller $U_r$ value for cases at $Re=500$}
\label{fig:Aymax_Ur}
\end{figure}
\begin{figure} \centering
	\begin{subfigure}[b]{0.5\textwidth}	
		\centering
		\hspace{-25pt}\includegraphics[trim=0.1cm 0.1cm 0.1cm 0.1cm,scale=0.097,clip]{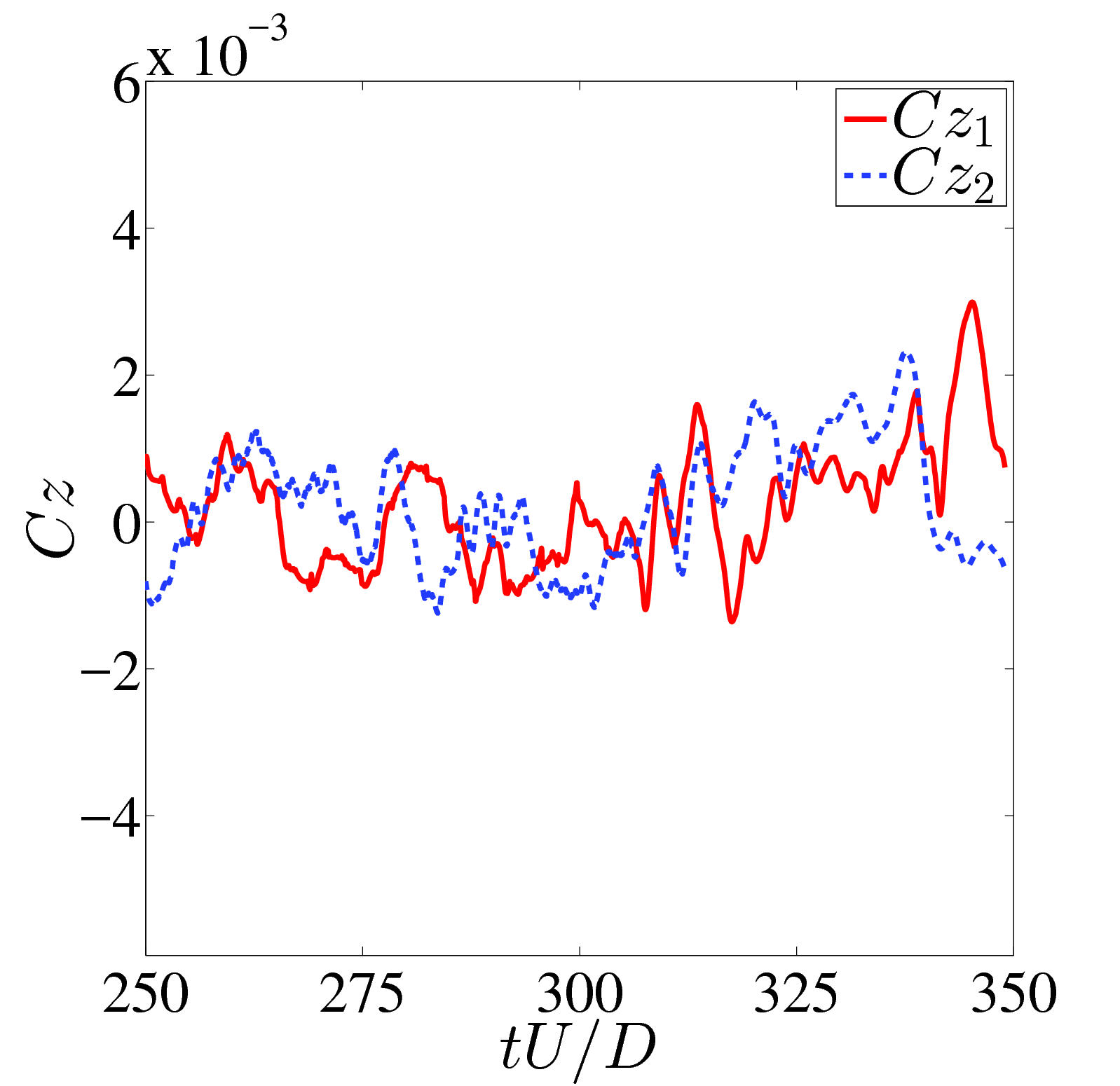}
		\caption{$\qquad$}
		\label{fig:cz_500td18r35}
	\end{subfigure}%
	\begin{subfigure}[b]{0.5\textwidth}
		\centering
		\hspace{-25pt}\includegraphics[trim=0.1cm 0.1cm 0.1cm 0.1cm,scale=0.097,clip]{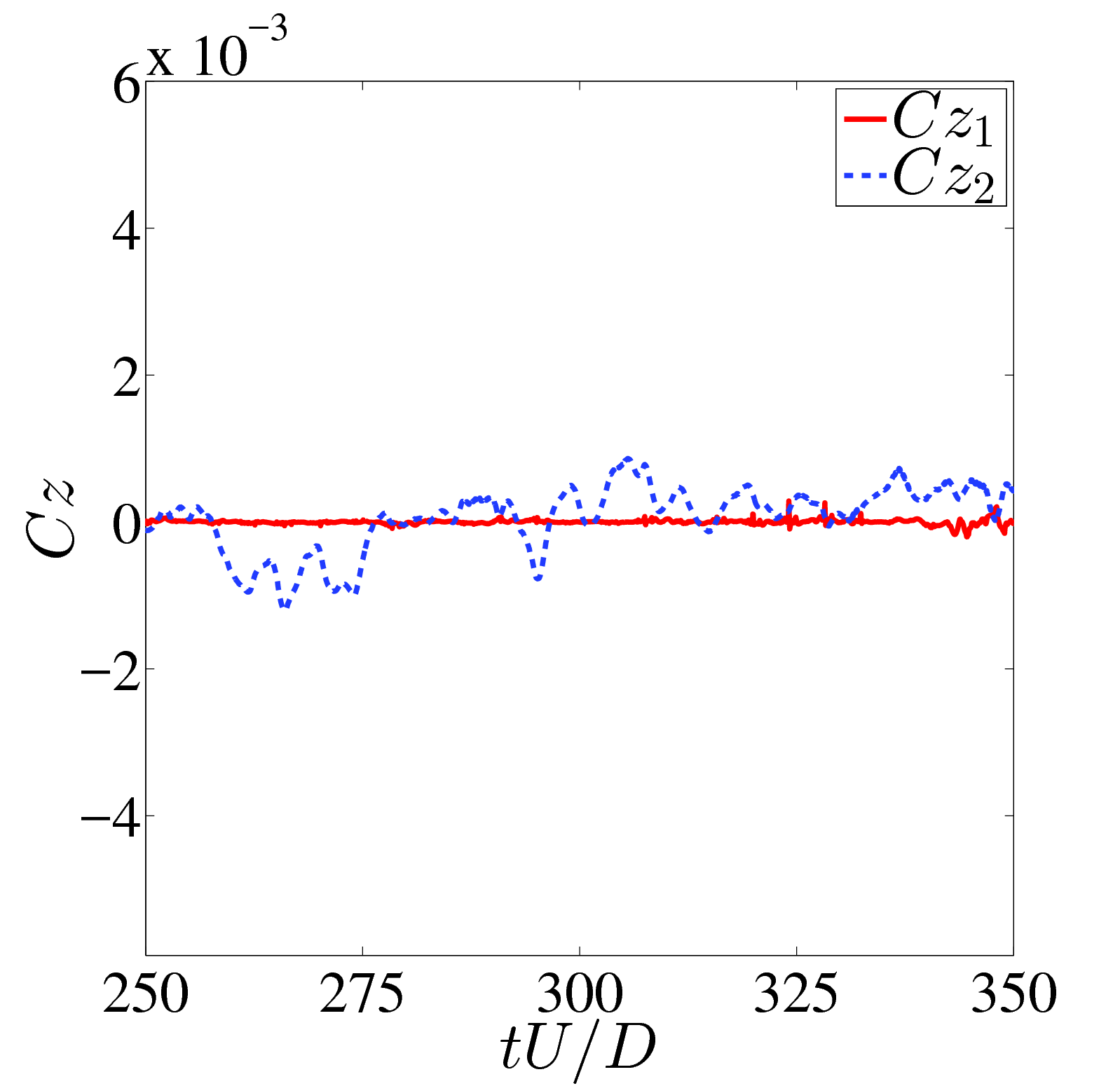}
		\caption{$\qquad$}
		\label{fig:cz_500td18r4}
	\end{subfigure}
	\begin{subfigure}[b]{0.5\textwidth}
		\centering
		\hspace{-25pt}\includegraphics[trim=0.1cm 0.1cm 0.1cm 0.1cm,scale=0.097,clip]{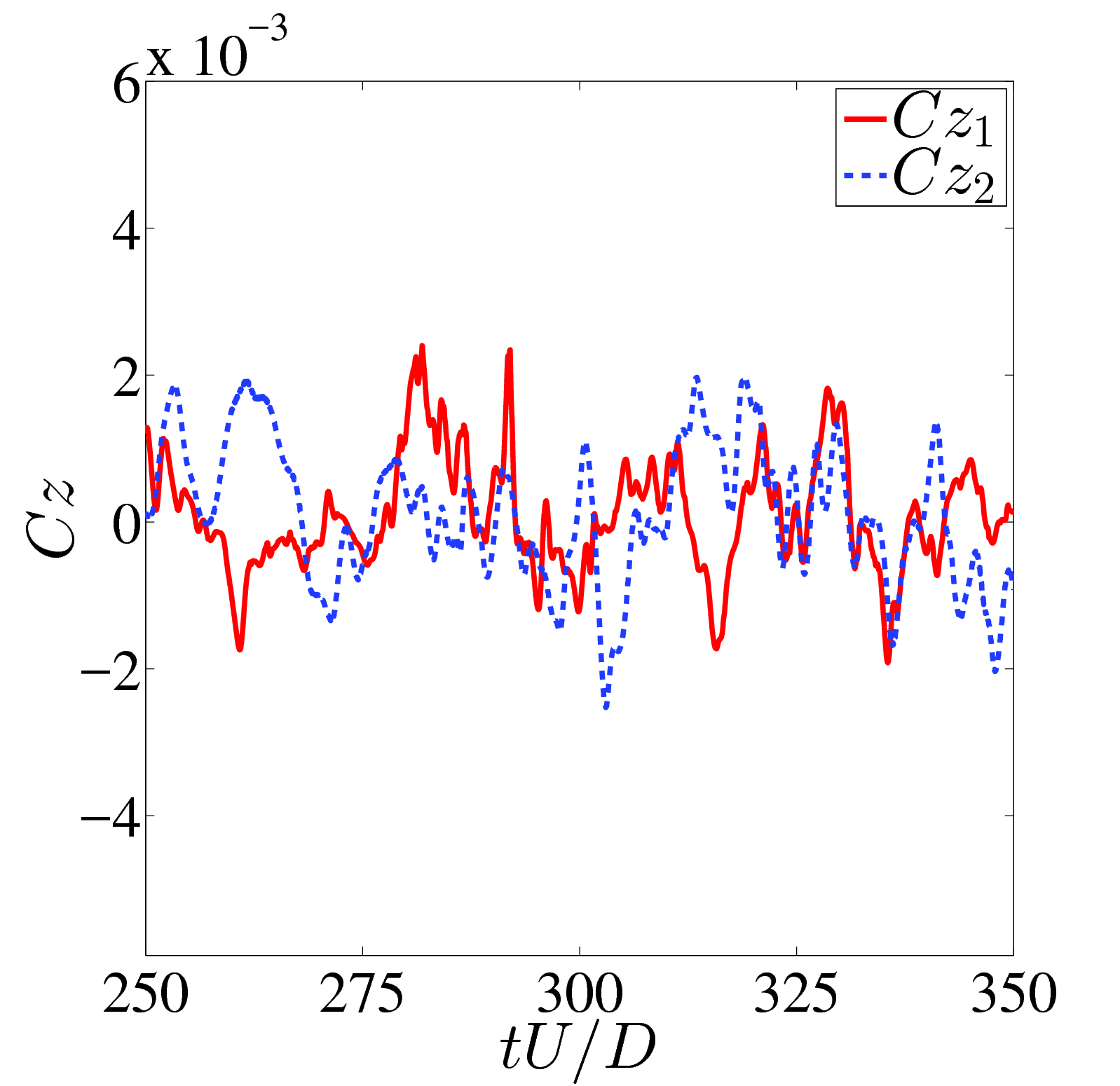}
		\caption{$\qquad$}
		\label{fig:cz_re500td18r5}
	\end{subfigure}%
	\begin{subfigure}[b]{0.5\textwidth}
		\centering
		\hspace{-25pt}\includegraphics[trim=0.1cm 0.1cm 0.1cm 0.1cm,scale=0.097,clip]{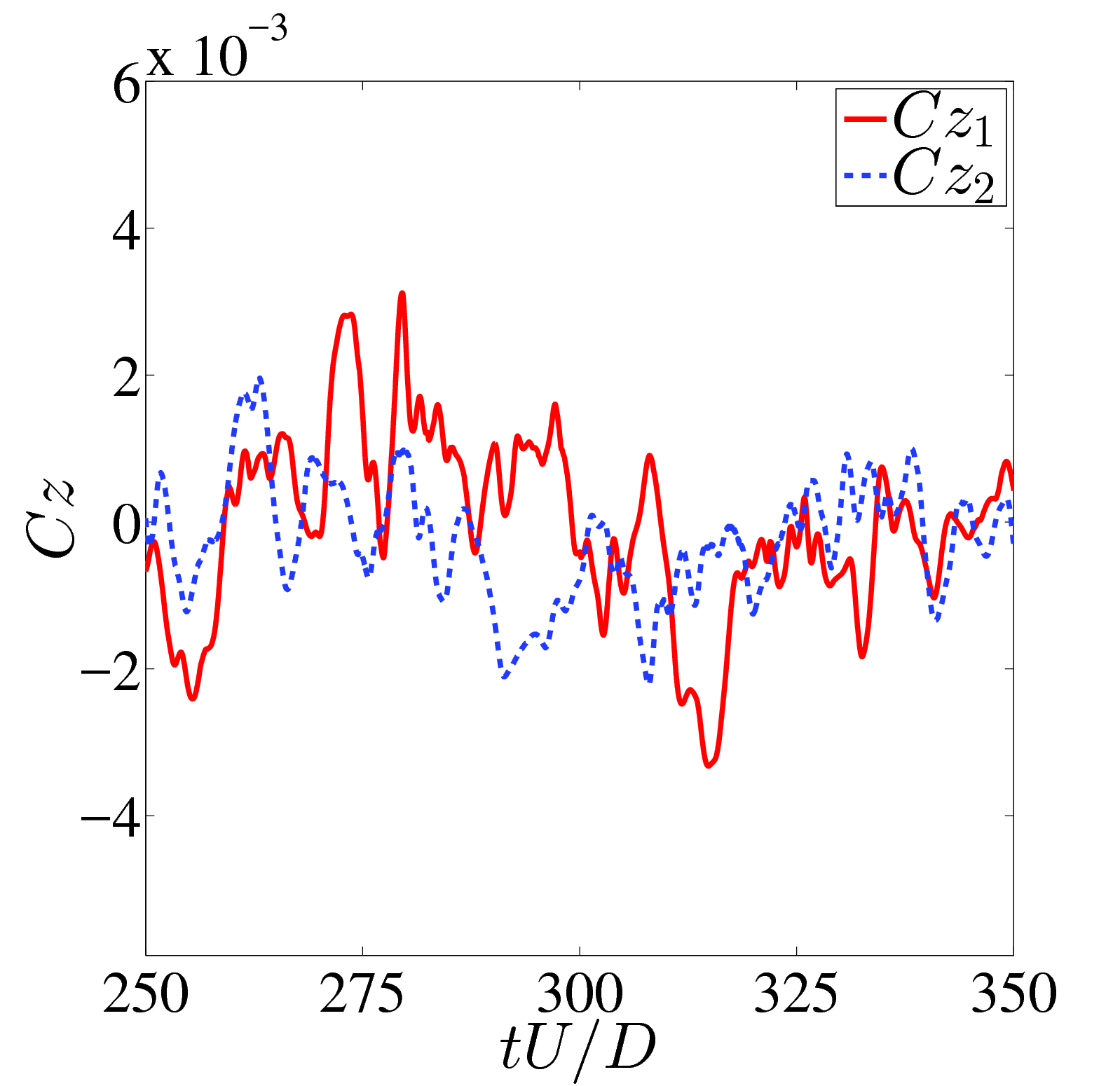}
		\caption{$\qquad$}
		\label{fig:cz_re500td18r6}
	\end{subfigure}
	\caption{Time traces of the spanwise hydrodynamic forces of the cylinders in the VSBS arrangements at $Re=500$, $g^*=0.8$, $m^*=10$ and $\zeta=0.01$: $U_r=$ (a) 3.5, (b) 4, (c) 5, (d) 6. A recovery of two-dimensional hydrodynamic responses is observed along the locked-in cylinder.}
	\label{fig:czs_500td18}
\end{figure}
\begin{figure} \centering
	\begin{subfigure}[b]{0.5\textwidth}	
		\centering
		\hspace{-25pt}
		\includegraphics[trim=0.1cm 0.5cm 0.1cm 0.1cm,scale=0.097,clip]{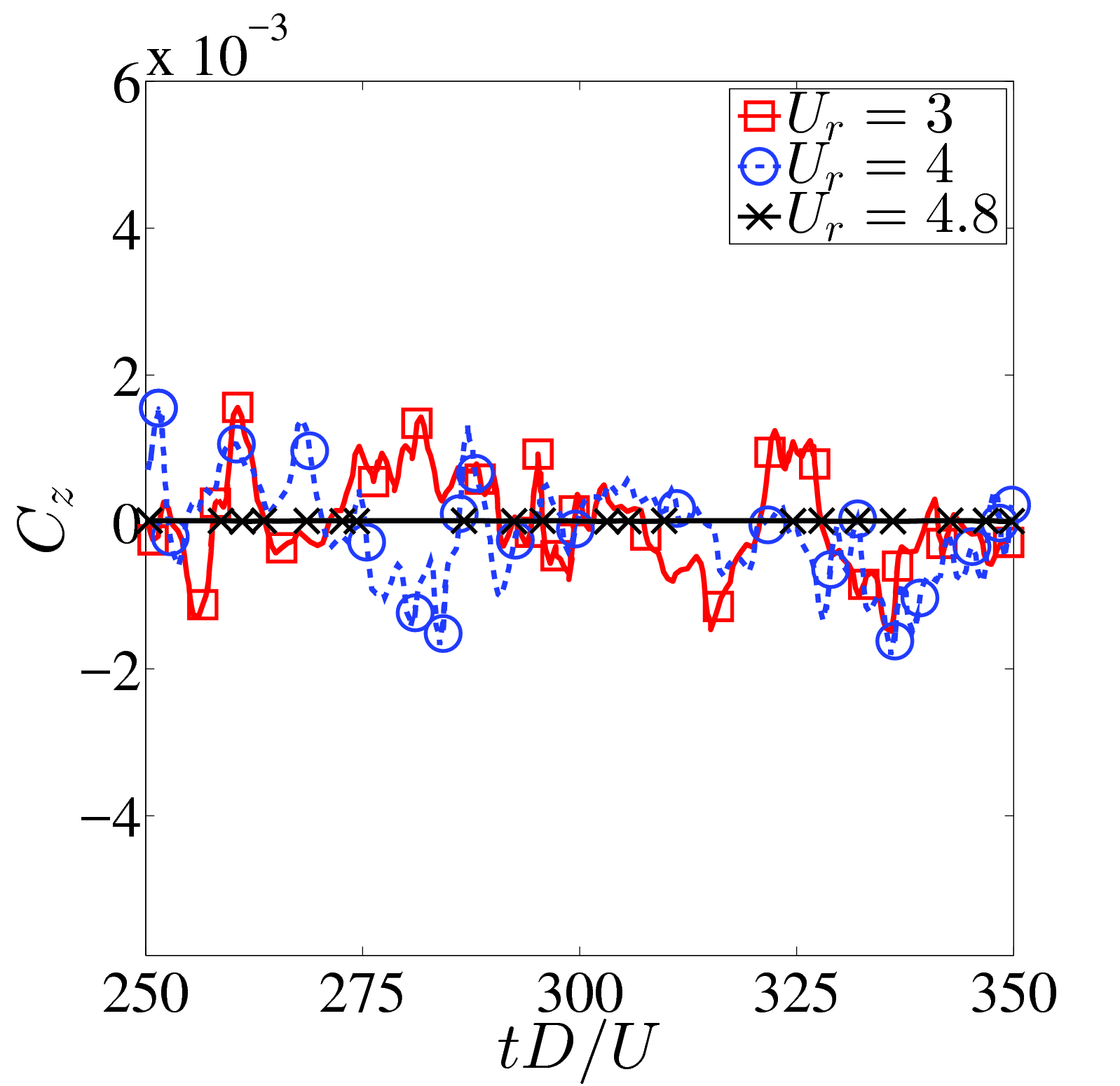}
		\caption{$\qquad$}
		\label{fig:re800r48_cz}
	\end{subfigure}%
	\begin{subfigure}[b]{0.5\textwidth}
		\centering
		\hspace{-25pt}\includegraphics[trim=0.1cm 0.2cm 0.1cm 0.1cm,scale=0.38,clip]{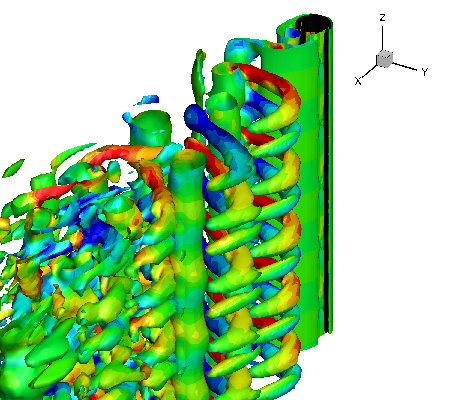}
		\caption{$\qquad$}
		\label{fig:con_re800r48}
	\end{subfigure}
	\caption{Recovery of two-dimensional hydrodynamic responses of a freely-vibrating cylinder at the peak lock-in: $Re=800$, $m^*=10$, $\zeta=0.01$ and $U_r=4.8$: (a) $C_z$ at $tU/D \in [250$, $350]$, where two dimensional hydrodynamic responses are observed along the cylinder span; (b) iso-surfaces using the Q-Criterion at $tU/D = 300$, $Q = 0.2$ $\omega_y = \pm 1$ (\emph{contours}).}
	\label{fig:re800r48}
\end{figure}
\begin{figure} \centering
    \begin{subfigure}[b]{0.5\textwidth}	
	\centering
	\hspace{-25pt}\includegraphics[trim=0.1cm 0.1cm 0.15cm 0.1cm,scale=0.125,clip]{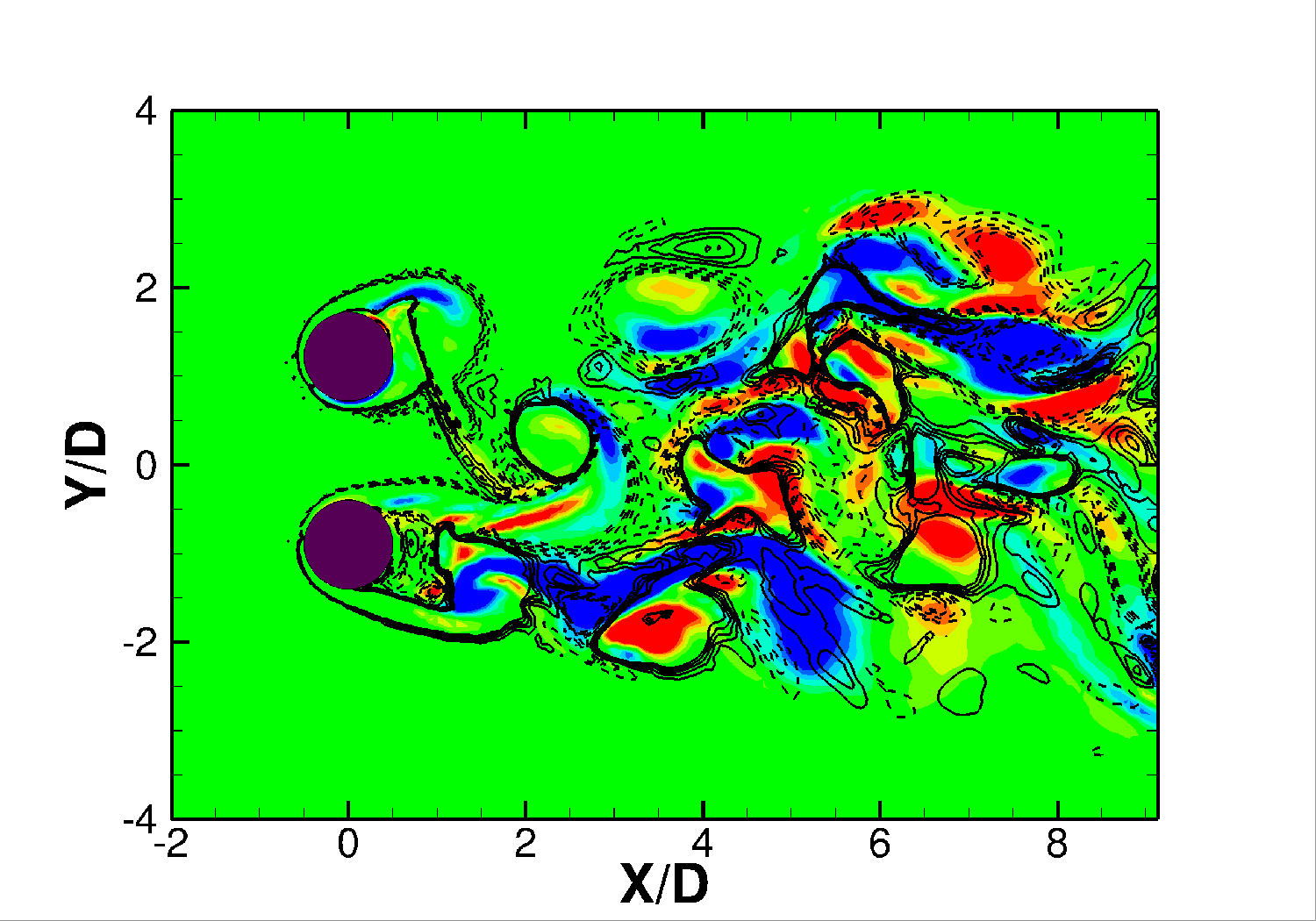}
    \caption{$\qquad$}
	\label{fig:td18r4_z4_wxz}
	\end{subfigure}%
	\begin{subfigure}[b]{0.5\textwidth}
	\centering
	\hspace{-25pt}\includegraphics[trim=0.1cm 0.1cm 0.15cm 0.1cm,scale=0.125,clip]{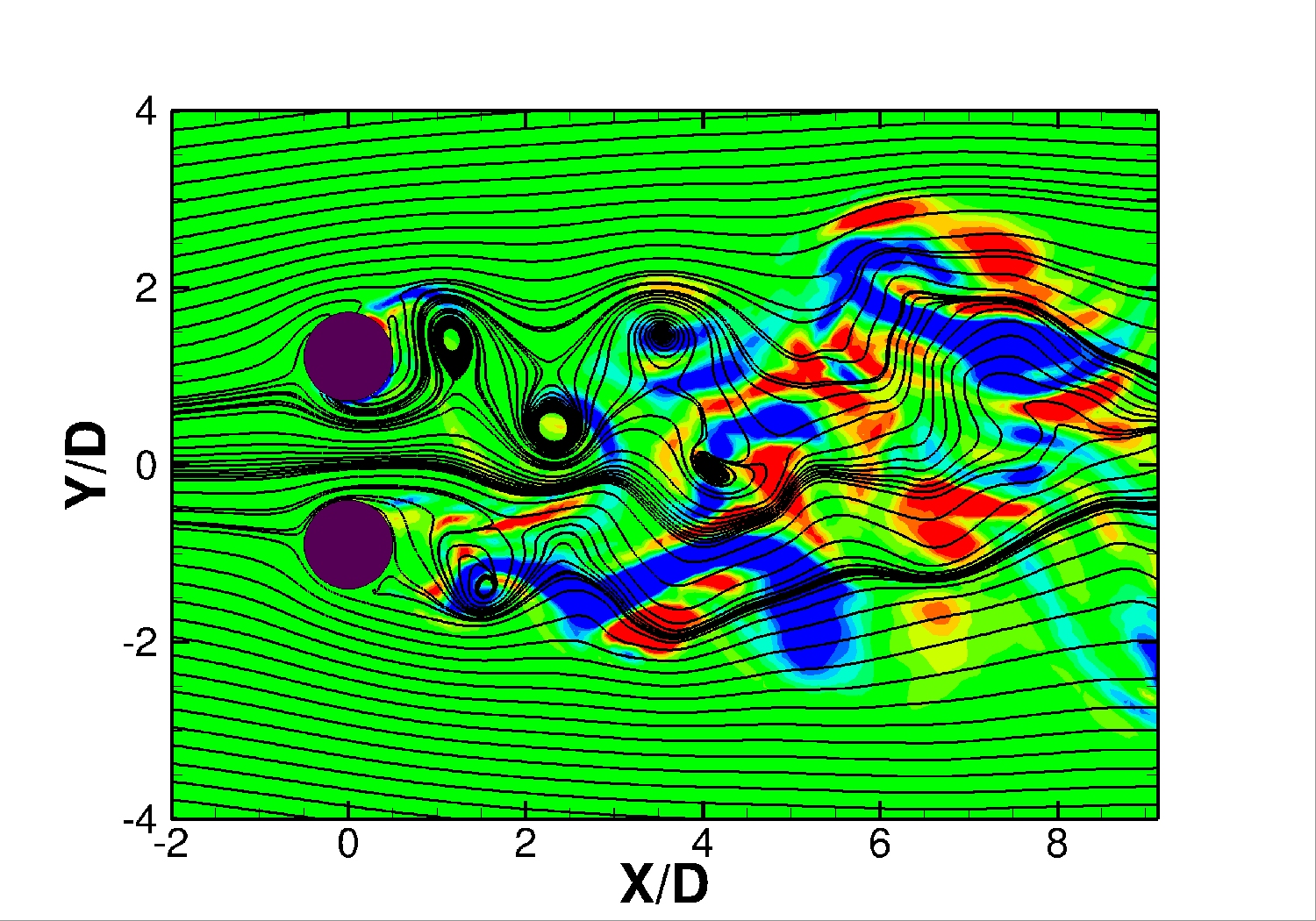}
    \caption{$\qquad$}
    \label{fig:td18r4_z4_swx}
    \end{subfigure}
    \begin{subfigure}[b]{0.5\textwidth}	
	\centering
	\hspace{-25pt}\includegraphics[trim=0.1cm 0.1cm 0.15cm 0.1cm,scale=0.125,clip]{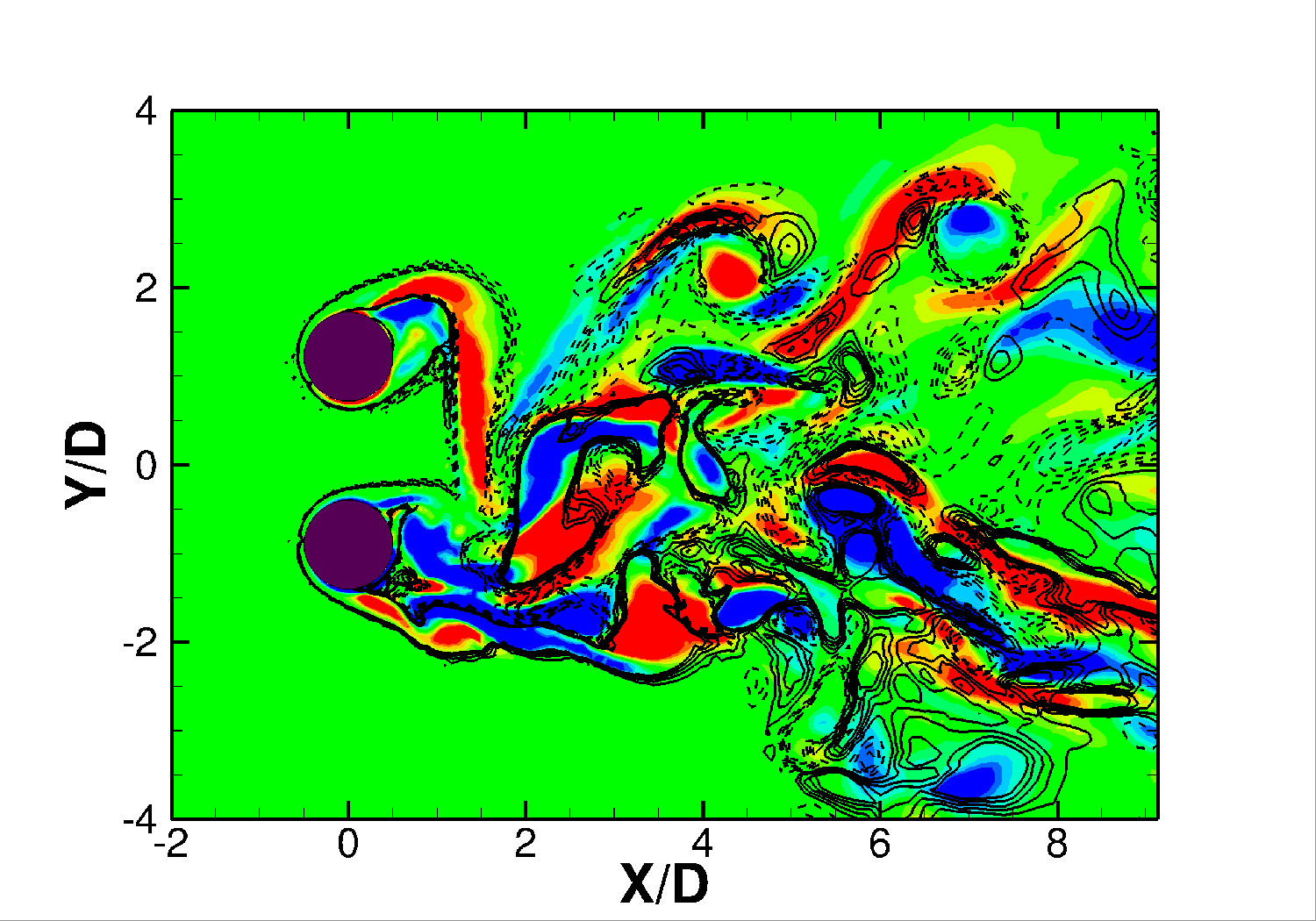}
    \caption{$\qquad$}
	\label{fig:td18r4_z8_wxz}
	\end{subfigure}%
	\begin{subfigure}[b]{0.5\textwidth}
	\centering
	\hspace{-25pt}\includegraphics[trim=0.1cm 0.1cm 0.15cm 0.1cm,scale=0.125,clip]{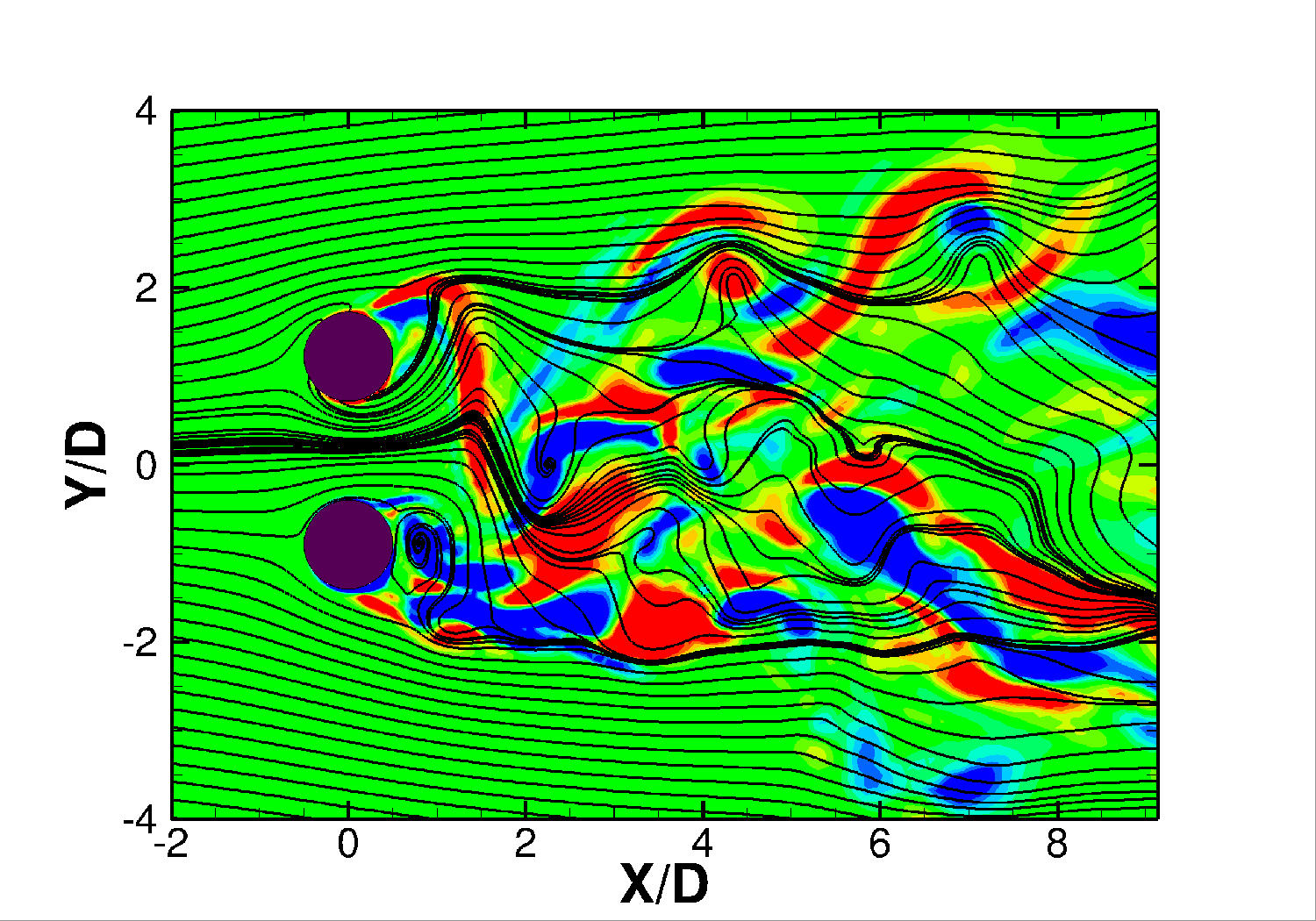}
    \caption{$\qquad$}
    \label{fig:td18r4_z8_swx}
    \end{subfigure}
\caption{Streamline and contours of the streamwise vorticity $\omega_x$ and the spanwise vorticity $\omega_z$  of the cylinders in VSBS arrangement at $(x,y)$-plane for $Re=500$, $g^*=0.8$, $m^*=10$, $\zeta=0.01$, $U_r=4.0$ \emph{(the peak lock-in stage)}, $\omega_x=\pm 1.0$ \emph{(colour contours)}, $\omega_z=\pm 1.0$ \emph{(solid-dash)} in (a, c), and streamline in (b, d): (a,b) $l^*=4$; (c,d) $l^*=8$}
\label{fig:td18r4_snapshots}
\end{figure}

The flip flop was frequently described as an intermittent deflection of the gap flow. As reported by \cite{lbjrk2016pof}, the switch-over of $C^{mean}_d$ from each cylinder in the SSBS arrangement could indicate the direction of gap-flow deflection. Since the vortex-to-vortex interaction is enhanced in the narrow near-wake region, the corresponding $f_{vs}$ value is higher than its wide near-wake-region counterpart. To investigate the characteristics related to gap flow features without and with the presence of 3D effects, figure~\ref{fig:clcd_100500td18} is plotted. In the two-dimensional laminar flow, figures~\ref{fig:cl_re100td18} and~\ref{fig:cd_re100td18}, $f_{vs}$ of the cylinder with the narrow near-wake region is observed possessing a larger value. However, this tendency is not confirmed in its three-dimensional counterpart, as shown in figure~\ref{fig:cl_re500td18} and figure~\ref{fig:cd_re500td18}. The figures show that there is no significant difference among the mean vortex-shedding frequency of two cylinders for the 3D flow, although the gap flow deflects. In addition, the flip flop is not observed in the selected time window $tU/D \in [250$, $350]$ in figure~\ref{fig:cd_re500td18}, since $f_{flip}$ is remarkably low for the 3D flow. A comparison of $C_z$ between the SSBS arrangement at $g^*=0.8$ and the stationary isolated cylinder at $Re=500$ is plotted in figure~\ref{fig:cz_re500_isosbs}. The fluctuation of $C_z$ is evidently amplified by a factor of $1.3$ along the cylinder with the narrow near-wake region, \emph{Cylinder2} at $tU/D \in [250$, $350]$, compared with its isolated and wide counterparts respectively. This amplification factor is practically constant in all other SSBS arrangements at $g^* \in [0.6$, $1.3]$. As discussed in Appendix~\ref{appB}, the near-wake instability is dependent upon the gap-flow proximity interference. Consequently, the three-dimensional structure is prevailing in the narrow near-wake region where the gap-flow proximity interference is significant, as visualized in figure~\ref{fig:con_iso&td18}. 

A further investigation on the velocity profile in figure~\ref{fig:profile_1} exhibits that the three-dimensional gap flow at higher Reynolds number possesses a much larger fluid shearing than its two-dimensional counterparts. 
This high velocity gradient arising from the gap flow promotes additional instability in the near-wake region. Considering the two critical factors identified in Section~\ref{sec:3D_viv}, we also observe another unstable factor in the narrow near-wake region, large adjoining interfaces of primary vorticity clusters, as shown in figure~\ref{fig:con_td18}.
Remarkable shear stresses are present along these adjoining interfaces and result in a significant streamwise vorticity concentration formed in the narrow near-wake region in figure~\ref{fig:td20_snapshots}. 
The locations with intensified streamwise vorticity clusters follow well along these interfaces in the near-wake region, and confirms the observation about SSP in Section~\ref{sec:3D_viv}, where the SSP lies right at the point with significant streamwise vorticity concentration along the interface of primary vorticity clusters. A more straightforward visualization is exhibited by $(y,z)$-sectional contour plots of $\omega_x$ in figure~\ref{fig:con_td18_175_350}. Figures~\ref{fig:con_td18_175a}
and~\ref{fig:con_td18_350a} refer to the $\omega_x$ contour plots of the SSBS arrangement, where the gap flow deflects to \emph{Cylinder1} and \emph{Cylinder2} at $tU/D=175$ and $tU/D=350$ respectively. Two distinct streamwise vorticity distributions were evidently shown in the narrow and wide near-wake regions.

This asymmetric distribution of streamwise vorticity has a profound influence to the hydrodynamic response. Here figure~\ref{fig:cd_vs_g} shows distinctively higher and lower $C^{mean}_d$ values for the cylinders with a narrow and wide near-wake regions, respectively. Furthermore the algebraic sum of $C^{mean}_{d}$ shows a base-bleeding type effect, which is well-documented for the SSBS arrangements in the literature, e.g., \cite{bearman1973interaction}. Hence the overall response of $C_d$ is diminished. However, this base-bleeding effect is weakened as the value of $g^*$ increases beyond the deflected gap-flow regime. To analyze the transverse hydrodynamic response, $C^{rms}_l$ is adopted to indicate the fluctuating extent of $C_l$ as a function of the gap ratio $g^*$ in figure~\ref{fig:cl_vs_g}.
Here $C^{rms}_{l}$ represents the fluctuation intensity (absolute value) of transverse force and is measured from the $C^{mean}_{l}$ value between a time interval, when the gap flow stably deflects to one particular side of the SSBS arrangements. It should be noted that the in-phase and anti-phase of $C_l$ from both cylinders have to be taken into account, when taking the measurement of the resultant transverse force fluctuation for SBS arrangements. 
Similar to figure~\ref{fig:cd_vs_g}, a drastic transverse fluctuation of the lift appears along the cylinder with the narrow near-wake region. The overall transverse fluctuation of $C_l$ is calculated as a sum of $C_l$ from each cylinder, the \emph{solid circle} in figure~\ref{fig:cl_vs_g}. A force modulation is clearly shown at the deflected gap flow regime, $g^* \in [0.8$, $1.5]$, where the overall transverse fluctuation of $C_l$ is excited by a factor of $2.4$. Since the gap flow is significantly suppressed at $g^* \lesssim 0.5$ with a remarkable proximity interference, the overall fluctuation of $C_l$ is much benign. On the other hand, albeit $C^{rms}_{l}$ along individual cylinder is drastically amplified beyond $g^{*} \gtrsim 1.5$, the overall value of the entire structure system is diminished and canceled out instead, due to the dominant anti-phase vortex shedding regime at these gap ratios. The above observations show that the gap-flow instability is critical to the overall stability of the SBS systems in engineering operations with a relatively small gap ratio, where a strong force modulation is observed.  

It is observed that the 3D flow not only modulates the hydrodynamic responses, but also the frequency $f_{flip}$ value. Here the $f_{flip}$ value appears to be small in three-dimensional flow, compared with its two-dimensional laminar flow counterparts. \cite{lbjrk2016pof} visualized the flip-flopping instant as a zero phase angle difference between $C_l$ in the SSBS arrangements. Different from a two-dimensional laminar flow, the existence of the streamwise vorticity clusters in the formation region varies $f_{vs}$ along the cylinder span and results in a repetitive temporal modulation of the $f_{vs}$ values along the span. To completely flip over the gap-flow direction, approximately half or one in-phase shedding of a primary vortex is necessary. 
However, due to the complex non-linear nature of the flow, this modulation of $f_{vs}$ on each cylinder is chaotic, swift and highly unstable. Consequently, $f_{flip}$ value is significantly influenced in a three-dimensional flow.
\section{Coupling of VIV and gap-flow kinematics}\label{sec:3D_VIV_gap}
\begin{figure} \centering
    \begin{subfigure}[b]{1.0\textwidth}	
	\centering
	\hspace{-25pt}
	\includegraphics[trim=0.1cm 0.1cm 0.1cm 0.1cm,scale=0.215,clip]{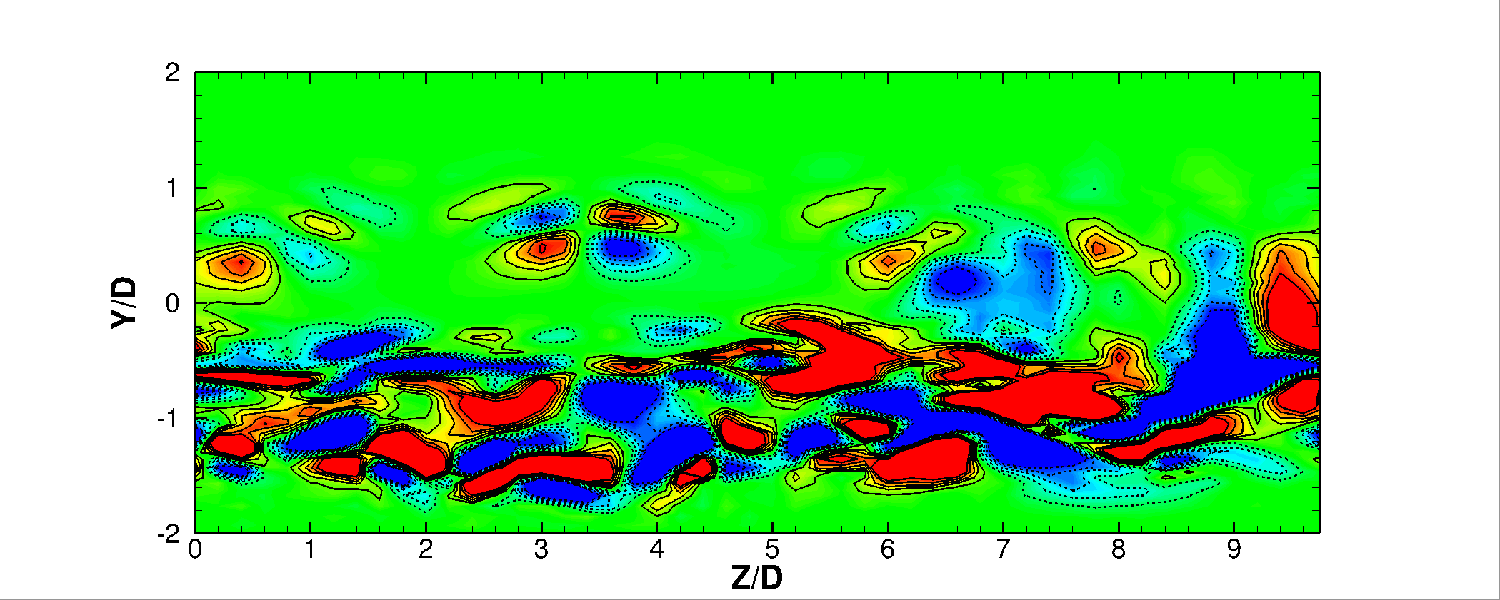}
    \caption{$\qquad$}
	\label{fig:con_td18r4_350_125}
	\end{subfigure}
	\begin{subfigure}[b]{1.0\textwidth}
	\centering
	\hspace{-25pt}\includegraphics[trim=0.1cm 0.1cm 0.1cm 0.1cm,scale=0.215,clip]{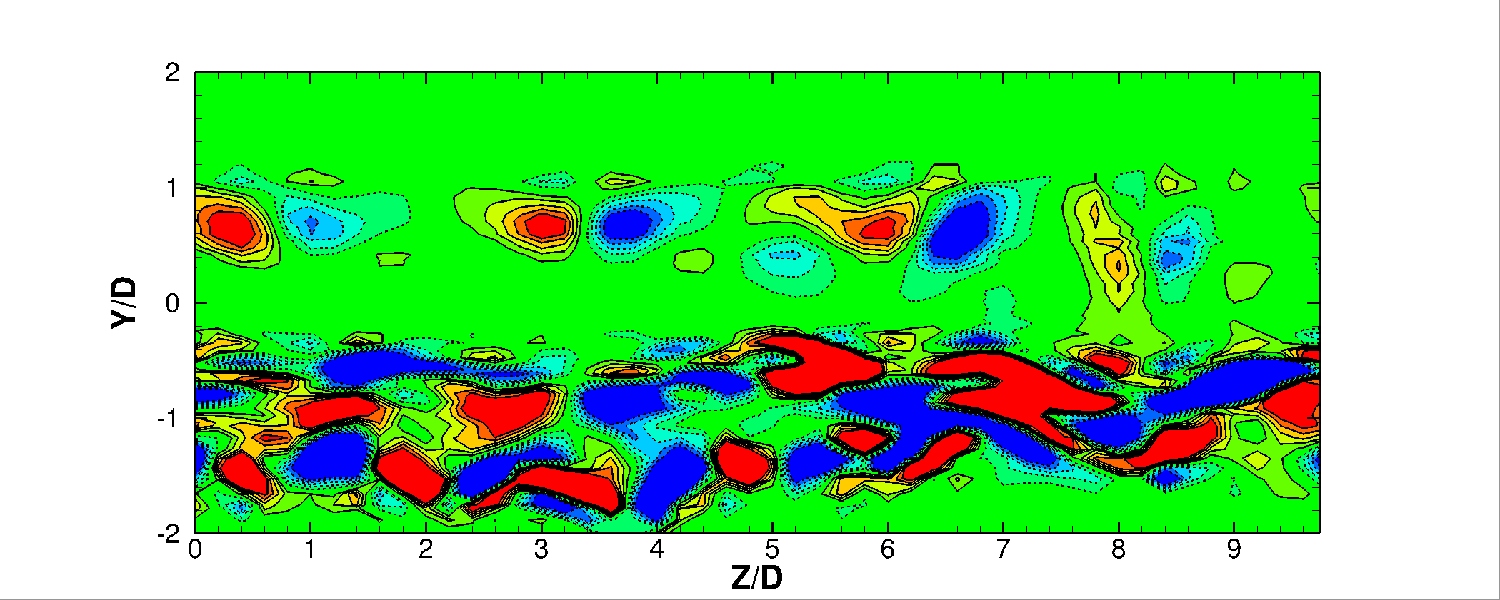}
    \caption{$\qquad$}
    \label{fig:con_td18r4_350_15}
    \end{subfigure}
\caption{$\omega_x$ contours in $(y,z)$-plane for the cylinders in VSBS arrangement at $Re=500$, $g^*=0.8$, $m^*=10.0$, $\zeta=0.01$, $U_r=4$ \emph{(the peak lock-in sage)}, $\omega_x=\pm 1.0$ \emph{(contours)} and $tU/D=350$: (a) $x^*=1.25$; (b) $x^*=1.5$.}
\label{fig:con_td18r4_350}
\end{figure}

In this section, the VIV kinematics and the gap flow instability are coupled in the VSBS arrangements, where \emph{Cylinder1} is elastically-mounted in the transverse direction.
Similar to the SSBS arrangements, the $f_{flip}$ is barely observed on the  VSBS arrangement at the off-lock-in stage in the selected time window in figure~\ref{fig:td18_cds}. On the contrary, the $f_{flip}$ values at the lock-in stage in figure~\ref{fig:td18r6_cd} surge remarkably. Based on the discussion from \cite{lbjrk2016pof}, $f_{flip}$ is related to the Reynolds number and gap ratio. Here $f_{flip}$ is also found susceptible to the influence of three-dimensionality. As the three-dimensional vortical structure is subjected to the modulation from the VIV kinematics,  $f_{flip}$ becomes VIV-dependent. Some phenomena reported by \cite{lbjrk2016pof} from the VSBS arrangements are also observed in a three-dimensional flow. For instance, a quasi-stable deflected gap flow regime occurs at the peak lock-in stage, where the gap flow permanently deflects toward the locked-in vibrating cylinder. The onset of VIV occurs at a lower $U_r$ value, as shown in figure~\ref{fig:Aymax_Ur}, owing to the enhanced vortex-to-vortex interaction in the narrow near-wake region. The aforementioned recovery of two-dimensional hydrodynamic responses is also found along the locked-in vibrating cylinder in the VSBS arrangements in figure~\ref{fig:czs_500td18}. 
It is worth noting that the amplitude and the frequency of $C_z$ are relatively insensitive to the variation of $U_r$ values, except the $C_z$ suppression at the peak lock-in stage. Figure~\ref{fig:cz_500td18r4} shows a trivial $C_z$ oscillation at the peak lock-in stage, compared with the aforementioned perfect $C_z$ suppression along isolated cylinder at present order of magnitude ($\times ~10^{-3}$) in Section~\ref{sec:3D_viv}. 
It means that a perfect \emph{two-dimensional recovery} may not be feasible when the fluid instability is intensified, e.g., proximity interference from a strong gap-flow jet or higher Reynolds number. To investigate the Reynolds number effect alone, the Reynolds number is increased slightly from 500 to 800 and the proximity interference is eliminated by considering very large gap ratio, i.e.  a stationary isolated cylinder. Figure~\ref{fig:re800r48_cz} shows that the $C_z$ is completely suppressed along an isolated cylinder at its peak lock-in stage ($U_r=4.8$, \emph{solid cross}) at $Re=800$. Nonetheless,  streamwise vorticity ribs are still prominently visible further downstream in figure~\ref{fig:con_re800r48}. This evidence not only supports that an intensified fluid momentum is detrimental to the stability of fluid, but also indicates that the streamwise vorticity clusters are originated from the vortex-to-vortex interaction, in which the fluid shearing is prominent along the vortical interfaces. This observation is supported by the discussion from \cite{Chantry2016JoFM}, in which the fluid shearing alone was reported to be essential to the flow transition, instead of the boundary layers from the walls. As a result, the corresponding sectional contour plots of aforementioned VSBS arrangement show a distinctive concentration difference of streamwise vorticity from the narrow and wide near-wake regions in figure~\ref{fig:td18r4_snapshots} on the $(x,y)$-plane and figure~\ref{fig:con_td18r4_350} on the $(y,z)$-plane. 

\begin{figure} \centering
	\begin{subfigure}[b]{0.5\textwidth}	
	\centering
	\hspace{-25pt}\includegraphics[trim=0.01cm 0.1cm 0.1cm 0.1cm,scale=0.3,clip]{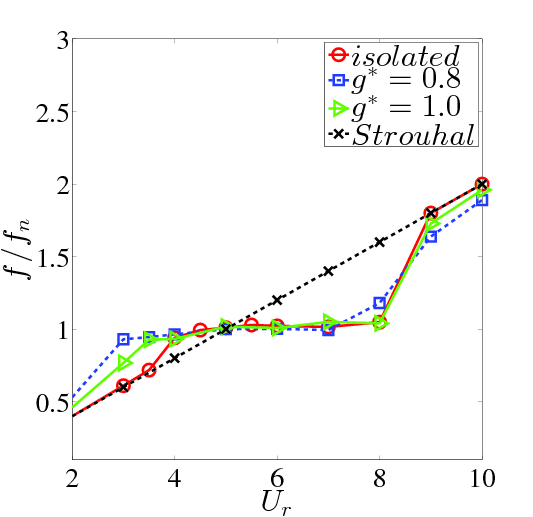}
    \caption{$\qquad$}
	\label{fig:f_vs_ur}
	\end{subfigure}%
	\begin{subfigure}[b]{0.5\textwidth}
	\centering
	\hspace{-25pt}\includegraphics[trim=0.01cm 0.1cm 0.1cm 0.1cm,scale=0.3,clip]{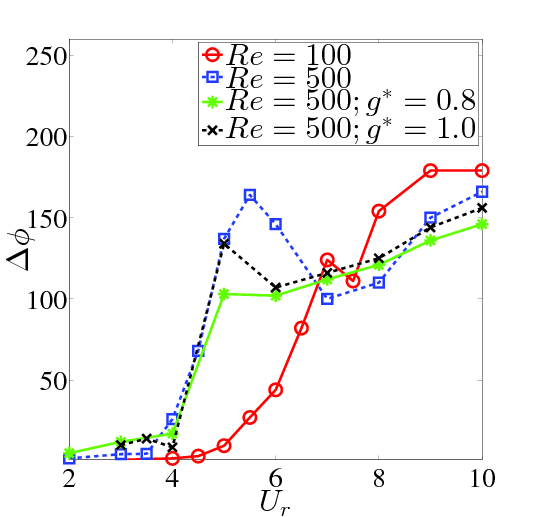}
    \caption{$\qquad$}
    \label{fig:phi_vs_ur}
    \end{subfigure}
	\begin{subfigure}[b]{0.5\textwidth}
    \centering
    \hspace{-25pt}\includegraphics[trim=0.1cm 0.1cm 0.1cm 0.1cm,scale=0.3,clip]{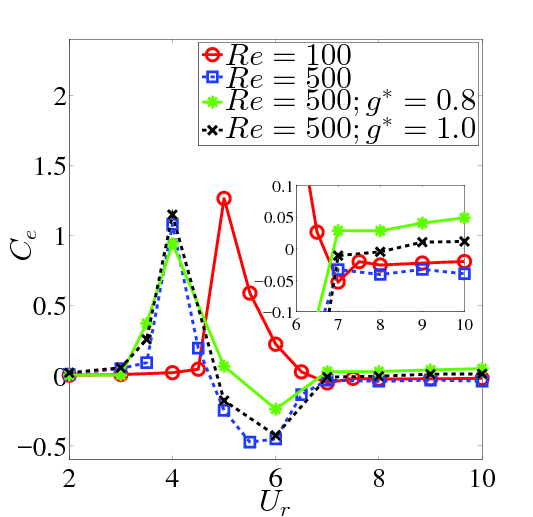}
    \caption{$\qquad$}
    \label{fig:Ce_vs_ur}
	\end{subfigure}
\caption{Frequency and energy transfer analysis at $Re \in [100,$ $500]$, $m^*=10$ and $\zeta=0.01$: (a) frequency ratio $f/f_{n}$ as a function of the reduced velocity at $Re=500$; (b) phase angle $\Delta \phi$ between $A_y$ and $C_l$ as a function of the reduced velocity; (c) averaged energy transfer in transverse direction for one primary vortex shedding cycle.}
\label{fig:f_phi_vs_ur}
\end{figure}

The VSBS arrangements at two typical gap ratios $g^*=0.8$ and $1.0$ in the deflected-gap flow regime are adopted to further investigate the characteristics of the VIV lock-in subjected to the gap-flow proximity interference. The frequency ratio plot in figure~\ref{fig:f_vs_ur} confirms the early onset of the VIV lock-in at a smaller $U_r$ value in figure~\ref{fig:Aymax_Ur}. Albeit the onsets of the VIV lock-in at various $g^*$ values are different, the ends of their VIV lock-in approximately occur at an identical $U_r$ value for the VSBS arrangements. The corresponding time-averaged phase angle difference $\Delta \phi$ between $A_y$ and $C_l$ is plotted in figure~\ref{fig:phi_vs_ur}. The mean $\Delta \phi$ is at about $110^{\circ}$ from $ 6.5 \lesssim U_r \lesssim 7.5$, which represents a relative equilibrium state where the energy transfer between the fluid and the structure is balanced. In addition, a gradient discontinuity is observed at about $U_r \approx 7.0$, which stabilizes the VIV lock-in stage. A similar discontinuity of phase angle difference was also reported by \cite{Leontini2006PoF} at VIV lock-in stage, which is correlated to the VIV kinematics and the vortex wakes.
Following that, as the $U_r$ value exceeds $8.0$, the $\Delta \phi$ value becomes completely anti-phase. The profile of $\Delta \phi$ for an isolated cylinder in a three-dimensional flow (\emph{dash square}) shows a similar profile to its two-dimensional counterpart (\emph{solid circle}). However its amplitude is excited at a smaller $U_r$ value.

Generally speaking, in the VSBS arrangements, as long as two cylinders are sufficiently close, the gap-flow proximity interference becomes remarkable and the corresponding $\Delta \phi$ values are more close to $90^{\circ}$ (\emph{the equilibrium state}). This can be seen from the relative positions of the $\Delta \phi$ profiles (\emph{dash-square, solid-star and dash-cross}) in figure~\ref{fig:phi_vs_ur}. This gap-flow proximity interference becomes very prominent at the peak lock-in stage, where the two cylinders are very close, because of the extensive VIV motion. Besides $\Delta \phi$, the energy transfer coefficient $C_e$  indicates not only the energy flow direction, but also the the amount of work done during the energy transfer, which is defined in table~\ref{tab:quan}. For $C_e$, $v$, $\tau = tU/D$ and $T$ respectively are the transverse velocity of a vibrating cylinder, dimensionless time scale and one (dimensionless) shedding cycle of primary vortex. The magnitude of $C_e$ quantifies the energy transfer between the fluid flow and  the structure. Its sign indicates the direction of energy transfer. For instance, a positive value means the energy is transferred from the fluid flow to structure, which corresponds to $0^\circ \lesssim \Delta \phi \lesssim 90^\circ$ (in-phase). The $C_e$ value, the work done between the fluid and the structure, is averaged and computed over the duration of one primary vortex shedding cycle, as shown in figure~\ref{fig:Ce_vs_ur}. The aforementioned earlier onset of VIV lock in is also confirmed for higher Reynolds number and VSBS arrangement cases, $U_r \approx 4$. Over the entire $U_r$ range of lock-in, the direction or the sign change of energy transfer is observed for both isolated and the VSBS arrangement cases. This sign change of energy transfer is associated with the aforementioned phase discontinuity observed by \cite{Leontini2006PoF} and in figure~\ref{fig:phi_vs_ur} of the present investigation. Although this phase discontinuity is relatively small for the  isolated cylinder at lower Reynolds number $Re=100$, it is still evidently observed in the zoomed-in plot (\emph{red line at $U_r=7$}) of figure~\ref{fig:Ce_vs_ur}. Furthermore, the amount of energy transferred over off-lock-in $U_r$ ranges is trivial compared to their lock-in counterparts.
To analyze the relationship among the near-wake instability, the gap flow and the VIV kinematics from another point of view, the spanwise wavelength $\lambda^*$ is discussed in the next section. The next section will show that the magnitudes of the $\lambda^*$ and the $\omega_x$ render excellent means to analyze the 3D vortical structures.
\section{Interference to the spanwise correlation}\label{lamda}
\begin{figure} \centering
	\begin{subfigure}[b]{1\textwidth}	
	\centering
	\hspace{-25pt}\includegraphics[trim=0.2cm 0.1cm 0.1cm 0.3cm,scale=0.38,clip]{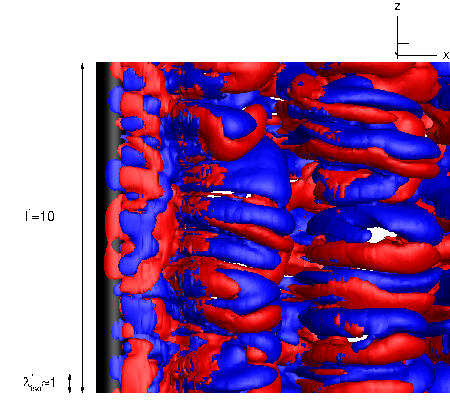}
    \caption{$stationary$, $\omega_x=\pm0.2$}
    \label{fig:lamda_re500}
	\end{subfigure}
	\begin{subfigure}[b]{0.5\textwidth}	
	\centering
	\hspace{-25pt}\includegraphics[trim=0.2cm 0.1cm 0.1cm 0.3cm,scale=0.38,clip]{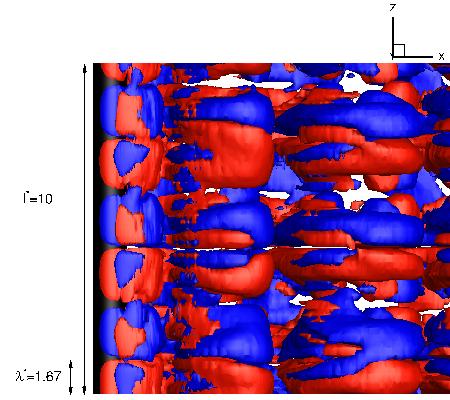}
    \caption{$U_r=3.5$, $\omega_x=\pm0.2$}
	\label{fig:lamda_re500r35}
	\end{subfigure}%
	\begin{subfigure}[b]{0.5\textwidth}
	\centering
	\hspace{-25pt}\includegraphics[trim=0.1cm 0.1cm 0.1cm 0.3cm,scale=0.38,clip]{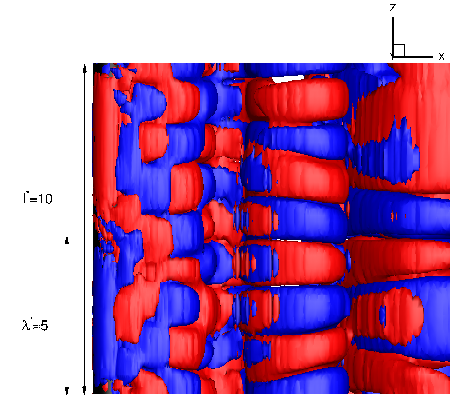}
    \caption{$U_r=4$, $\omega_x=\pm0.0002$}
    \label{fig:lamda_re500r4}
    \end{subfigure}
    \begin{subfigure}[b]{0.5\textwidth}	
	\centering
	\hspace{-25pt}\includegraphics[trim=0.2cm 0.1cm 0.1cm 0.3cm,scale=0.38,clip]{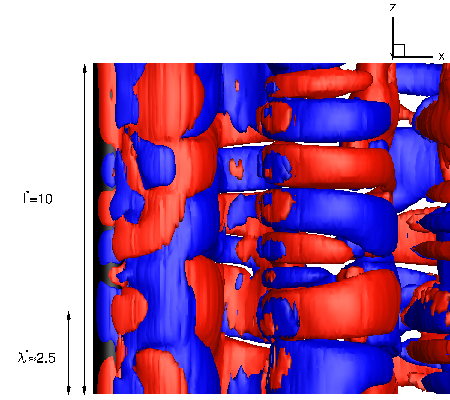}
    \caption{$U_r=6$, $\omega_x=\pm0.02$}
	\label{fig:lamda_re500r6}
	\end{subfigure}%
	\begin{subfigure}[b]{0.5\textwidth}
	\centering
	\hspace{-25pt}\includegraphics[trim=0.1cm 0.1cm 0.1cm 0.3cm,scale=0.38,clip]{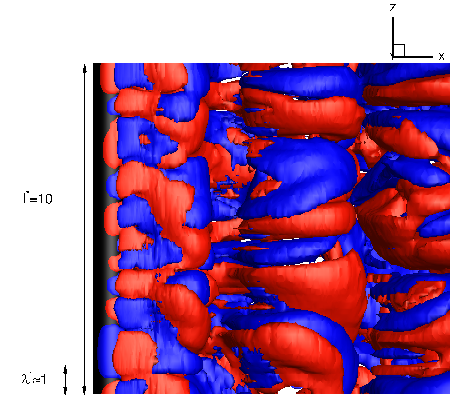}
    \caption{$U_r=8$, $\omega_x=\pm0.2$}
    \label{fig:lamda_re500r8}
    \end{subfigure}
    \caption{Contours of $\omega_x$ for an isolated cylinder at $tU/D = 200$: (a) stationary cylinder at $Re=500$ and $\omega_x \in [\pm 0.0002, \pm 0.2]$ with $\lambda^* \approx 1$ for the mode-B; (b,c,d,e) a transverse-vibrating cylinder at $Re=500$, $m^*=10$, $\zeta=0.01$ and $\omega_x \in [\pm 0.0002, \pm 0.2]$. The $\lambda^*$ value significantly enlarges at the peak lock-in stage.}
   \label{fig:lamdas_re500}
\end{figure}
\begin{figure} \centering
	\begin{subfigure}[b]{0.5\textwidth}	
	\centering
	\hspace{-25pt}\includegraphics[trim=0.2cm 0.1cm 0.1cm 0.3cm,scale=0.38,clip]{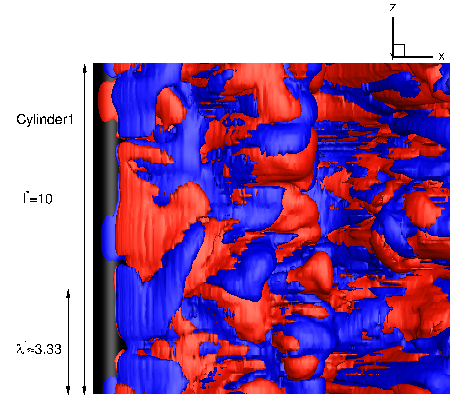}
    \caption{$tU/D=175$, \emph{Cylinder1}}
	\label{fig:lamda_re500td18_175a}
	\end{subfigure}%
	\begin{subfigure}[b]{0.5\textwidth}
	\centering
	\hspace{-25pt}\includegraphics[trim=0.1cm 0.1cm 0.1cm 0.3cm,scale=0.38,clip]{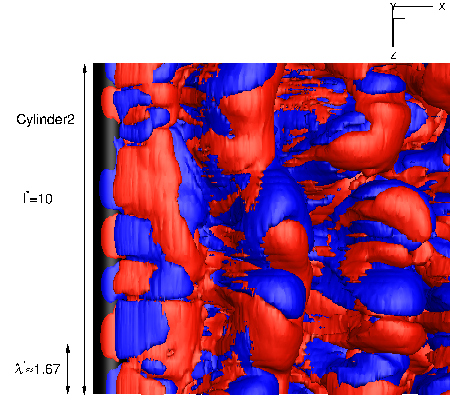}
    \caption{$tU/D=175$, \emph{Cylinder2}}
    \label{fig:lamda_re500td18_175b}
    \end{subfigure}
    \begin{subfigure}[b]{0.5\textwidth}	
	\centering
	\hspace{-25pt}\includegraphics[trim=0.2cm 0.1cm 0.1cm 0.3cm,scale=0.38,clip]{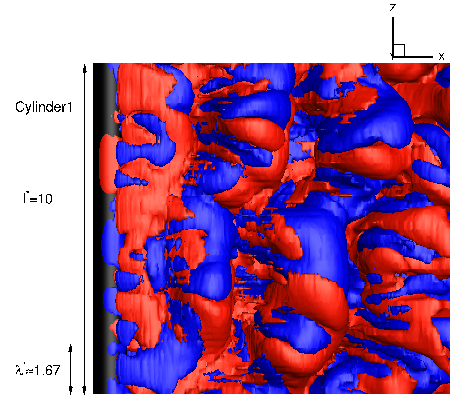}
    \caption{$tU/D=300$, \emph{Cylinder1}}
	\label{fig:lamda_re500td18_300a}
	\end{subfigure}%
	\begin{subfigure}[b]{0.5\textwidth}
	\centering
	\hspace{-25pt}\includegraphics[trim=0.1cm 0.1cm 0.1cm 0.3cm,scale=0.38,clip]{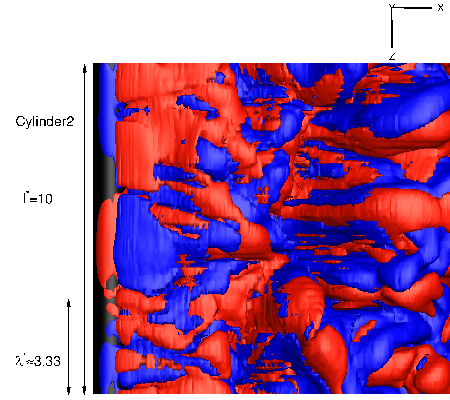}
    \caption{$tU/D=300$, \emph{Cylinder2}}
    \label{fig:lamda_re500td18_300b}
    \end{subfigure}
\caption{Contours of $\omega_x$ in SSBS arrangement: the dimensionless spanwise wavelength length at $Re=500$, $g^*=0.8$ and $\omega_x=\pm 0.2$. The wavelength $\lambda^*$ increases in the SSBS arrangements.}
   \label{fig:lamdas_re500td18}
\end{figure}
To quantitatively describe the interference of the VIV and the gap-flow kinematics on the three-dimensional vortical structure, the dimensionless spanwise correlation wavelength $\lambda^*$ is discussed herein. The wavelength $\lambda^*$ is a good representation of the number of the streamwise vortex pair formed along the cylinder span. \cite{Williamson1996ARoFM} documented that the mode-A was typically associated with $\lambda^* \in (3$, $4)$ and the mode-B which supersedes the mode-A after the flow transition possess one wavelength. In the present investigation, the measurement of $\lambda^*$ is undertaken along the cylinder span, since the aforementioned two-dimensional recovery and the gap-flow proximity interference could significantly change $\lambda^*$ value in the near-wake. Based on the proposed formula of spanwise wavelength in table~\ref{tab:quan}, the observed $\lambda^*$ value of the mode-B in figure~\ref{fig:lamda_re500} conforms well with the results reported by \cite{Williamson1996ARoFM}.  

\begin{figure} \centering
	\begin{subfigure}[b]{0.5\textwidth}	
	\centering
	\hspace{-25pt}\includegraphics[trim=0.2cm 0.1cm 0.1cm 0.3cm,scale=0.38,clip]{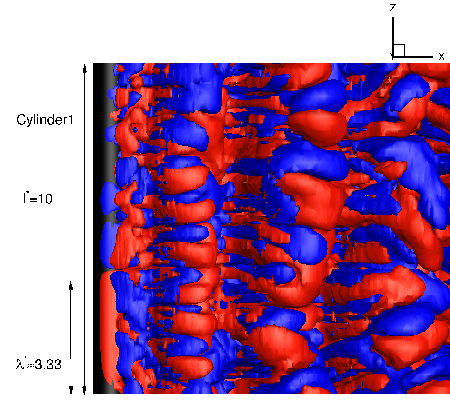}
    \caption{$tU/D=325$, \emph{Cylinder1}}
	\label{fig:lamda_re500td18r4_325a}
	\end{subfigure}%
	\begin{subfigure}[b]{0.5\textwidth}
	\centering
	\hspace{-25pt}\includegraphics[trim=0.1cm 0.1cm 0.1cm 0.3cm,scale=0.38,clip]{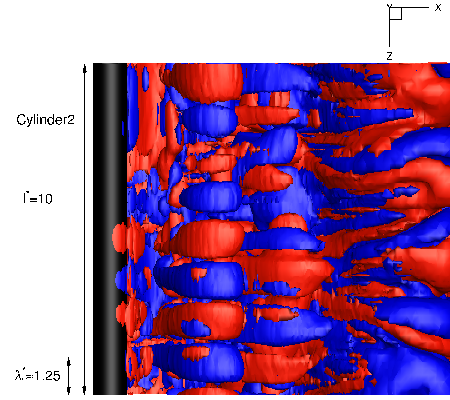}
    \caption{$tU/D=325$, \emph{Cylinder2}}
    \label{fig:lamda_re500td18r4_325b}
    \end{subfigure}
\caption{Contours of $\omega_x$ and the dimensionless spanwise wavelength length in VSBS arrangement at $Re=500$, $g^*=0.8$, $U_r=4.0$ and $\omega_x=\pm0.2$.}
   \label{fig:lamdas_re500td18r4}
\end{figure}

For the isolated cylinders in figure~\ref{fig:lamdas_re500}, the values of $\lambda^*$  at the off-lock-in stage are similar, except the lock-in stage where the wavelength $\lambda^*$ value gradually strides over the entire cylinder span until the peak lock-in stage, whereby the magnitude of $\omega_x$ is reduced by three orders of magnitudes. The $\lambda^*$ value at the peak lock-in stage quantitatively exhibits the aforementioned recovery of two-dimensional hydrodynamic responses along a circular cylinder via the VIV kinematics in Section~\ref{sec:3D_viv}.
On the other hand, the wavelength  $\lambda^*$ value is also subjected to the influence from the gap-flow kinematics in the SBS arrangements. To investigate further, a pair of cylinders in the SSBS arrangement at $Re=500$ and $g^*=0.8$ is plotted in figure~\ref{fig:lamdas_re500td18} as an example, where figures~\ref{fig:lamda_re500td18_175a} and~\ref{fig:lamda_re500td18_175b} are plotted at $tU/D=175$ when the gap flow deflects to \emph{Cylinder1}, and figures~\ref{fig:lamda_re500td18_300a} and~\ref{fig:lamda_re500td18_300b} show that the gap flow deflects to \emph{Cylinder2} at $tU/D = 300$. The wavelength $\lambda^*$ value is noticed to be relatively large $\lambda^* \approx 3.33$ along the cylinder with a narrow near-wake region in the SSBS arrangement at $g^*=0.8$. A similar tendency of the $\lambda^*$ augmentation is also observed for all other SSBS arrangements with a deflected gap flow. 
In the VSBS arrangements, the primary focus is on the quasi-stable gap flow regime. Hence the iso-surfaces of $\omega_x$ for the VSBS arrangement at the peak lock-in stage are taken as an example and visualized in
figure~\ref{fig:lamdas_re500td18r4} for analysis. Similar to the SSBS arrangements, the $\lambda^*$ value increases along the locked-in vibrating cylinder to which the gap flow permanently deflects in figure~\ref{fig:lamdas_re500td18r4}. The scaling factor of $\lambda^*$, approximately 3.33, is similar to the aforementioned SSBS arrangement at an identical $g^*$ value. Furthermore the wavelength $\lambda^*$ value of its stationary counterpart in figure~\ref{fig:lamda_re500td18r4_325b} is very close to that of an isolated cylinder case $\lambda^* \approx 1$.

Overall the above observations show that the spanwise wavelength is closely related to the VIV kinematics and the gap-flow proximity interference. In the SBS arrangements, when $g^*$ decreases, the gap-flow proximity interference becomes intense. As a result, the $\lambda^*$ value increases along the cylinder with a narrow near-wake region. On the contrary, the $\lambda^*$ value is relatively small in the wide near-wake region, because of a weakened vortex-to-vortex interaction.
\section{Three-dimensional modal analysis}\label{3D_DMD_VD}
\begin{figure} \centering
	\centering
	\hspace{-25pt}\includegraphics[trim=0.1cm 0.1cm 0.1cm 0.1cm,scale=0.105,clip]{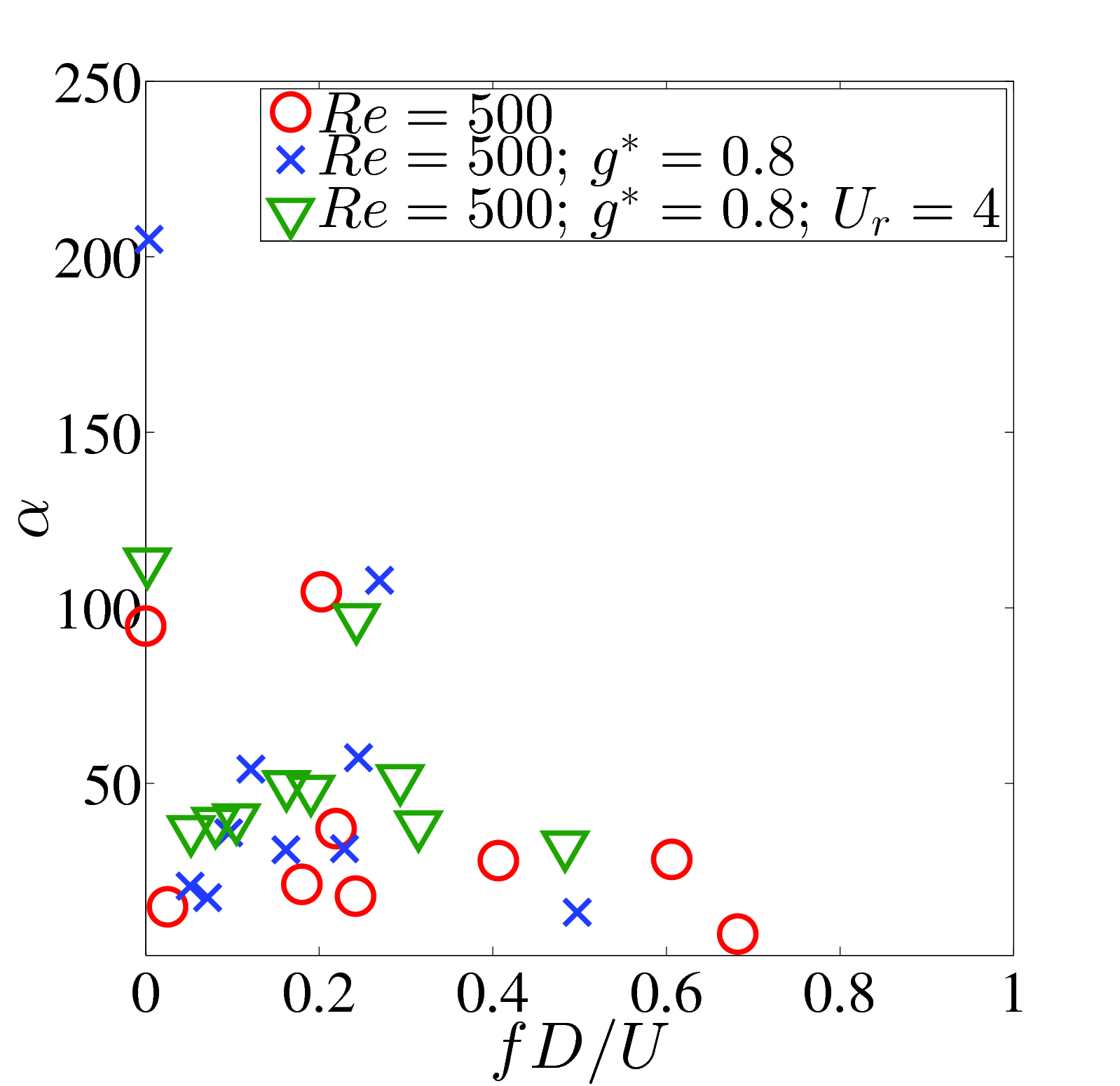}
	\caption{Dependence of DMD amplitude $\alpha$ on frequency at $Re=500$ for stationary cylinder, SSBS arrangement at $g^*=0.8$, and VSBS arrangement at $g^*=0.8$, $m^*=10$, $\zeta=0.01$ and $U_r=4$. Here $\alpha$ value is the optimal amplitude of each DMD mode and obtained from an optimization process in the sparsity-promoting DMD analysis.}
   \label{fig:3D-DMD}
\end{figure}
\begin{figure} \centering
	\begin{subfigure}[b]{0.5\textwidth}	
	\centering
	\hspace{-25pt}\includegraphics[trim=0.15cm 3cm 0.2cm 0.1cm,scale=0.1,clip]{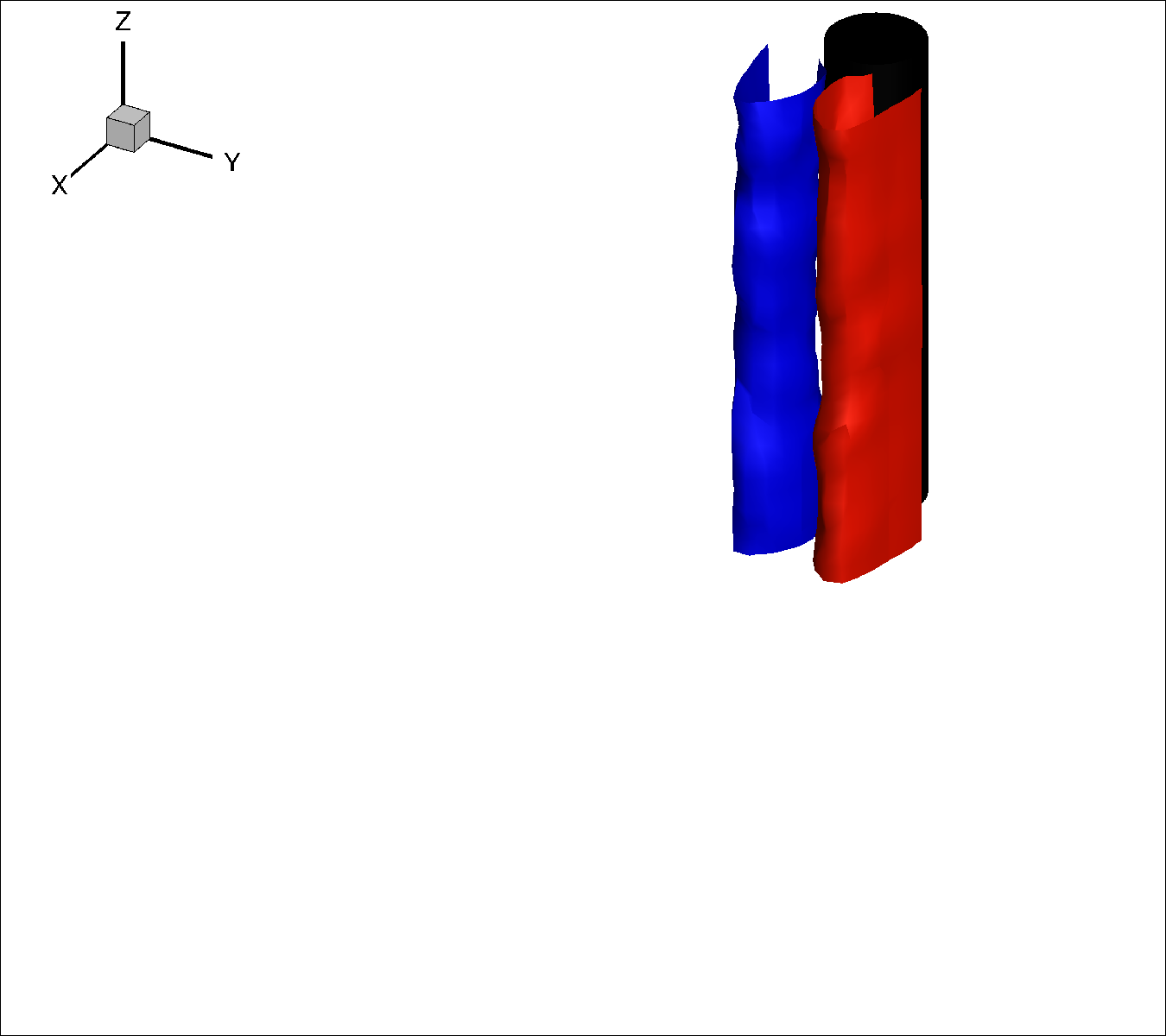}
    \caption{$fD/U=0$, $\alpha=97$}
	\label{fig:re500_mode_0}
	\end{subfigure}%
	\begin{subfigure}[b]{0.5\textwidth}
	\centering
	\hspace{-25pt}\includegraphics[trim=0.15cm 3cm 0.2cm 0.1cm,scale=0.1,clip]{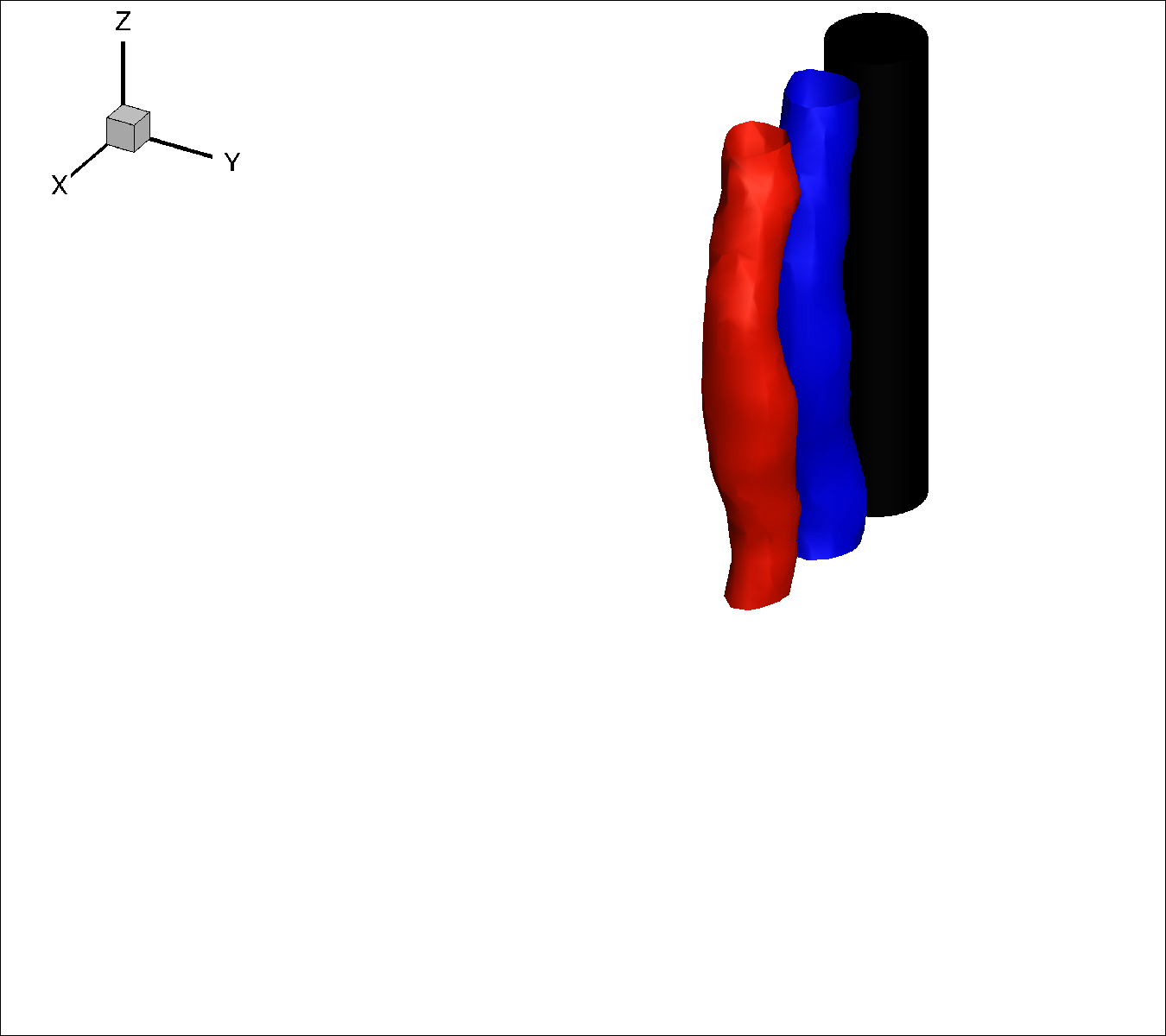}
    \caption{$fD/U=0.2$, $\alpha=39$}
    \label{fig:re500_mode_2018}
    \end{subfigure}
    \begin{subfigure}[b]{0.5\textwidth}	
	\centering
	\hspace{-25pt}\includegraphics[trim=0.15cm 3cm 0.2cm 0.1cm,scale=0.1,clip]{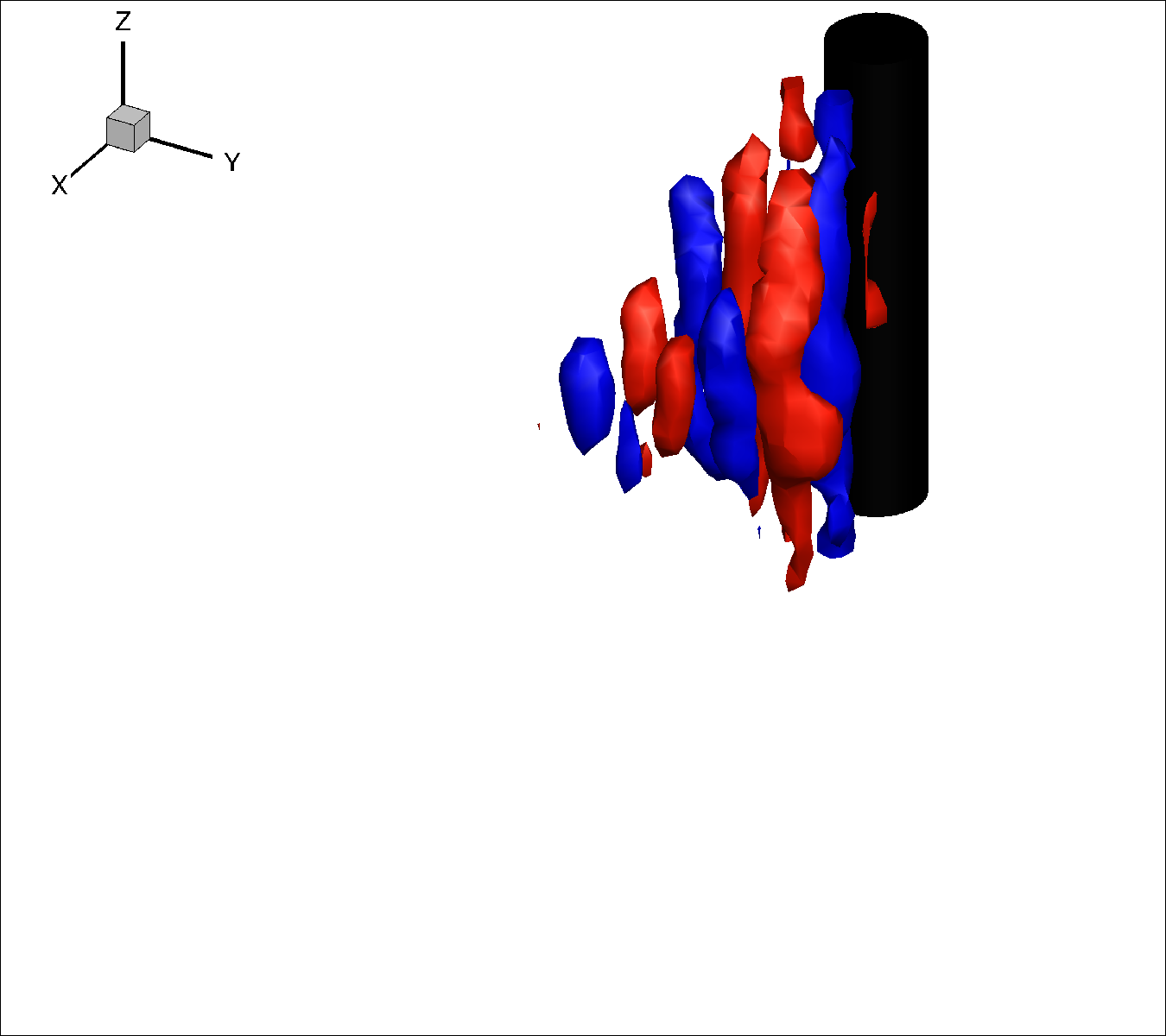}
    \caption{$fD/U=0.41$, $\alpha=31$}
	\label{fig:re500_mode_4056}
	\end{subfigure}%
	\begin{subfigure}[b]{0.5\textwidth}
	\centering
	\hspace{-25pt}\includegraphics[trim=0.15cm 3cm 0.2cm 0.1cm,scale=0.1,clip]{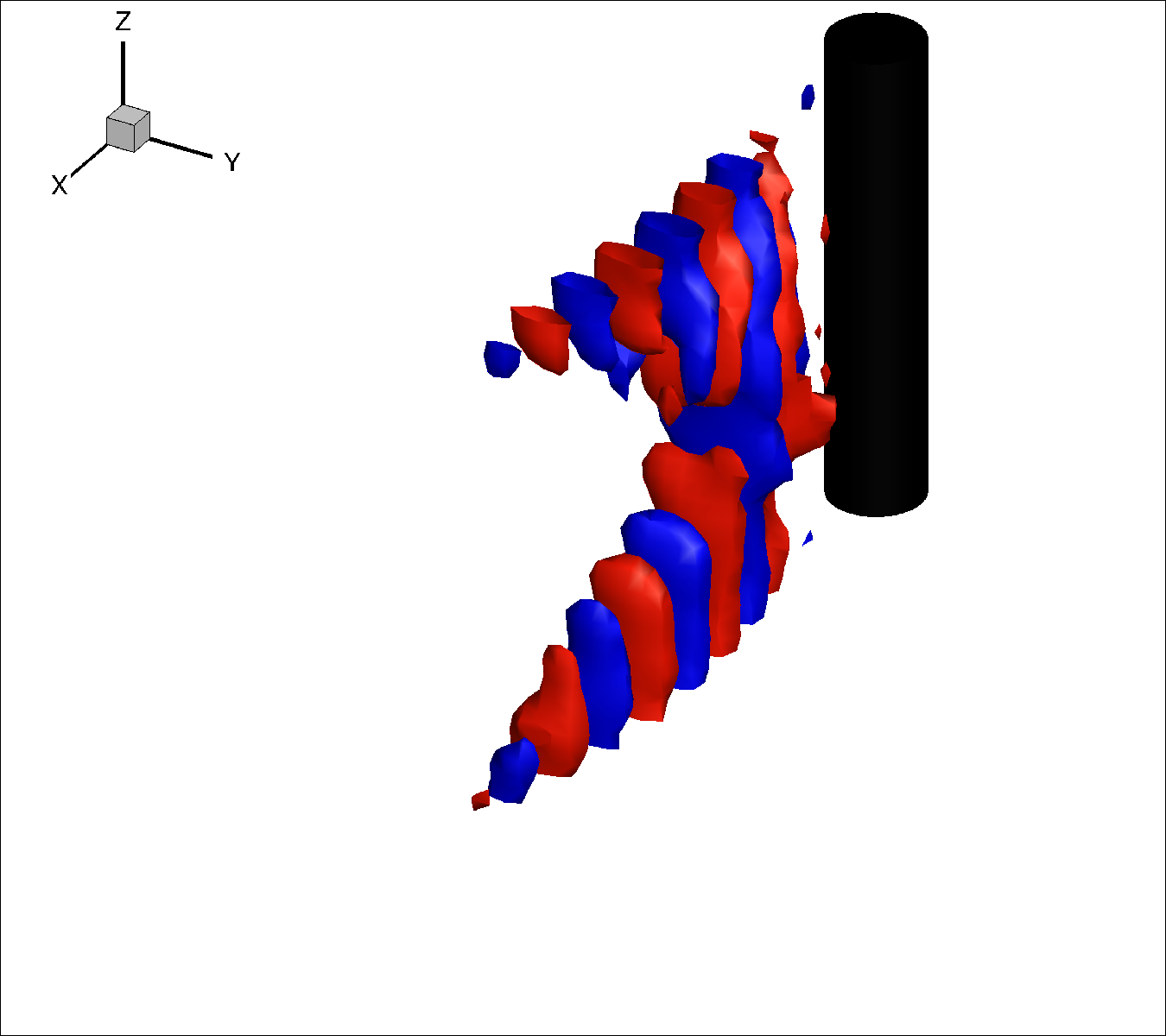}
    \caption{$fD/U=0.61$, $\alpha=32$}
    \label{fig:re500_mode_6064}
    \end{subfigure}
\caption{Iso-surface plots of the primary vortex modes of a stationary cylinder at $Re=500$, $tU/D \in [250,$ $350]$ and $\omega_z= \pm 0.01$. A strong third-order harmonic vortex mode is decomposed at $fD/U \approx 0.61$.}
   \label{fig:3D-DMD_re500}
\end{figure}
\begin{figure} \centering   
	\begin{subfigure}[b]{0.5\textwidth}	
	\centering
	\hspace{-25pt}\includegraphics[trim=0.15cm 3cm 0.2cm 0.1cm,scale=0.1,clip]{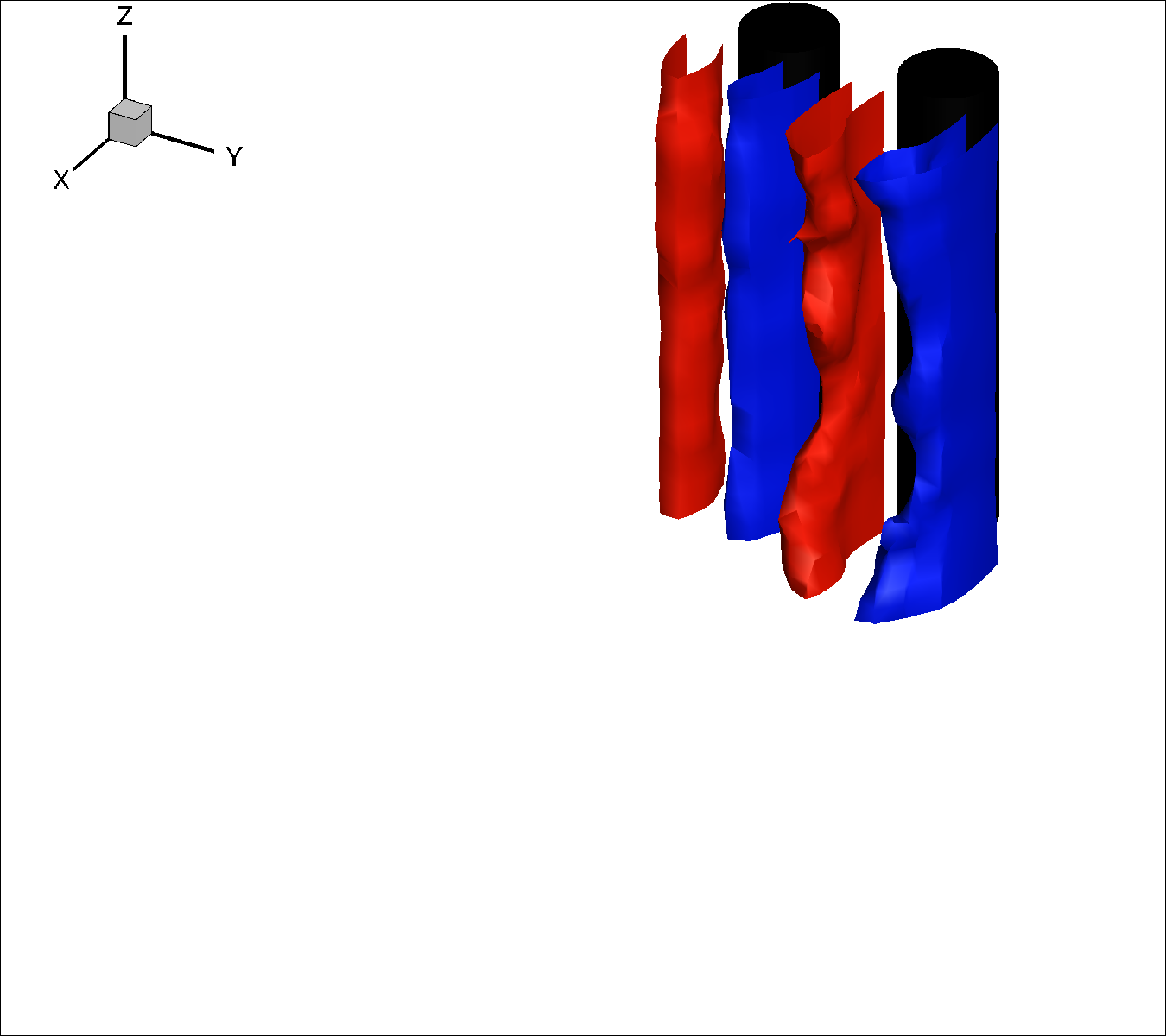}
    \caption{$fD/U=0$, $\alpha=206$}
	\label{fig:re500td18_0}
	\end{subfigure}%
	\begin{subfigure}[b]{0.5\textwidth}
	\centering
	\hspace{-25pt}\includegraphics[trim=0.15cm 3cm 0.2cm 0.3cm,scale=0.1,clip]{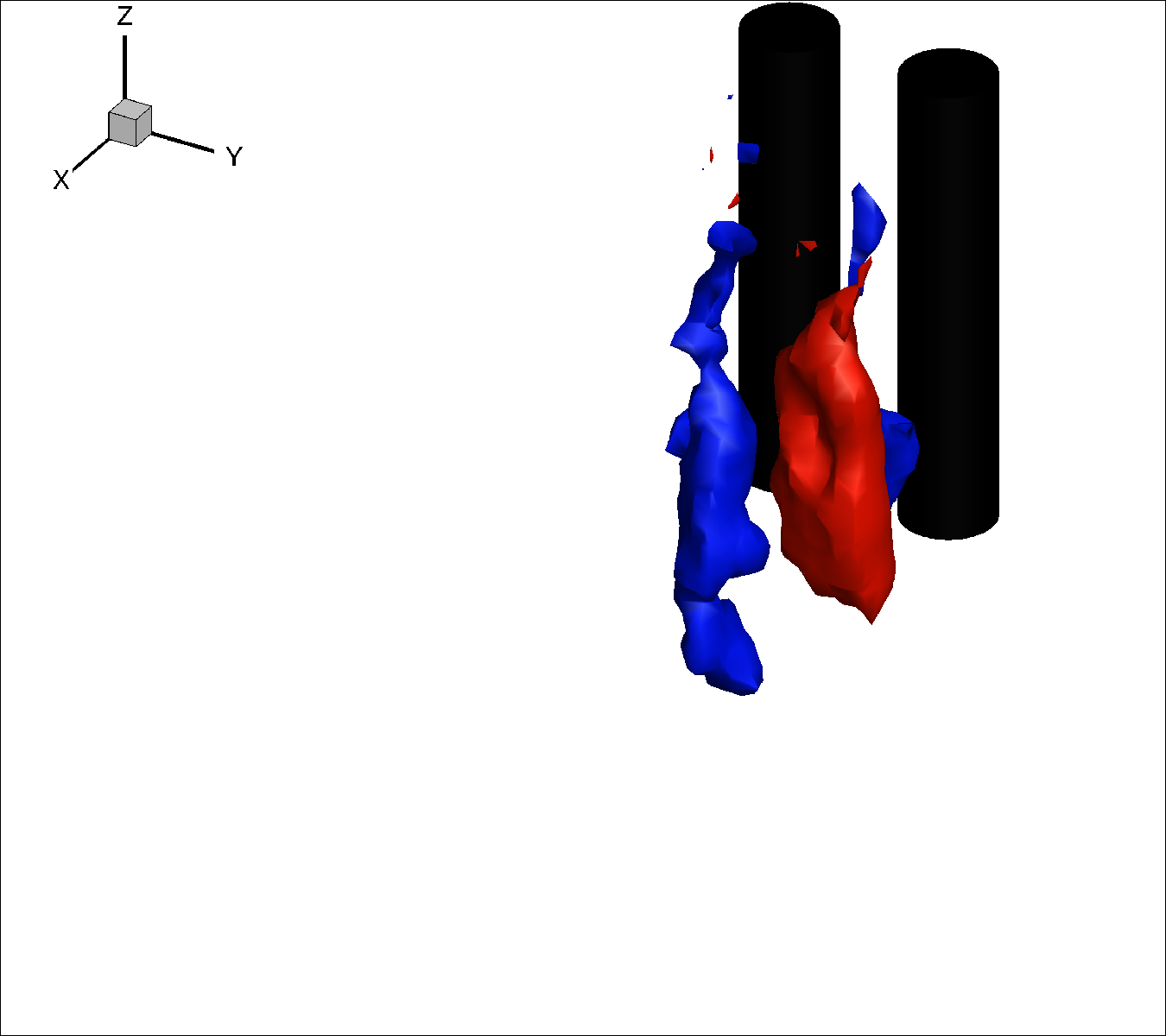}
    \caption{$fD/U=0.12$, $\alpha=56$}
    \label{fig:re500td18_1207}
    \end{subfigure}
    \begin{subfigure}[b]{0.5\textwidth}	
	\centering
	\hspace{-25pt}\includegraphics[trim=0.15cm 3cm 0.2cm 0.3cm,scale=0.1,clip]{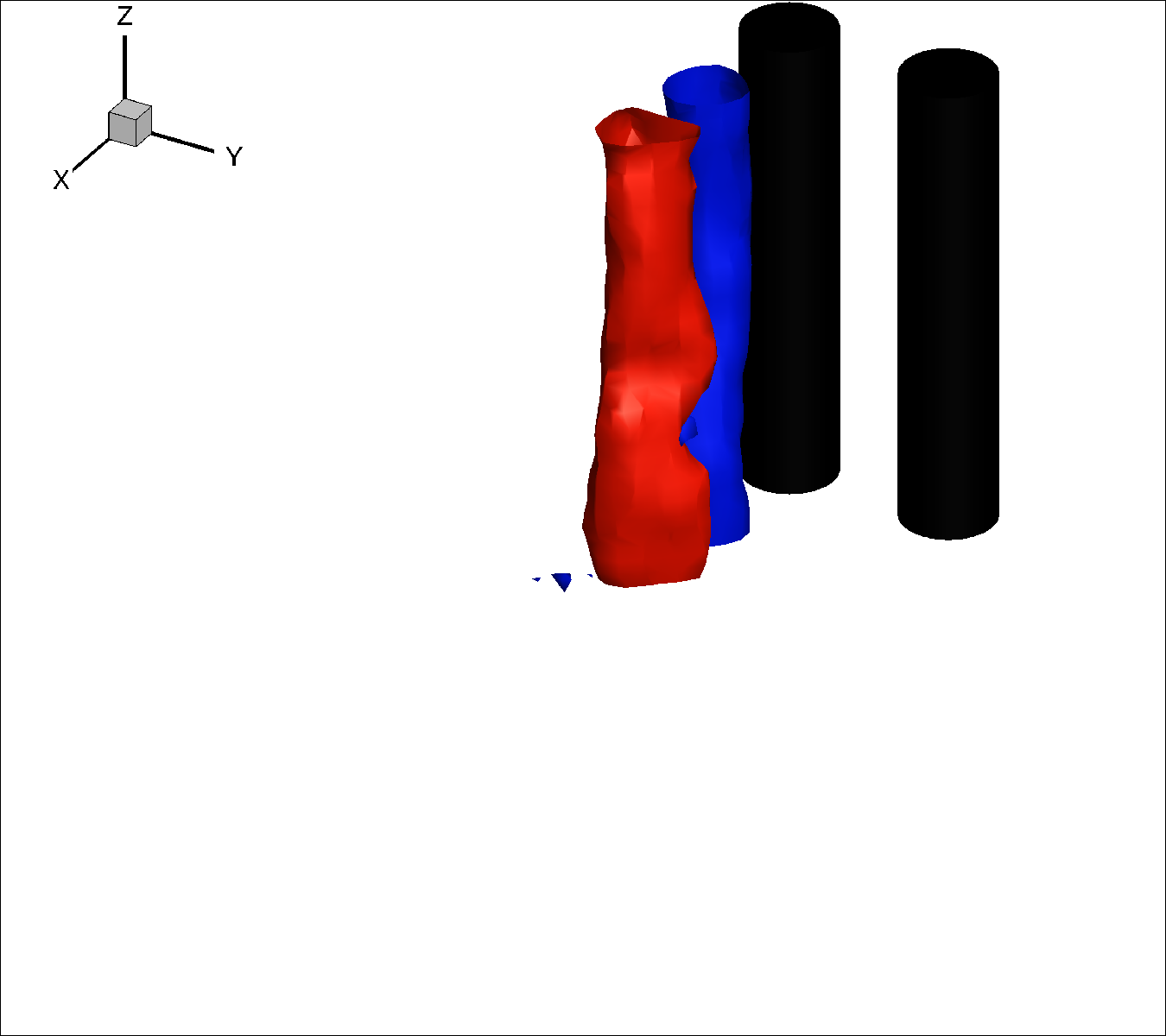}
    \caption{$fD/U=0.24$, $\alpha=57$}
	\label{fig:re500td18_2471}
	\end{subfigure}%
	\begin{subfigure}[b]{0.5\textwidth}
	\centering
	\hspace{-25pt}\includegraphics[trim=0.15cm 3cm 0.2cm 0.3cm,scale=0.1,clip]{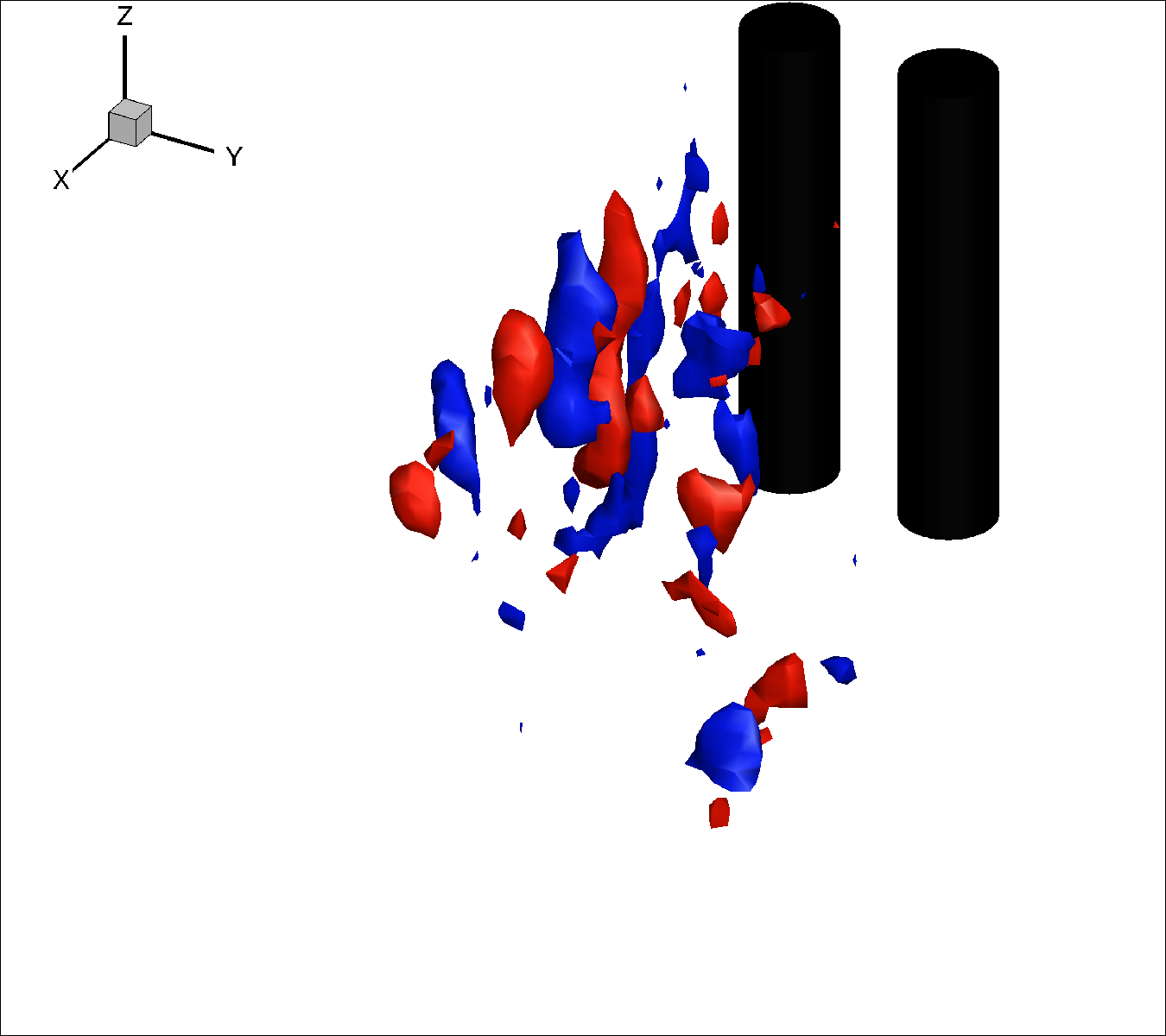}
    \caption{$fD/U=0.5$, $\alpha=22$}
    \label{fig:re500td18_4975}
    \end{subfigure}
\caption{Iso-surface plots of the primary vortex modes of SSBS arrangement at $Re=500$, $g^*=0.8$, $tU/D \in [250,$ $350]$ and $\omega_z= \pm 0.01$. A vortex discontinuity is observed in a wide near-wake region behind the right-hand-side cylinder in figure~\ref{fig:re500td18_1207}. A strong third-order harmonic vortex mode is decomposed in a narrow near-wake region behind the left-hand-side cylinder in figure~\ref{fig:re500td18_4975}}
   \label{fig:3D-DMD_re500td18}
\end{figure}
\begin{figure} \centering   
	\begin{subfigure}[b]{0.5\textwidth}	
	\centering
	\hspace{-25pt}\includegraphics[trim=0.15cm 0.1cm 0.2cm 0.1cm,scale=0.12,clip]{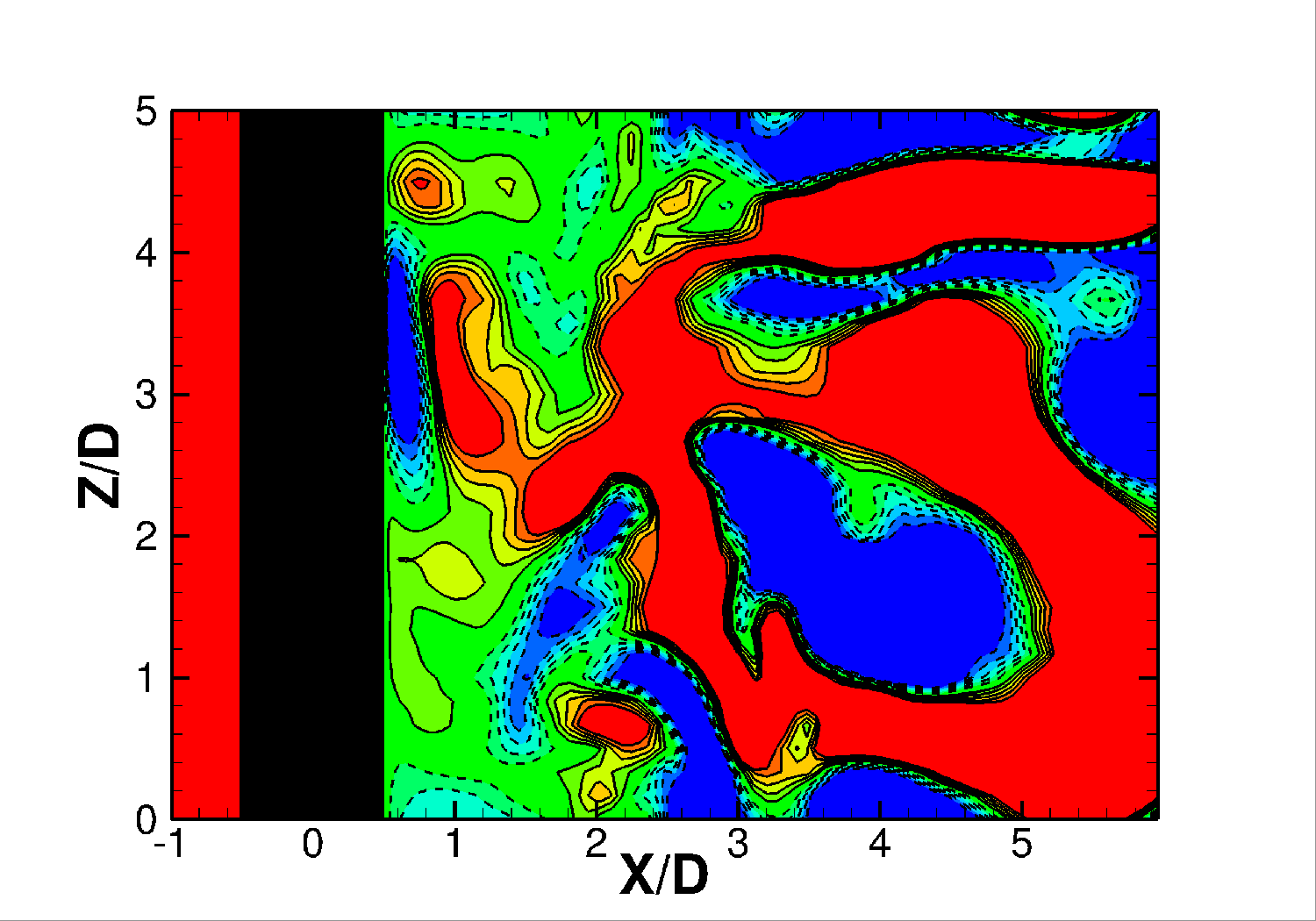}
    \caption{\emph{Cylinder1}}
	\label{fig:dis_C1}
	\end{subfigure}%
	\begin{subfigure}[b]{0.5\textwidth}
	\centering
	\hspace{-25pt}\includegraphics[trim=0.15cm 0.1cm 0.2cm 0.3cm,scale=0.12,clip]{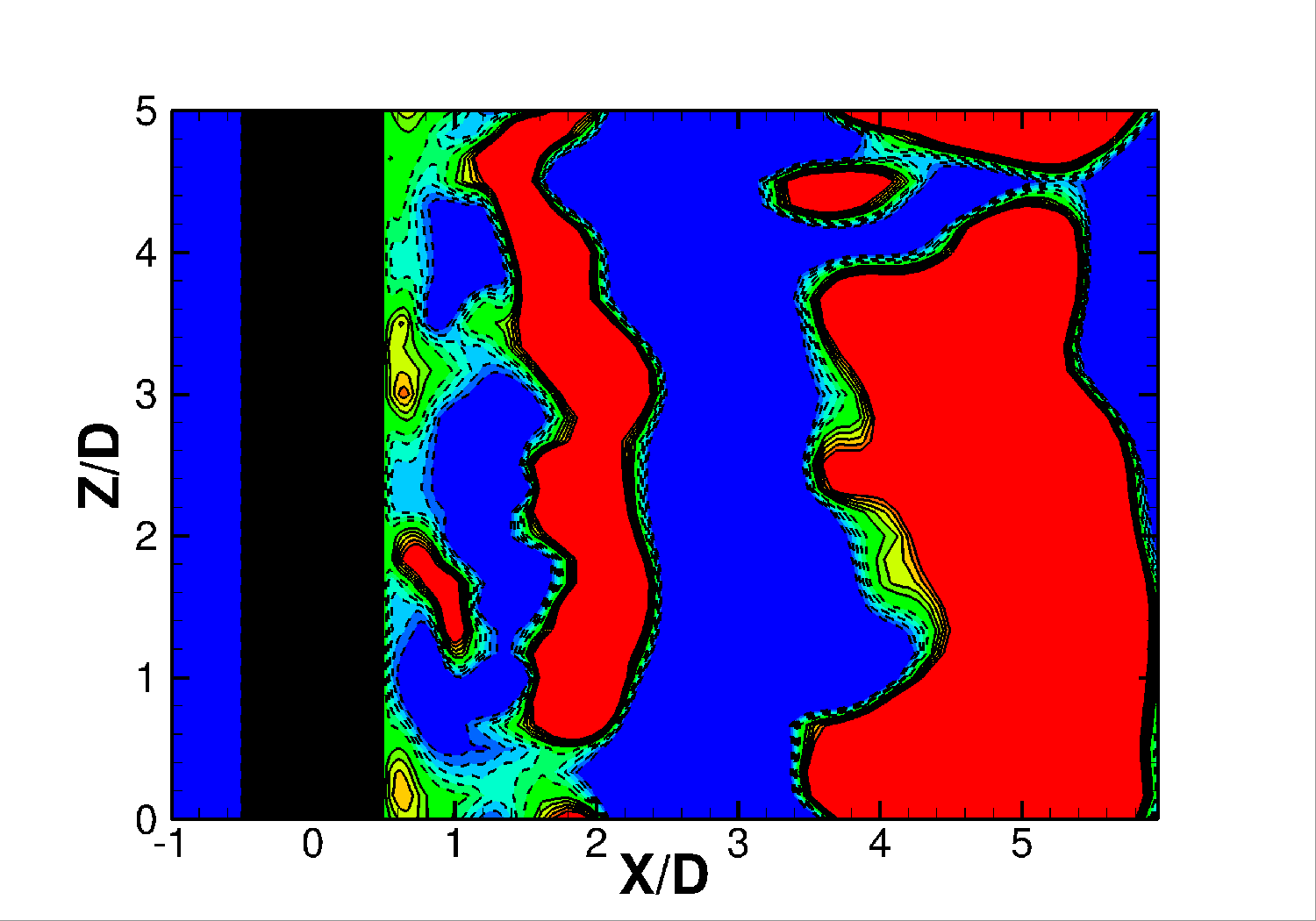}
    \caption{\emph{Cylinder2}}
    \label{fig:dis_C2}
    \end{subfigure}
\caption{Contours of the transverse velocity $v$ at $(z,x)$-plane for SSBS arrangement at $Re=500$, $g^*=0.8$, $tU/D=320$ and $v= \pm 0.1$: (a) wide near-wake region of \emph{Cylinder1} at $y=0.9D$, where the gap flow deflects away momentarily, (b) narrow near-wake region of \emph{Cylinder2} at $y=-0.9D$ with the gap flow deflection.}
   \label{fig:re500td18_dis}
\end{figure}
\begin{figure} \centering   
	\begin{subfigure}[b]{0.5\textwidth}	
	\centering
	\hspace{-25pt}\includegraphics[trim=0.15cm 3cm 0.2cm 0.3cm,scale=0.1,clip]{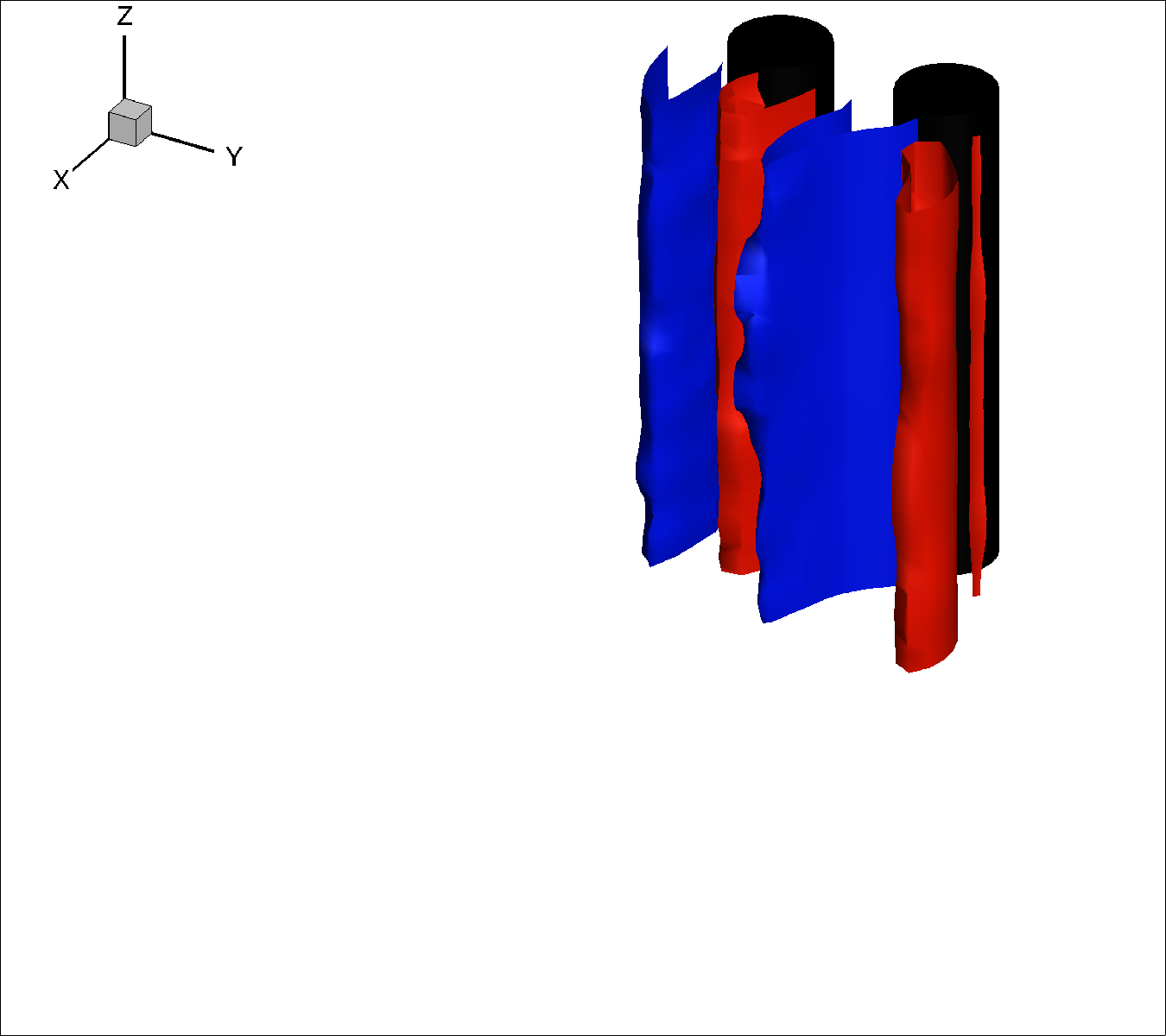}
    \caption{$fD/U=0$, $\alpha=111$}
	\label{fig:re500td18r4_0}
	\end{subfigure}%
	\begin{subfigure}[b]{0.5\textwidth}
	\centering
	\hspace{-25pt}\includegraphics[trim=0.15cm 3cm 0.2cm 0.3cm,scale=0.1,clip]{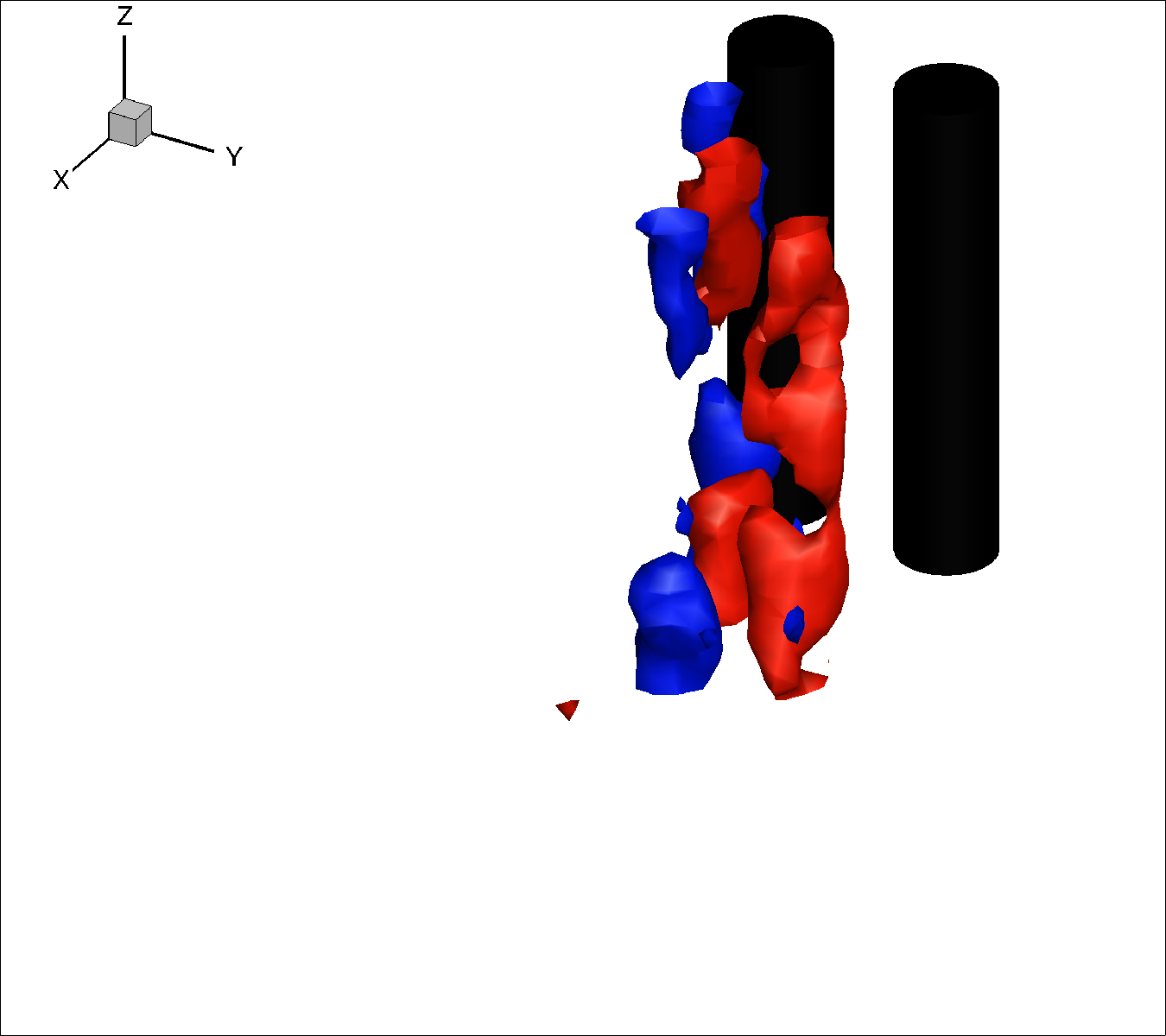}
    \caption{$fD/U=0.16$, $\alpha=51$}
    \label{fig:re500td18r4_16}
    \end{subfigure}
    \begin{subfigure}[b]{0.5\textwidth}	
	\centering
	\hspace{-25pt}\includegraphics[trim=0.15cm 3cm 0.2cm 0.3cm,scale=0.1,clip]{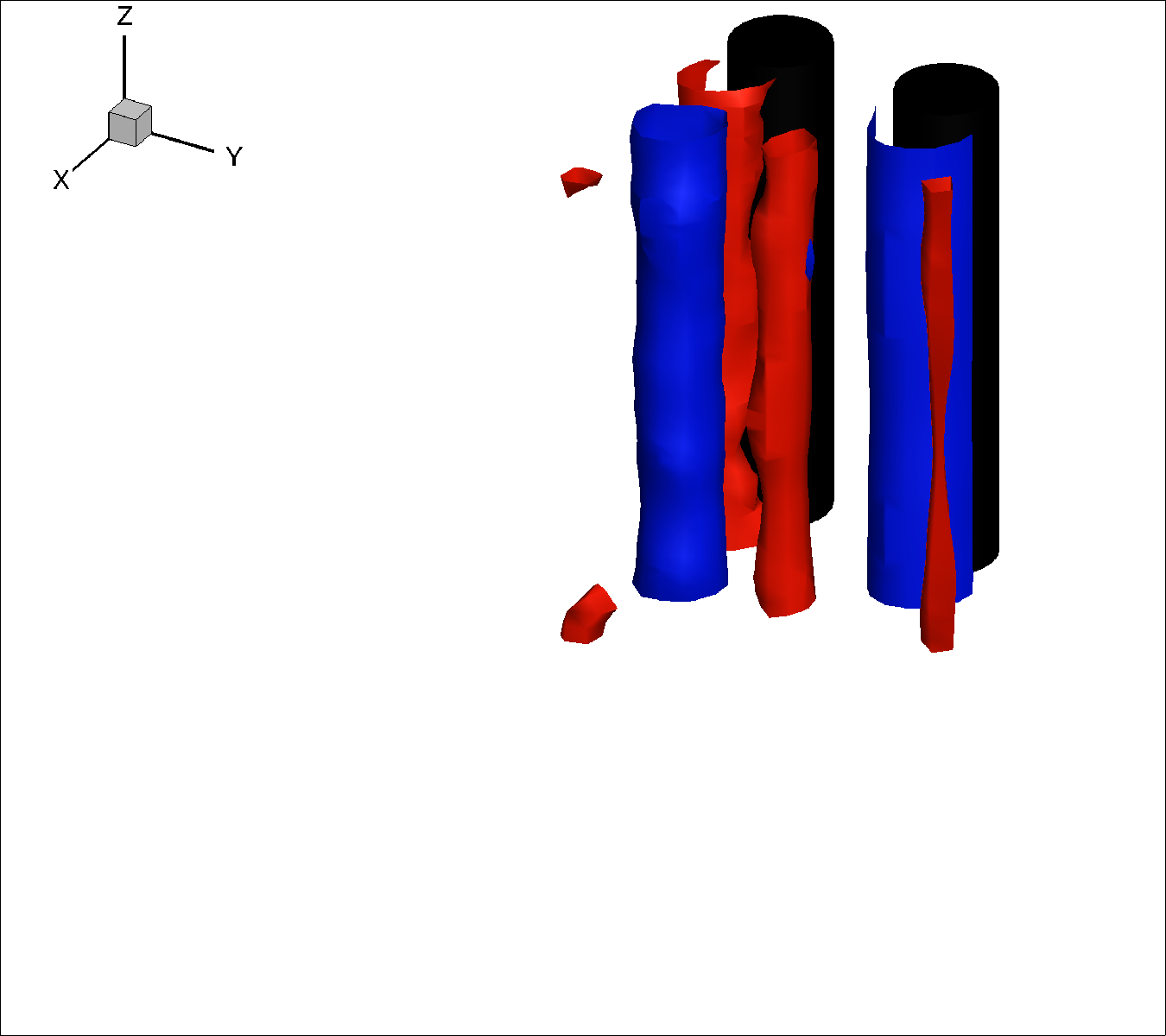}
    \caption{$fD/U=0.24$, $\alpha=101$}
	\label{fig:re500td18r4_24}
	\end{subfigure}%
	\begin{subfigure}[b]{0.5\textwidth}
	\centering
	\hspace{-25pt}\includegraphics[trim=0.15cm 3cm 0.2cm 0.3cm,scale=0.1,clip]{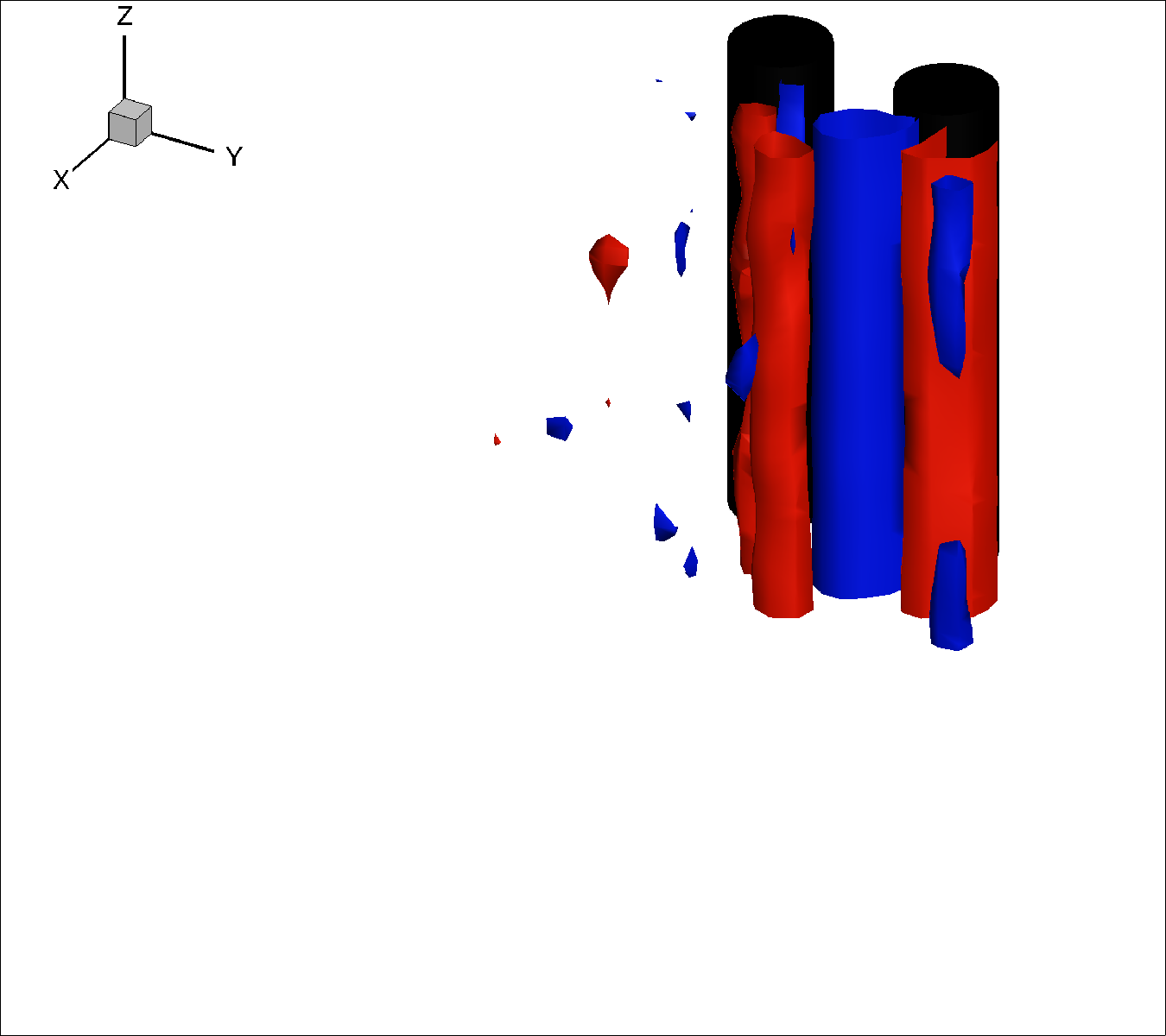}
    \caption{$fD/U=0.48$, $\alpha=41$}
    \label{fig:re500td18r4_48}
    \end{subfigure}
\caption{Iso-surface plots of the primary vortex modes of VSBS arrangement at $Re=500$, $g^*=0.8$, $m^*=10$, $\zeta=0.01$, $tU/D \in [250,$ $350]$ and $\omega_z= \pm 0.01$. The vortex modes are well-distributed at the peak lock-in stage. A strong low frequency vortex mode is observed in the middle path of the gap flow.}
   \label{fig:3D-DMD_re500td18r4}
\end{figure}
The three-dimensional wakes behind the multi-body systems at even a moderate Reynolds number can exhibit complex temporal and spatial flow features. A modal analysis has become a common practice to decompose physical important features or modes as a first step for the subsequent analysis.
Similar to the modal analysis in \cite{lbjrk2016pof}, SP-DMD is employed to investigate the primary vortices behind the SBS arrangements in a three-dimensional flow. The primary focus is to investigate the presentation of the complex coupling between the three-dimensional wakes in the space and the frequency domains. For the computational efficiency viewpoint, $l^*$ value is reduced to five diameters. 
In accordance to the discussion in Section~\ref{sec:Val}, the adopted $l^*$ value should be sufficient to extract and manifest all aforementioned phenomena. A summary of the DMD amplitude as a function 
of frequency for the representative cases is shown in figure~\ref{fig:3D-DMD}. 

The decomposed DMD modes are selected via the sparsity-promoting process in the standard DMD technique, as demonstrated in the examples from \cite{Jovanovic2014PoF1}. Here all selected cases show a strong mean flow vortex mode at $fD/U \approx 0$. In particular, a stationary isolated circular cylinder case shows three clusters of modal frequencies at about $0.2$, $0.4$ and $0.6$. Based on the discussion in \cite{lbjrk2016pof} and comparison with the other spectral analyses in literature, the DMD modes at the modal frequency $0.0$, $0.2$ and $0.4$ in figures~\ref{fig:3D-DMD} account for the mean flow, lift and drag characteristics respectively. Their corresponding DMD modes are visualized in figures~\ref{fig:3D-DMD_re500}(a), (b) and (c), where $fD/U$ refers to a dimensionless frequency for a particular DMD mode. $f$ represents the vortex-wake frequency in each DMD mode with SI unit of Hz. The observed DMD mode with $fD/U = 0$ manifests the mean flow across the cylinder from where the kinetic energy is extracted. In addition, a DMD mode at $fD/U=0.61$ is not observed by \cite{lbjrk2016pof} in the corresponding two-dimensional flow. Taking the mode in figure~\ref{fig:re500_mode_2018} as a fundamental one, the mode at $fD/U=0.61$ can be treated as a third-order harmonic mode. On the other hand, the SSBS arrangement case in figure~\ref{fig:3D-DMD} shows more concentrated vortex modes at low frequency $fD/U \in (0.05$, $0.2)$, because of the gap-flow proximity interference.

For further discussion, their corresponding DMD vortex modes are visualized in figure~\ref{fig:3D-DMD_re500td18}. In the SSBS arrangement, two particular vortex DMD modes are extracted, which possess different modal frequencies and associate with the lift forces over each cylinder. One is at $fD/U \approx 0.12$ behind \emph{Cylinder1} and the other one $fD/U \approx 0.24$ is for \emph{Cylinder2} with a narrow near-wake region\footnote{The gap flow deflects to \emph{Cylinder2} at $tU/D \in (250,$ $350)$ and induces a narrow near-wake region}. The aforementioned $f_{vs}$ augmentation is simply visualized to be induced from the enhanced vortex-to-vortex interaction in a narrow near-wake region. Furthermore a similar third-order harmonic vortex mode is also extracted in this SSBS arrangement, while taking the mean of $0.12$ and $0.24$ as a normalized modal frequency. This third-order harmonic vortex mode is represented by an isolated vortex mode at $fD/U \approx 0.5$ (\emph{crosses}) in figure~\ref{fig:3D-DMD} and is highly concentrated in the narrow near-wake region, as visualized in figure~\ref{fig:re500td18_4975}. 

It is known that the third-order harmonic mode is strongly related to the instability of a dynamical system, since it breaks down the equilibrium of the fundamental base mode and distorts the wave forms. These third-order harmonic modes of the primary vortex are only observed after the flow transition and 
highly concentrated in the region with enhanced three-dimensional 
flow structures. The footprints of these modal vortex patterns are found forming further away from the cylinders in both the isolated and the SBS configurations. 
Consistent with the discussion on the recovery of 2D hydrodynamic responses in Section~\ref{sec:3D_VIV_gap}, it confirms again that the three-dimensional vortical structure is originated from the vortex interactions. 
In addition, figure~\ref{fig:re500td18_1207} shows a discontinuous 
vortex roller mode behind \emph{Cylinder1} with a wide near-wake region 
momentarily. The shedding cell strength is about three diameters 
along the spanwise direction. On the contrary, \emph{Cylinder2} with 
the narrow near-wake region is followed by a continuous vortex roller 
mode in figure~\ref{fig:re500td18_2471} in the same time window. 
The above observation is confirmed from the contour plots of 
the $y$-component velocity in figure~\ref{fig:re500td18_dis}, 
where a vortex discontinuity is observed behind \emph{Cylinder1} 
with a wide near-wake region in figure~\ref{fig:dis_C1}. 
The primary vortex roller of the same sign was dislocated in the 
streamwise direction and represents a discontinuity in $f_{vs}$. 
Since the primary vortex roller momentarily or partially deflects away 
from the wide near-wake region, $f_{vs}$ of \emph{Cylinder1} 
is dynamically varied by the motion of the gap flow and entails 
a vortex discontinuity. 

In the VSBS arrangement, the $f_{vs}$ on both cylinders are synchronized 
with the natural frequency $f_{n}$ of the locked-in vibrating cylinder. 
The vortex modes at the same modal frequencies (\emph{triangles}) 
in figure~\ref{fig:3D-DMD_re500td18r4} is well distributed behind 
both cylinders. Owing to the recovery of 2D hydrodynamic response, 
the primary vortex modes in figures~\ref{fig:re500td18r4_24} and~\ref{fig:re500td18r4_48} are rather similar to their two-dimensional counterparts in figures~12(b) and 12(d) from \cite{lbjrk2016pof} at relatively higher modal frequencies 
respectively. Nonetheless, different from its two-dimensional laminar flow counterparts, an influential vortex mode at $fD/U \approx 0.16$ is observed in 
the middle path of the gap flow in figure~\ref{fig:re500td18r4_16}. The three-dimensional effect is minimized to the most extent along the locked-in vibrating cylinder, thus the three-dimensional vortical structure is merely coupled between the gap flow and the stationary cylinder. By Fast Fourier Transform (FFT) of $C_z$ from the stationary cylinder in this VSBS arrangement, we find a resemblance between the spanwise force $C_z$ frequency and the modal frequency of this particular vortex mode. Further analyzing the position of its vortex pattern, this vortex mode is believed to be originated from the secondary vortex-to-vortex interaction from the gap flow. The undulation of the spanwise hydrodynamic response along the cylinder is strongly influenced by this particular vortex mode induced by the gap flow. 

\section{Conclusions}\label{sec:conclusion}
The dynamics of three-dimensional gap flow and VIV interaction is 
numerically investigated in the side-by-side circular cylinder arrangements 
at moderate Reynolds numbers ranging $ 100 \le Re \le 800$. 
A body-conforming Eulerian-Lagrangian technique based on the variational finite-element formulation has been applied for the fluid-structure interaction.
The in-determinant streamline saddle point was found forming along the interface between the imbalanced counter-signed vorticity clusters. The saddle-point regions are intrinsically associated with high local strain rates and contribute to the formation of three-dimensional vortical structures.
From the analysis of flow field around the saddle-point regions, 
two factors are found to be important for the near-wake stability, 
namely the intensity of the fluid momentum and the fluid shearing induced along the vortical interfaces. 
Consistent with the analysis of near-wake instability, 
the VIV regulation effect on the three-dimensional aspects of organized motion in the near-wake and a recovery of two-dimensional hydrodynamic response
at the peak VIV lock-in are inter-linked. 

In the SBS arrangements, the streamwise vorticity concentration was found having a strong dependency on the gap-flow proximity interference. 
A high streamwise vorticity concentration was observed behind the cylinder with the narrow near-wake region. Furthermore, the $C_z$ response was amplified in both SSBS and VSBS arrangements along the cylinder with a narrow near-wake region. 
This asymmetry of hydrodynamic responses induced from the deflected gap flow caused a spatial modulation, about a factor of 2.4 amplification of the $C^{rms}_{l}$ response.
The flip flop was also subjected to the influence by the three-dimensionality of flow, 
while it was remarkably suppressed in both SSBS and VSBS arrangements at the off-lock-in 
and restored together with two-dimensional response at the lock-in stage for 
the VSBS arrangements.
The gap flow was also found to promote the three-dimensional flow feature through enhanced fluid shearing and mixing, which 
exerts a strong proximity interference stabilizing the energy transfer between the fluid and structure.
In addition to the three-dimensional interference, the VIV kinematics and the gap flow are mutually influenced by each other.
The onset of the VIV lock-in in the VSBS arrangement was observed at a smaller $U_r$ value. 
A quasi-stable deflected gap-flow regime was observed in the VSBS arrangements at the peak lock-in, in which the gap flow deflected steadily behind the locked-in vibrating cylinder.
The spanwise wavelength $\lambda^*$ value was significantly affected by both the VIV and the gap-flow proximity interference. 
Due to the regulation of energy transfer between 
the fluid flow and the structure, the wavelength $\lambda^*$ become more uniform along the cylinder and resulted into increased $\lambda^*$ at the peak lock-in.

Through the modal analysis, the third-order harmonic primary vortex modes were dynamically decomposed for the isolated cylinder and the SSBS arrangements. 
Owing to their odd-order mathematical characteristics, the third-order vortex modes are crucial to the stability of the dynamic fluid-structure system and represents an unstable factor. 
A vortex discontinuity originated from the gap-flow kinematics was observed using the DMD technique in the wide near-wake region. An additional influential primary vortex mode along the middle path of the gap flow in the VSBS arrangement was observed, which was related to the periodic undulation 
of spanwise hydrodynamic response along the cylinder and represented the promoted gap-flow instability.

Overall, it was found that the vortex-to-vortex interaction between the imbalanced counter-signed vorticity clusters plays an important role in the near-wake stability, because of the significant fluid shearing along the vortical interfaces. In general, the intensive fluid shearings along the vortical interfaces were associated with the in-determinant streamline saddle-point regions. The saddle-point region is found in all range of Reynolds number and is interlinked with various flow dynamic events, e.g. the vortex shedding, the flip flop, the streamwise vorticity clusters. Furthermore, the near-wake instability is found to be closely interlinked with the gap flow and the VIV kinematics. In particular, as the VIV kinematics increases and stretches the vortices, the vorticity clusters are more separated to weaken the vortex-to-vortex interaction in the near-wake region. As a result, the two-dimensional hydrodynamic responses are significantly restored along the cylinder. On the contrary, the interaction dynamics between the gap-flow proximity interference and the gap-flow instability enhances the vortex-to-vortex interaction. 
%
These observations and findings are important in multi-body systems, from both operations and design viewpoints, found in offshore and aeronautical engineering. 

The authors would like to thank Singapore Maritime Institute Grant (SMI-2014-OF-04) for the financial support. 

\bibliographystyle{jfm}
\bibliography{refs}

\begin{thebibliography}{47}
\expandafter\ifx\csname natexlab\endcsname\relax\def\natexlab#1{#1}\fi
\def\au#1{#1} \def\ed#1{#1} \def\yr#1{#1}\def\at#1{#1}\def\jt#1{\textit{#1}}
  \def\bt#1{#1}\def\bvol#1{\textbf{#1}} \def\vol#1{#1} \def\pg#1{#1}
  \def\publ#1{#1}\def\arxiv#1{#1}\def\org#1{#1}\def\st#1{\textit{#1}}

\bibitem[Agrawal {\em et~al.\/}(2006)Agrawal, Djenidi \&
  Antonia]{Agrawal2006Cf}
{\sc \au{Agrawal, Amit}, \au{Djenidi, Lyazid} \& \au{Antonia, RA}} \yr{2006}
  \at{Investigation of flow around a pair of side-by-side square cylinders
  using the lattice boltzmann method}.  \jt{Computers \& fluids}
  \bvol{35}~(10),  \pg{1093--1107}.

\bibitem[Alam \& Sakamoto(2005)]{alam2005investigation}
{\sc \au{Alam, M.~M.} \& \au{Sakamoto, H.}} \yr{2005}  \at{Investigation of
  strouhal frequencies of two staggered bluff bodies and detection of
  multistable flow by wavelets}.  \jt{Journal of Fluids and Structures}
  \bvol{20}~(3),  \pg{425--449}.

\bibitem[Bearman \& Wadcock(1973)]{bearman1973interaction}
{\sc \au{Bearman, P.~W.} \& \au{Wadcock, A.~J.}} \yr{1973}  \at{The interaction
  between a pair of circular cylinders normal to a stream}.  \jt{Journal of
  Fluid Mechanics}  \bvol{61}~(03),  \pg{499--511}.

\bibitem[Behara \& Mittal(2010)]{behara2010wake}
{\sc \au{Behara, Suresh} \& \au{Mittal, Sanjay}} \yr{2010}  \at{Wake transition
  in flow past a circular cylinder}.  \jt{Physics of Fluids}  \bvol{22}~(11),
  \pg{114104}.

\bibitem[Blackburn {\em et~al.\/}(2001)Blackburn, Govardhan \&
  Williamson]{Blackburn2001JoFaS}
{\sc \au{Blackburn, H.~M.}, \au{Govardhan, R.~N.} \& \au{Williamson, C. H.~K.}}
  \yr{2001}  \at{A complementary numerical and physical investigation of
  vortex-induced vibration}.  \jt{Journal of Fluids and Structures}
  \bvol{15}~(3),  \pg{481--488}.

\bibitem[Borazjani \& Sotiropoulos(2009)]{Borazjani2009Jofm}
{\sc \au{Borazjani, Iman} \& \au{Sotiropoulos, Fotis}} \yr{2009}
  \at{Vortex-induced vibrations of two cylinders in tandem arrangement in the
  proximity--wake interference region}.  \jt{Journal of fluid mechanics}
  \bvol{621},  \pg{321--364}, three dimensionality suppression in the gap
  between two tandem cylinders.

\bibitem[Carini {\em et~al.\/}(2014)Carini, Giannetti \&
  Auteri]{carini2014origin}
{\sc \au{Carini, M.}, \au{Giannetti, F.} \& \au{Auteri, F.}} \yr{2014}  \at{On
  the origin of the flip--flop instability of two side-by-side cylinder wakes}.
   \jt{Journal of Fluid Mechanics}  \bvol{742},  \pg{552--576}.

\bibitem[Chantry {\em et~al.\/}(2016)Chantry, Tuckerman \&
  Barkley]{Chantry2016JoFM}
{\sc \au{Chantry, M.}, \au{Tuckerman, L.~S.} \& \au{Barkley, D.}} \yr{2016}
  \at{Turbulent--laminar patterns in shear flows without walls}.  \jt{Journal
  of Fluid Mechanics}  \bvol{791},  \pg{R8}.

\bibitem[Huang(2014)]{huang2014hilbert}
{\sc \au{Huang, Norden~Eh}} \yr{2014} {\em Hilbert-Huang transform and its
  applications\/}, ,  \vol{vol.~16}.  \publ{World Scientific}.

\bibitem[Huang {\em et~al.\/}(2010)Huang, Narasimhamurthy \&
  Andersson]{Huang2010PoF}
{\sc \au{Huang, Zhiyong}, \au{Narasimhamurthy, Vagesh~D} \& \au{Andersson,
  Helge~I}} \yr{2010}  \at{Oblique and cellular vortex shedding behind a
  circular cylinder in a bidirectional shear flow}.  \jt{Physics of Fluids}
  \bvol{22}~(11),  \pg{114105}.

\bibitem[Hunt {\em et~al.\/}(1988)Hunt, Wray \& Moin]{hunt}
{\sc \au{Hunt, J. C.~R.}, \au{Wray, A.} \& \au{Moin, P.}} \yr{1988}
  \bt{Eddies, stream, and convergence zones in turbulent flows}. {\em Tech.
  Rep.\/}.  \org{Center for Turbulence Research Report}.

\bibitem[Ishigai {\em et~al.\/}(1972)Ishigai, Nishikawa, Nishimura \&
  Cho]{ishigai1972}
{\sc \au{Ishigai, S.}, \au{Nishikawa, E.}, \au{Nishimura, K.} \& \au{Cho, K.}}
  \yr{1972}  \at{Experimental study on structure of gas flow in tube banks with
  tube axes normal to flow: Part 1, k\'arm\'an vortex flow from two tubes at
  various spacings}.  \jt{Bulletin of JSME}  \bvol{15}~(86),  \pg{949--956}.

\bibitem[Jaiman {\em et~al.\/}(2016{\natexlab{{\em a\/}}})Jaiman, Guan \&
  Miyanawala]{jaiman_caf2016}
{\sc \au{Jaiman, R.K.}, \au{Guan, M.Z.} \& \au{Miyanawala, T.~P.}}
  \yr{2016{\natexlab{{\em a\/}}}}  \at{Partitioned iterative and dynamic
  subgrid-scale methods for freely vibrating square-section structure in
  turbulent flow}.  \jt{Computers and Fluids}  \bvol{133},  \pg{68--89}.

\bibitem[Jaiman {\em et~al.\/}(2015)Jaiman, Sen \&
  Gurugubelli]{jaiman_ficf2015}
{\sc \au{Jaiman, R.K.}, \au{Sen, S.} \& \au{Gurugubelli, P.}} \yr{2015}  \at{A
  fully implicit combined field scheme for freely vibrating square cylinders
  with sharp and rounded corners}.  \jt{Computers and Fluids}  \bvol{112},
  \pg{1--18}.

\bibitem[Jaiman {\em et~al.\/}(2016{\natexlab{{\em b\/}}})Jaiman, Pillalamarri
  \& Guan]{Jaiman2016CMiAMaE}
{\sc \au{Jaiman, R.~K.}, \au{Pillalamarri, N.~R.} \& \au{Guan, M.~Z.}}
  \yr{2016{\natexlab{{\em b\/}}}}  \at{A stable second-order partitioned
  iterative scheme for freely vibrating low-mass bluff bodies in a uniform
  flow}.  \jt{Computer Methods in Applied Mechanics and Engineering}
  \bvol{301},  \pg{187--215}.

\bibitem[Jovanovic {\em et~al.\/}(2014)Jovanovic, Schmid \&
  Nichols]{Jovanovic2014PoF1}
{\sc \au{Jovanovic, Mihailo~R}, \au{Schmid, Peter~J} \& \au{Nichols, Joseph~W}}
  \yr{2014}  \at{Sparsity-promoting dynamic mode decomposition}.  \jt{Physics
  of Fluids (1994-present)}  \bvol{26}~(2),  \pg{024103}.

\bibitem[Kang(2003)]{kang2003characteristics}
{\sc \au{Kang, S.}} \yr{2003}  \at{Characteristics of flow over two circular
  cylinders in a side-by-side arrangement at low reynolds numbers}.
  \jt{Physics of Fluids (1994-present)}  \bvol{15}~(9),  \pg{2486--2498}.

\bibitem[Kerswell(2002)]{Kerswell2002Arofm}
{\sc \au{Kerswell, R.~R.}} \yr{2002}  \at{Elliptical instability}.  \jt{Annual
  review of fluid mechanics}  \bvol{34}~(1),  \pg{83--113}.

\bibitem[Kim(1988)]{kim1988investigation}
{\sc \au{Kim, H.~J.}} \yr{1988}  \at{Investigation of the flow between a pair
  of circular cylinders in the flopping regime}.  \jt{Journal of Fluid
  Mechanics}  \bvol{196},  \pg{431--448}.

\bibitem[Lankadasu \& Vengadesan(2008)]{lankadasu2008onset}
{\sc \au{Lankadasu, A} \& \au{Vengadesan, S}} \yr{2008}  \at{Onset of vortex
  shedding in planar shear flow past a square cylinder}.  \jt{International
  Journal of Heat and Fluid Flow}  \bvol{29}~(4),  \pg{1054--1059}.

\bibitem[Le~Dizes \& Laporte(2002)]{LeDizes2002JoFM}
{\sc \au{Le~Dizes, St{\'e}phane} \& \au{Laporte, Florent}} \yr{2002}
  \at{Theoretical predictions for the elliptical instability in a two-vortex
  flow}.  \jt{Journal of Fluid Mechanics}  \bvol{471},  \pg{169--201}.

\bibitem[Leblanc(1997)]{leblanc1997stability}
{\sc \au{Leblanc, S.}} \yr{1997}  \at{Stability of stagnation points in
  rotating flows}.  \jt{Physics of Fluids (1994-present)}  \bvol{9}~(11),
  \pg{3566--3569}.

\bibitem[Lei {\em et~al.\/}(2001)Lei, Cheng \& Kavanagh]{lei2001spanwise}
{\sc \au{Lei, C.}, \au{Cheng, L.} \& \au{Kavanagh, K.}} \yr{2001}  \at{Spanwise
  length effects on three-dimensional modelling of flow over a circular
  cylinder}.  \jt{Computer methods in applied mechanics and engineering}
  \bvol{190}~(22),  \pg{2909--2923}.

\bibitem[Leontini {\em et~al.\/}(2006)Leontini, Stewart, Thompson \&
  Hourigan]{Leontini2006PoF}
{\sc \au{Leontini, JS}, \au{Stewart, BE}, \au{Thompson, MC} \& \au{Hourigan,
  K}} \yr{2006}  \at{Wake state and energy transitions of an oscillating
  cylinder at low reynolds number}.  \jt{Physics of Fluids}  \bvol{18}~(6),
  \pg{067101}.

\bibitem[Li {\em et~al.\/}(2016)Li, Yao, Yang, Jaiman \& Khoo]{li2016vortex}
{\sc \au{Li, Z.}, \au{Yao, W.}, \au{Yang, K.}, \au{Jaiman, R.~K.} \& \au{Khoo,
  B.~C.}} \yr{2016}  \at{On the vortex-induced oscillations of a freely
  vibrating cylinder in the vicinity of a stationary plane wall}.  \jt{Journal
  of Fluids and Structures}  \bvol{65},  \pg{495--526}.

\bibitem[Lifschitz \& Hameiri(1991)]{lifschitz1991local}
{\sc \au{Lifschitz, A.} \& \au{Hameiri, E.}} \yr{1991}  \at{Local stability
  conditions in fluid dynamics}.  \jt{Physics of Fluids A: Fluid Dynamics
  (1989-1993)}  \bvol{3}~(11),  \pg{2644--2651}.

\bibitem[Lin {\em et~al.\/}(2002)Lin, Yang \& Rockwell]{lin2002}
{\sc \au{Lin, J.~C.}, \au{Yang, Y.} \& \au{Rockwell, D.}} \yr{2002}  \at{Flow
  past two cylinders in tandem: instantaneous and averaged flow structure}.
  \jt{Journal of Fluids and Structures}  \bvol{16}~(8),  \pg{1059--1071}.

\bibitem[Liu \& Jaiman(2016)]{lbjrk2016pof}
{\sc \au{Liu, B.} \& \au{Jaiman, R.~K.}} \yr{2016}  \at{Interaction dynamics of
  gap flow with vortex-induced vibration in side-by-side cylinder arrangement}.
   \jt{Physics of Fluids (1994-present)}  \bvol{28}~(12),  \pg{127103}.

\bibitem[Meunier {\em et~al.\/}(2005)Meunier, Le~Dizes \&
  Leweke]{Meunier2005CRP}
{\sc \au{Meunier, P.}, \au{Le~Dizes, S.} \& \au{Leweke, T.}} \yr{2005}
  \at{Physics of vortex merging}.  \jt{Comptes Rendus Physique}  \bvol{6}~(4),
  \pg{431--450}.

\bibitem[Mizushima \& Ino(2008)]{Mizushima2008JoFM}
{\sc \au{Mizushima, J} \& \au{Ino, Y}} \yr{2008}  \at{Stability of flows past a
  pair of circular cylinders in a side-by-side arrangement}.  \jt{Journal of
  Fluid Mechanics}  \bvol{595},  \pg{491--507}.

\bibitem[Mysa {\em et~al.\/}(2016)Mysa, Kaboudian \& Jaiman]{ravi2016}
{\sc \au{Mysa, R.~C.}, \au{Kaboudian, A.} \& \au{Jaiman, R.~K.}} \yr{2016}
  \at{On the origin of wake-induced vibration in two tandem circular cylinders
  at low reynolds number}.  \jt{Journal of Fluids and Structures}  \bvol{61},
  \pg{76--98}.

\bibitem[Perry \& Chong(1987)]{Perry1987ARoFM}
{\sc \au{Perry, A.~E.} \& \au{Chong, M.~S.}} \yr{1987}  \at{A description of
  eddying motions and flow patterns using critical-point concepts}.  \jt{Annual
  Review of Fluid Mechanics}  \bvol{19}~(1),  \pg{125--155}.

\bibitem[Persillon \& Braza(1998)]{persillon1998physical}
{\sc \au{Persillon, Helene} \& \au{Braza, Marianna}} \yr{1998}  \at{Physical
  analysis of the transition to turbulence in the wake of a circular cylinder
  by three-dimensional navier--stokes simulation}.  \jt{Journal of Fluid
  Mechanics}  \bvol{365},  \pg{23--88}.

\bibitem[Peschard \& Le~Gal(1996)]{Peschard1996PRL}
{\sc \au{Peschard, I.} \& \au{Le~Gal, P.}} \yr{1996}  \at{Coupled wakes of
  cylinders}.  \jt{Physical Review Letters}  \bvol{77}~(15),  \pg{3122}.

\bibitem[Schmid(2010)]{Schmid2010JoFM}
{\sc \au{Schmid, Peter~J}} \yr{2010}  \at{Dynamic mode decomposition of
  numerical and experimental data}.  \jt{Journal of Fluid Mechanics}
  \bvol{656},  \pg{5--28}.

\bibitem[Schmid {\em et~al.\/}(2011)Schmid, Li, Juniper \&
  Pust]{Schmid2011TaCFD}
{\sc \au{Schmid, Peter~J}, \au{Li, L}, \au{Juniper, MP} \& \au{Pust, O}}
  \yr{2011}  \at{Applications of the dynamic mode decomposition}.
  \jt{Theoretical and Computational Fluid Dynamics}  \bvol{25}~(1-4),
  \pg{249--259}.

\bibitem[Sumner(2010)]{Sumner2010JoFaS}
{\sc \au{Sumner, D.}} \yr{2010}  \at{Two circular cylinders in cross-flow: A
  review}.  \jt{Journal of Fluids and Structures}  \bvol{26}~(6),
  \pg{849--899}.

\bibitem[Sumner {\em et~al.\/}(2000)Sumner, Price \& Paidoussis]{sumner2000}
{\sc \au{Sumner, D.}, \au{Price, S.~J.} \& \au{Paidoussis, M.~P.}} \yr{2000}
  \at{Flow-pattern identification for two staggered circular cylinders in
  cross-flow}.  \jt{Journal of Fluid Mechanics}  \bvol{411},  \pg{263--303}.

\bibitem[Sumner {\em et~al.\/}(1999)Sumner, Wong, Price \&
  Paidoussis]{sumner1999}
{\sc \au{Sumner, D.}, \au{Wong, S. S.~T.}, \au{Price, S.~J.} \& \au{Paidoussis,
  M.~P.}} \yr{1999}  \at{Fluid behaviour of side-by-side circular cylinders in
  steady cross-flow}.  \jt{Journal of Fluids and Structures}  \bvol{13}~(3),
  \pg{309--338}.

\bibitem[Szepessy \& Bearman(1992)]{Szepessy1992JoFM}
{\sc \au{Szepessy, S} \& \au{Bearman, PW}} \yr{1992}  \at{Aspect ratio and end
  plate effects on vortex shedding from a circular cylinder}.  \jt{Journal of
  Fluid Mechanics}  \bvol{234},  \pg{191--217}.

\bibitem[Williamson(1985)]{williamson1985evolution}
{\sc \au{Williamson, C. H.~K.}} \yr{1985}  \at{Evolution of a single wake
  behind a pair of bluff bodies}.  \jt{Journal of Fluid Mechanics}  \bvol{159},
   \pg{1--18}.

\bibitem[Williamson(1996{\natexlab{{\em a\/}}})]{williamson1996three}
{\sc \au{Williamson, C. H.~K.}} \yr{1996{\natexlab{{\em a\/}}}}
  \at{Three-dimensional wake transition}.  \bt{In {\em Advances in Turbulence
  VI\/}},  \pg{pp. 399--402}.  \publ{Springer}.

\bibitem[Williamson(1996{\natexlab{{\em b\/}}})]{Williamson1996ARoFM}
{\sc \au{Williamson, C. H.~K.}} \yr{1996{\natexlab{{\em b\/}}}}  \at{Vortex
  dynamics in the cylinder wake}.  \jt{Annual Review of Fluid Mechanics}
  \bvol{28}~(1),  \pg{477--539}.

\bibitem[Zdravkovich(1987)]{Zdravkovich1987Jofas}
{\sc \au{Zdravkovich, M.~M.}} \yr{1987}  \at{The effects of interference
  between circular cylinders in cross flow}.  \jt{Journal of fluids and
  structures}  \bvol{1}~(2),  \pg{239--261}.

\bibitem[Zhang {\em et~al.\/}(1995)Zhang, Fey, Noack, K{\"o}nig \&
  Eckelmann]{zhang1995transition}
{\sc \au{Zhang, Hong-Quan}, \au{Fey, Uwe}, \au{Noack, Bernd~R}, \au{K{\"o}nig,
  Michael} \& \au{Eckelmann, Helmut}} \yr{1995}  \at{On the transition of the
  cylinder wake}.  \jt{Physics of Fluids}  \bvol{7}~(4),  \pg{779--794}.

\bibitem[Zhao {\em et~al.\/}(2014)Zhao, Cheng, An \& Lu]{Zhao2014JoFaS}
{\sc \au{Zhao, Ming}, \au{Cheng, Liang}, \au{An, Hongwei} \& \au{Lu, Lin}}
  \yr{2014}  \at{Three-dimensional numerical simulation of vortex-induced
  vibration of an elastically mounted rigid circular cylinder in steady
  current}.  \jt{Journal of Fluids and Structures}  \bvol{50},  \pg{292--311},
  good investigation of VIV of 3D isolated cylinder in turbulent flow. It also
  proves that 2D DNS is not capable to solve turbulent VIV. In 2D, RANS can
  improve this.

\bibitem[Zhou \& Antonia(1994)]{Zhou1994JoFM}
{\sc \au{Zhou, Y} \& \au{Antonia, RA}} \yr{1994}  \at{Critical points in a
  turbulent near wake}.  \jt{Journal of Fluid Mechanics}  \bvol{275},
  \pg{59--81}.

\end{thebibliography}

\appendix
\setcounter{equation}{0} 
\setcounter{figure}{0}
\renewcommand{\theequation}{A.\arabic{equation}}
\renewcommand{\thefigure}{A.\arabic{figure}}

\section{Streamline saddle point and $(x,y)$-plane velocity profile}\label{appA}
A streamline saddle point is located in a two-dimensional incompressible flow far from any boundaries, where the velocity components are defined from stream function $\psi(x,y)$ as
\begin{align}
u =  \frac{\p \psi}{\p y}, \qquad v  =  -\frac{\p \psi}{\p x} \label{eq:def}
\end{align}
A fourth-order biquartic polynomial of $\psi(x,y)$ surface patch\footnote{where the fluid shear stresses are approximated linearly in the field} is employed to approximate the local continuous $\psi(x,y)$ field around an SSP. Assuming that the parametric surface of two-dimensional stream function is smooth, continuous and their spatial derivatives are everywhere well-defined up to the highest order of the approximating function, a local flow field can be represented by a general form as follows
\begin{eqnarray}
\psi(x,y) = a_0 + a_1 x + a_2 y + a_3 xy + a_4 x^2 + a_5 y^2 + a_6 x^3  + a_7 x^2 y + a_8 x y^2 + a_9 y^3 \nonumber\\ 
+ a_{10} x^4 + a_{11} x^3 y + a_{12} x^2 y^2 + a_{13} xy^3  + a_{14}  y^4 + O(x^5,y^5)\label{eq:saddle_1}
\end{eqnarray}
Here $a_i$ ($i=0, ..., 14$) are arbitrary scalar constants and $O(x^5,y^5)$ is the truncation error. Based on the \emph{second derivative test for local extreme values}, Eq. (\ref{eq:saddle_1}) has to satisfy a criterion Eq. (\ref{eq:cri}) to approximate a (non-degenerated) two-dimensional SSP at ($0$,$0$). 
\begin{eqnarray}
\frac{\p \psi}{\p x} = 0; \qquad \frac{\p \psi}{\p y} = 0; \qquad 
\frac{\p^2 \psi}{\p x^2}\frac{\p^2 \psi}{\p y^2}-\left[\frac{\p^2 \psi}{\p x \p y}\right]^2 < 0 \qquad at \quad (x,y)=(0, 0) \label{eq:cri}
\end{eqnarray}
By imposing Eq. (\ref{eq:cri}) on Eq. (\ref{eq:saddle_1}), $a_1 = a_2 = 0$ and $4 a_4 \cdot a_5 < (a_3)^2$. The approximating function reduces to
\begin{eqnarray}
\psi(x,y) = a_0 + a_3 xy + a_4 x^2 + a_5 y^2 + a_6 x^3  + a_7 x^2 y + a_8 x y^2 + a_9 y^3 \nonumber\\ 
+ a_{10} x^4 + a_{11} x^3 y + a_{12} x^2 y^2 + a_{13} xy^3  + a_{14}  y^4 + O(x^5,y^5)\label{eq:saddle_2}
\end{eqnarray}
To investigate the fluid field characteristics in a saddle-point region, $x=0$ and $y=0$ are substituted into the first and second derivatives of Eq. (\ref{eq:saddle_2}) as follows
\begin{eqnarray}
u(0,0) & = & 0; \quad v(0,0) = 0  \nonumber\\
\frac{\p u}{\p y} |_{(0,0)} & = & 2 a_{5} ; \quad \frac{\p v}{\p x} |_{(0,0)} = -2 a_{4} \quad \,  \nonumber\\
\frac{\p^2 u}{\p y^2} |_{(0,y)} & = & 6 a_{9} + 24 a_{14} y; \quad \frac{\p^2 v}{\p x^2} |_{(x,0)} = -6 a_{6} - 24 a_{10} x \nonumber\\
\frac{\p^2 u}{\p y^2} |_{(0,0)} & = & 6 a_{9}; \quad \frac{\p^2 v}{\p x^2} |_{(0,0)} = -6 a_{6} \label{eq:00}
\end{eqnarray}
Analyzing Eq. (\ref{eq:00}), the velocity is zero at a saddle point and generate a local stagnant region. The directional gradients of $u$ and $v$ are non-zero scalar constants at the saddle point. The second-order derivatives of $u$ and $v$ are also found having a linear relationship with respect to $y$ and $x$ variables respectively. Hence there is a point along $y$ and $x$ axes respectively across the streamline saddle point, where $\frac{\p^2 u}{\p y^2}$ and $\frac{\p^2 v}{\p x^2}$ switch their signs. The zero values of $\frac{\p^2 u}{\p y^2}$ and $\frac{\p^2 v}{\p x^2}$ are the locations of the inflection points of $u$ and $v$. As the values of $a_9$ and $a_6$ approach zero, the velocity profile inflection points translate toward the streamline saddle point. Since the linear odd-order terms in the approximating biquartic polynomial tend to destroy the symmetry of $\psi(x,y)$ about the saddle point, the amplification of their coefficient values is detrimental to the formation of the saddle point. Therefore, the locations of the velocity profile inflection points is expected to be not far from the streamline saddle point.

\section{Stability analysis of streamline saddle point}\label{appB} 
This appendix is concerned with the near-wake instability in a saddle-point region behind cylinders in the SSBS arrangements. The goal is to investigate the correlation between the near-wake stability and its stability parameters, e.g., Reynolds number $Re$, fluid shear ratio $S = U_1 / U_2$ and gap ratio $g^*$, using DMD technique. Here $U_1$ and $U_2$ are respectively x-component inlet velocities in front of the gap flow and the free-side of cylinders in the SSBS arrangement. The development of the well-known vortex shedding \emph{Hopf bifurcation} at a low Reynolds number $Re \lesssim 100$ is taken as an indication of the near-wake instability. Its relationship with $Re$ and $S$ values are used to analyze the importance of the fluid momentum and the fluid shearing ratio induced from the imbalanced counter-signed vorticity concentration to the near-wake stability. Here a large $g^*$ value, $g^* > 3.5$ at $Re \lesssim 100$, is adopted to eliminate the gap-flow proximity interference. The gap-flow proximity interference to the near-wake stability is subsequently analyzed by further reducing $g^*$ value. The rest of parameters is identical to the $(x,y)$-section of the three-dimensional computational setup.

\begin{figure} \centering   
	\begin{subfigure}[b]{0.5\textwidth}	
		\centering
		\hspace{-25pt}\includegraphics[trim=0.1cm 0.1cm 0cm 0.1cm,scale=0.097,clip]{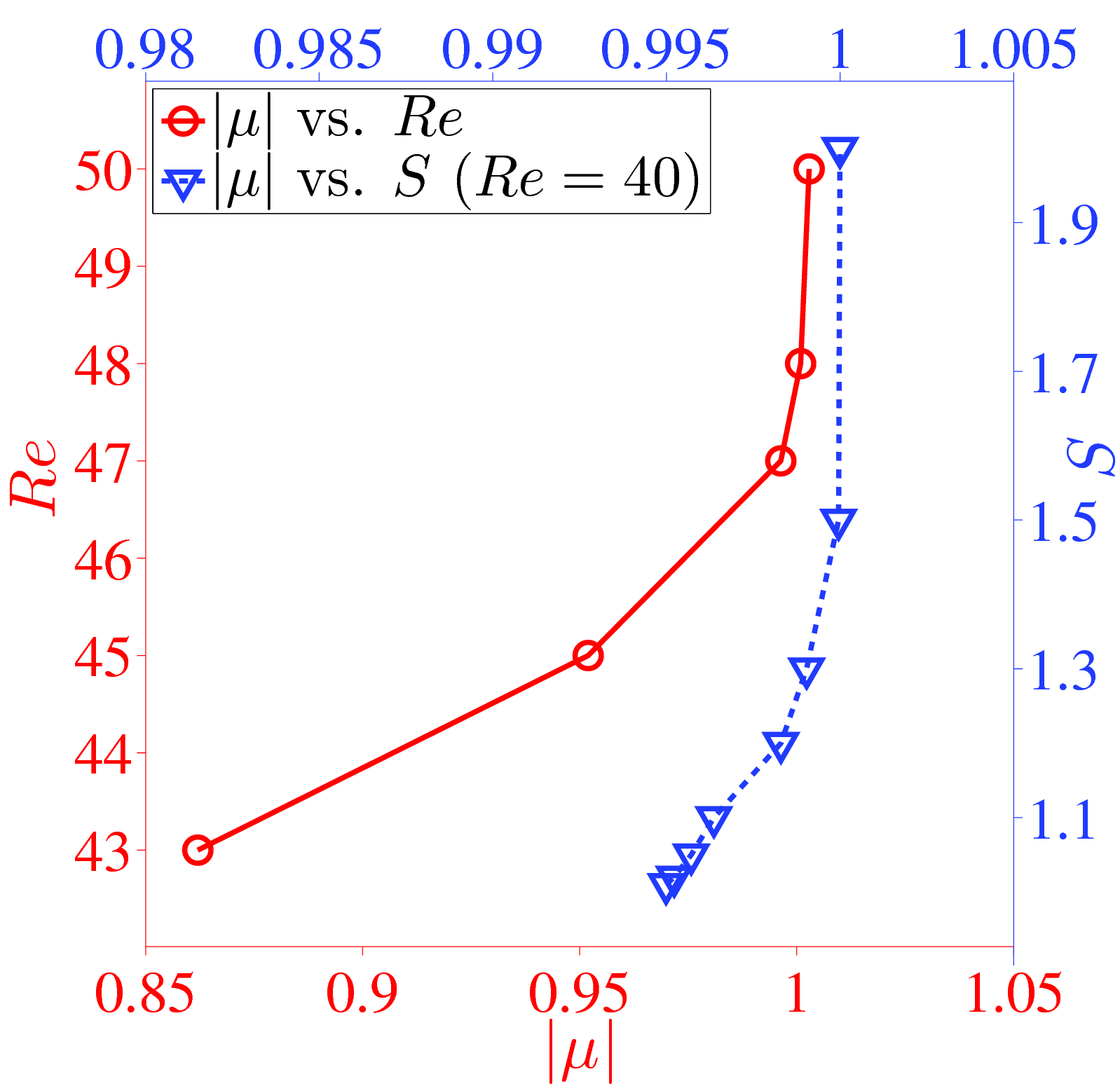}
		\caption{$\qquad$}
		\label{fig:RES_Mu}
	\end{subfigure}%
	\begin{subfigure}[b]{0.5\textwidth}
		\centering
		\hspace{-25pt}\includegraphics[trim=0.1cm 0.1cm 0cm 0.1cm,scale=0.097,clip]{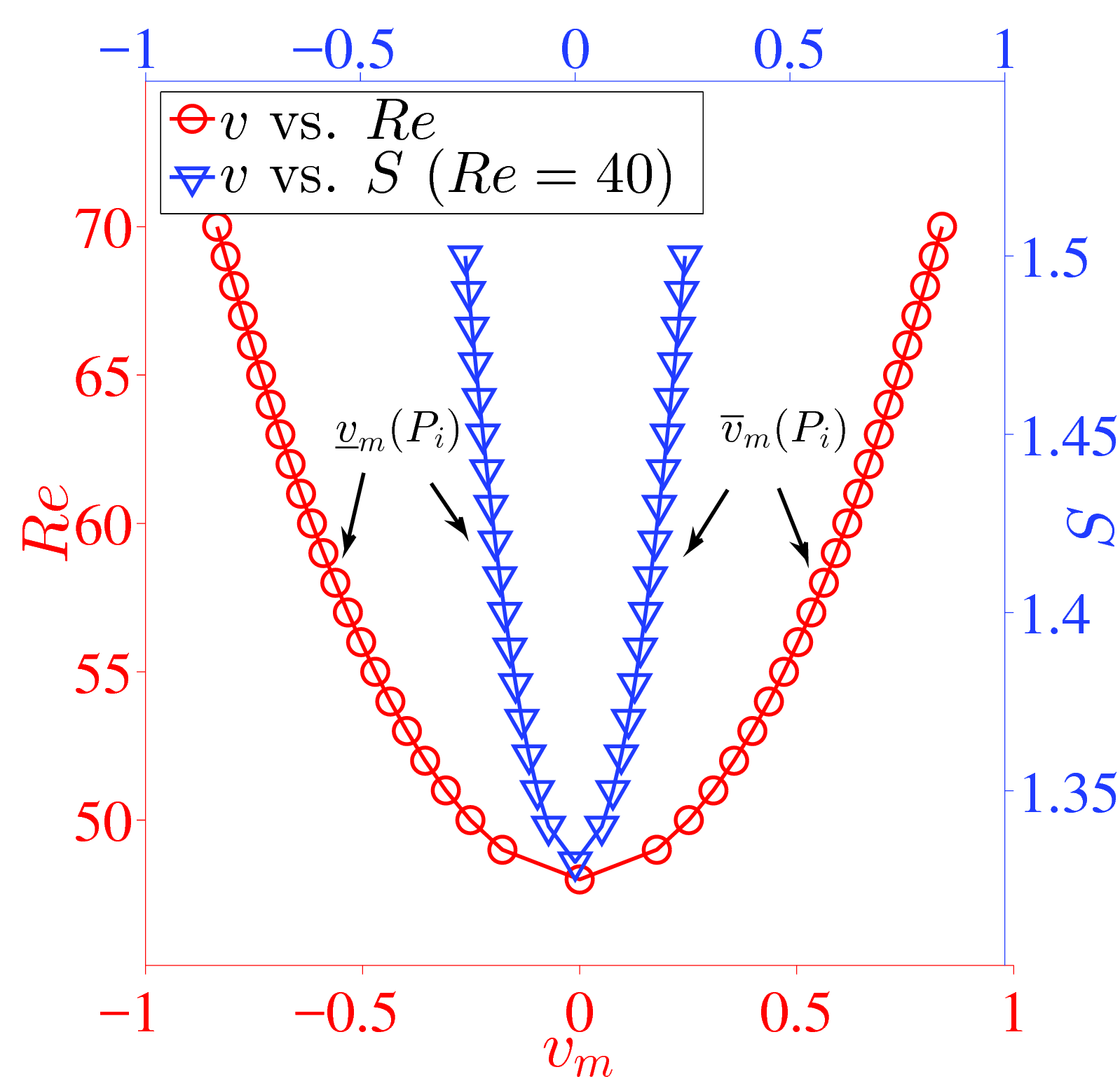}
		\caption{$\qquad$}
		\label{fig:RES_Hopf}
	\end{subfigure}
	\caption{Stability analysis of  wake around a saddle-point region behind in 
        the SSBS arrangements: 
        (a) $|\mu|$ vs. $Re$ and $|\mu|$ vs. $S$; 
        (b) bifurcation diagram at $Re=40$ for $v$ vs. $Re$  and $v$ vs. $S$, 
        where $v$ is an averaged peak value ($tU/D \in [300,$ $400]$) of 
        $y$-component velocity at the location ($2D$, $0D$) behind a cylinder ($0D$, $0D$).}
	\label{fig:MuHopf}
\end{figure}

To analyze the dependency of the near-wake instability on Reynolds number, the DMD mode which is account for the vortex shedding is identified in a saddle-point region in figure~\ref{fig:RES_Mu}. At smaller $Re$ and $S$ values, the unstable modes are supposed to decays as the fluid flow develops, since $|\mu| < 1$. Here $\mu = \mu_r + j \mu_i$ is the eigenvalue of the decomposed primary vortex mode using DMD technique, where $\mu_r$, $\mu_i$ and $j$ are its real parts, imaginary parts and imaginary unit respectively. The $|\mu|$ is its corresponding magnitude of $\mu$. As the values of $Re$ and $S$ increase, this primary vortex mode reaches an equilibrium state $|\mu| = 1$. Similar to the stability analysis in \cite{Mizushima2008JoFM}, a bifurcation diagram of y-component velocity at ($2D$, $0D$) is plotted in figure~\ref{fig:RES_Hopf}. The manifested fluid instabilities from both $Re$ and $S$ variation are identified as \emph{Hopf bifurcation}. This conclusion is based on the correlation $(v_+(P_i)-v_-(P_i)) \propto (P_i-P_s)^{0.5}$, where $P_i$ and $P_s$ are the stability parameter, e.g., $Re$ and $S$, and its critical value respectively. Here $v_+(P_i)$ and $v_-(P_i)$ are respectively the time-averaged maximum and minimum of y-component velocity $v(P_i)$. Owing to the imbalanced fluid shearing $S \neq 1$, the curve of original $v$ value is asymmetric about $v=0$ in the stability analysis of $S$. To present a symmetric curve, the time-averaged maximum and minimum $v$ values are modulated as $v_{m}$ and plotted in figure~\ref{fig:RES_Hopf}, where $\overline{v}_m(P_i) = 0.5(v_+(P_i) - v_-(P_i))$ and $\underline{v}_m(P_i) = -0.5(v_+(P_i) - v_-(P_i))$. Observing from figures~\ref{fig:RES_Mu} and~\ref{fig:RES_Hopf}, the onset of vortex shedding occurs exactly at $Re \approx 48$ as $Re$ value increases. Its modal frequency and vortex pattern at saturated state conform with those of vortex shedding documented in numerical and experimental investigations of a stationary isolated circular cylinder. Furthermore it also shows a similar saturation of the velocity oscillation at $S \approx 1.33$. 
\begin{figure}
	\begin{center}
		\centering
		\includegraphics[trim=0.1cm 0.1cm 0.1cm 0.1cm,scale=0.105,clip]{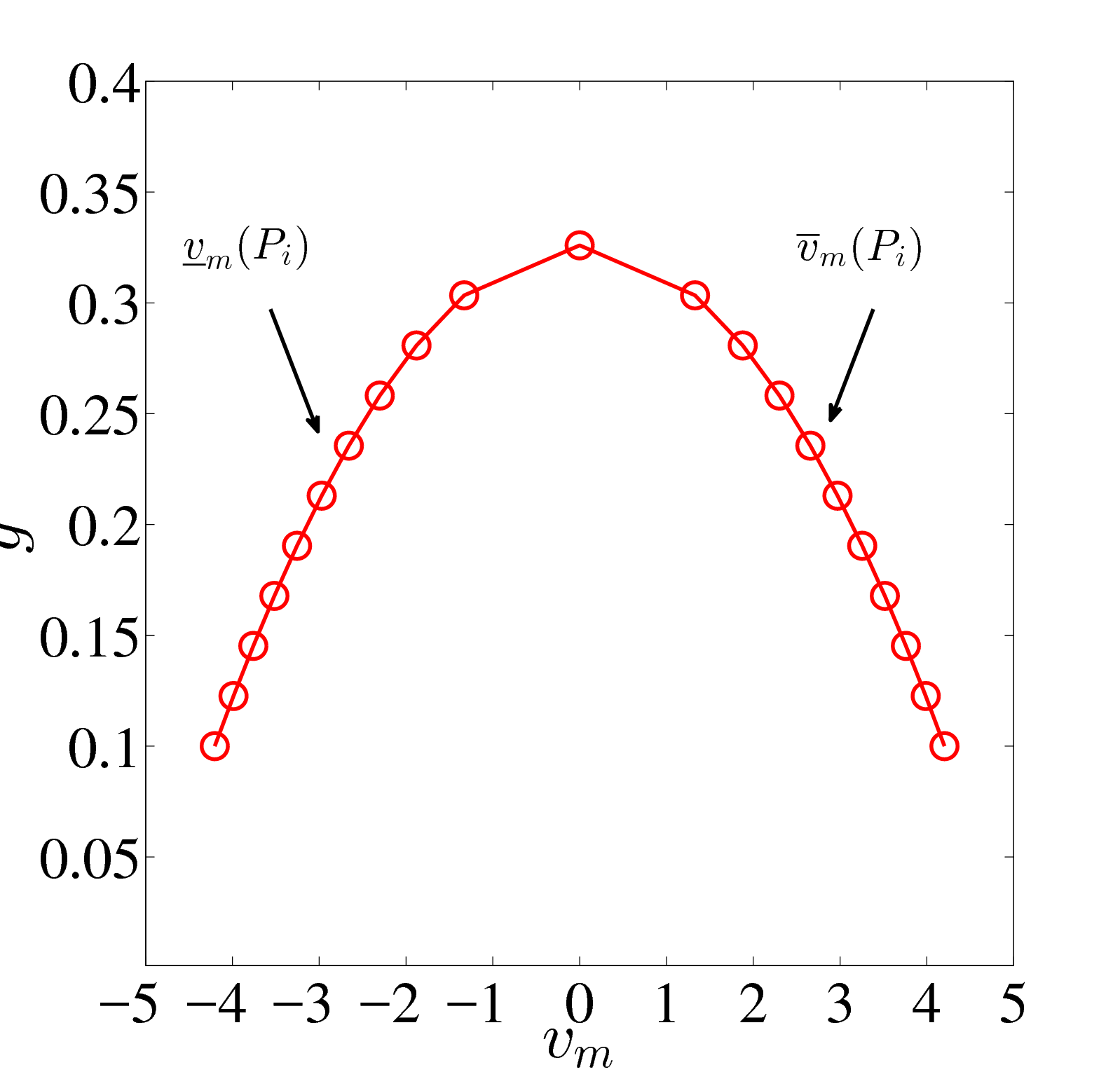}
	\end{center}
	\caption{Bifurcation diagram of $v$ with respect to the gap ratio $g^*$ at $Re=45$, $g^* \in [0.1$, $0.33]$, where $v$ is an averaged peak value ($tU/D \in [300,$ $400]$) of y-component velocity at location ($2D$, $0D$) behind a cylinder ($0D$, $0D$).}
	\label{fig:g_v}
\end{figure}

The stability analysis results indicate that the near-wake instability is not only significantly correlated with the fluid momentum, but also the fluid shearing ratio. This conclusion agrees with the observation of an earlier flow transition induced from a planar shear flow reported by \cite{lankadasu2008onset}, in which the planar shear was reported with an enhancement effect on the flow transition. As cylinders of the SSBS arrangement are brought close enough, the gap flow interference becomes remarkable. Similarly the DMD technique is used to investigate the correlation between the gap ratio $g^*$ and the near-wake stability. Here figure~\ref{fig:g_v} shows that the Hopf bifurcation starts to develop as $g^* \lesssim 0.33$. It means the gap-flow proximity interference is critical to the near-wake stability in the SSBS arrangements as well, as $g^*$ value decreases.  
To summarize, the fluid shearing ratio $S$, the fluid momentum intensity $Re$ and the gap-flow proximity interference $g^*$ are significant to the near-wake stability. In multi-body systems, e.g., SBS arrangements, either or all of these parameters could be influential at a particular fluid domain. To exceed their corresponding critical values are destructive to the near-wake stability.  
\end{document}